\documentclass{JHEP3}
\usepackage{graphicx}
\usepackage{amsmath}
\usepackage{psfrag}
\usepackage[sort&compress,numbers]{natbib}

\def\eps{\epsilon}

\def\Poles{{\cal P}oles}

\def\e{\epsilon}
\def\d{\hbox{d}}

\def\ln{\hbox{ln}}

\preprint{
ZU--TH 07/10\\  IPPP/10/26\\ SI-HEP-2010-09}

\title{Calculation of the quark and gluon form factors to three loops in QCD}

\author{T.\  Gehrmann$^a$, E.W.N. Glover$^b$, T.\ Huber$^{c}$,
   N.\ Ikizlerli$^b$, C.\ Studerus$^a$
	\\
$^a$ Institut f\"ur Theoretische Physik, Universit\"at Z\"urich, Winterthurerstrasse 190, CH-8057 Z\"urich, 
Switzerland\\
	$^b$Institute for Particle Physics Phenomenology, University of Durham,
South Road,\\ Durham DH1 3LE, England\\
$^c$ Fachbereich 7, Universit\"at Siegen, Walter-Flex-Strasse 3, D-57068 Siegen, Germany
}

\abstract{We describe the calculation of the three-loop QCD corrections to quark and gluon form factors. 
The relevant three-loop Feynman diagrams are evaluated and the resulting three-loop Feynman integrals 
are reduced to a small set of known master integrals by using integration-by-parts relations. 
Our calculation confirms the recent results by Baikov et al.\ for the three-loop form factors. In addition, we 
derive the subleading ${\cal O}(\e)$ terms for the fermion-loop type contributions to 
the three-loop form factors which are required for the extraction of the fermionic contributions to the four-loop 
quark and gluon collinear anomalous dimensions. The finite parts of the form factors are used to determine
the hard matching coefficients for the Drell-Yan process and inclusive Higgs-production in soft-collinear 
effective theory. }

\begin{document}

\section{Introduction}
The form factors are basic vertex functions, and are as such fundamental ingredients for many precision 
calculations in QCD. They couple an external, colour-neutral off-shell current to a pair of partons: 
the quark form factor is the coupling of a virtual photon to a quark-antiquark pair, while the gluon
form factor is the coupling of a Higgs boson to a pair of gluons through an effective Lagrangian. 
They appear as virtual higher-order corrections in coefficient functions for the inclusive Drell-Yan 
process~\cite{dy1l,dy2l,hk1} 
and the inclusive Higgs production cross section~\cite{hk1,higgs1l,higgsfull,higgs2l}. In these observables, the
infrared poles of the form factors
cancel with infrared singularities from real radiation corrections. Consequently, it is possible to 
relate the coefficients of the infrared poles of the form factors to the coefficients of large 
logarithmic terms in the corresponding real radiation processes~\cite{magnea1,mv}. A framework for combining 
the resummation of logarithmically enhanced terms at all orders with fixed-order results is 
provided in an effective field theory expansion~\cite{cusp2} of QCD, which is systematized by  
 soft-collinear effective theory~\cite{scet1}. 
In this context, the pole terms of the form factors 
yield the anomalous dimensions of the effective operators, while their finite terms determine 
the matching coefficients to a given order~\cite{Idilbi:2005ky,Idilbi:2006dg,Ahrens:2008nc}. 

The form factors are actually the simplest QCD objects that display a non-trivial infrared pole structure. 
As such, their infrared pole coefficients can be used to 
extract fundamental constants: the cusp anomalous dimensions~\cite{cusp1} 
which control the structure of soft divergences and the collinear quark and gluon anomalous dimensions. While the cusp anomalous dimensions were 
first obtained to three loops from 
the asymptotic behaviour of splitting 
functions~\cite{Moch:2004pa,Vogt:2004mw}, it is 
 the calculation~\cite{MMV1,MMV2} of the 
pole terms of the three-loop form factors 
(and finite plus subleading terms in the 
two-loop and one-loop form factors~\cite{vn,harlander,GHM}), which 
led to the derivation of the three-loop collinear 
anomalous dimensions~\cite{MMV1,Becher:2006mr,Becher:2009qa}. 
 An important observation is the agreement 
(up to an overall colour factor) of the cusp anomalous dimension for the quark and gluon, the so-called 
Casimir scaling~\cite{casimir}. Casimir scaling has been verified to 
three loops \cite{Moch:2004pa,Vogt:2004mw}, but 
it is an open question whether it holds at four loops and beyond \cite{Alday:2007mf}.
From non-perturbative arguments, the Casimir scaling is expected to break down at some 
loop order~\cite{ft}. 

Based on the observation that infrared singularities of massless on-shell amplitudes in QCD are related to 
ultraviolet singularities of operators in soft-collinear effective field theory~\cite{cusp1,Becher:2009cu}, the 
pole structure of these amplitudes can be analyzed using operator renormalization. 
The singularity structure of 
arbitrary multi-leg massless QCD amplitudes is determined by an anomalous dimension matrix. 
The terms allowed in this anomalous dimension matrix are strongly constrained by relations between 
soft and collinear terms, from non-abelian exponentiation and from soft and collinear factorization. 
Independently, Becher and Neubert \cite{Becher:2009qa}  and
Gardi and Magnea \cite{Gardi:2009qi} have proposed a 
remarkable all-loops conjecture that describes 
the pole structure of massless on-shell multi-loop multi-leg 
QCD amplitudes (generalizing earlier results at two~\cite{catani} and three loops~\cite{sterman})
in terms of the cusp anomalous dimensions and the collinear anomalous dimensions.
In this conjecture, the colour matrix structure of the soft
anomalous dimension generated by soft gluons is simply a sum over two-body interactions between hard partons, and thus
the matrix structure at any loop order is the same as at one loop. 
This result builds on the earlier work of Refs.~\cite{Aybat:2006mz,Aybat:2006wq} which showed  
the colour matrix structure of the soft
anomalous dimension at two loops is identical
to that at one loop.
There may be additional colour correlations at three loops or beyond, which cannot be excluded at present~\cite{Dixon:2009ur}.
However 
strong arguments for the absence 
of these terms are given in Refs.~\cite{Becher:2009qa}. 
If the all-order conjecture \cite{Gardi:2009qi,Becher:2009qa} holds, the calculation of 
the pole parts of the form factors to a given loop order 
(and of the finite and subleading parts at fewer loops) would be sufficient to determine the 
infrared poles of all massless on-shell QCD amplitudes to this order.

The calculation of the three-loop form factors requires two principal ingredients: the 
algebraic reduction of 
all three-loop integrals appearing in the relevant Feynman diagrams to master integrals, and the 
analytical calculation of these master integrals. The reduction of integrals to master integrals exploits 
linear relations among different integrals, and is done based on a lexicographic ordering of the 
integrals (the Laporta algorithm~\cite{laporta}). Several dedicated computer-algebra implementations 
of the Laporta algorithm are available~\cite{laporta,air,fire,reduze}. 
The reduction of the integrals relevant to the three-loop form 
factors is among the most challenging applications of the Laporta algorithm to 
date: due to the 
very large number of interconnected integrals to be reduced, the linear systems to be solved are 
often containing tens of thousand equations with a similar number of unknowns. 

The master integrals in the three-loop form factors were identified already several years ago~\cite{masterA}. 
Their analytical 
calculation proved to be a major computational challenge, which was completed only in several steps.
The one-loop bubble insertions into two-loop vertex integrals
as well as the two-loop bubble insertions into one-loop vertex integrals were derived using standard 
Feynman parameter integrals~\cite{masterA}, while the genuine three-loop integrals required 
an extensive use of Mellin-Barnes integration techniques~\cite{masterB,masterC,masterD}. 

A first calculation of the three-loop form factors (based in part on numerical results for some of the 
expansion coefficients of the master integrals) was accomplished by Baikov et al.~\cite{BCSSS} 
in 2009. The analytical calculation of the last remaining 
master integrals was only completed recently~\cite{masterD}. It is the purpose of this paper 
to validate the three-loop form factor 
results of Ref.~\cite{BCSSS,masterD} by an independent calculation, and to extend them in 
in part to a higher order in the expansion in the dimensional regularization parameter
$\e=2-d/2$. These further expansion terms will be needed for an extraction of the quark and gluon collinear 
anomalous dimensions from the single pole pieces of the four-loop form factors. 

We define the quark and gluon form factors in Section~\ref{sec:renorm}, where we also discuss their 
UV-renomalization and summarize existing results at one- and two-loops. The reduction of the 
form factors to master integrals is described in Section~\ref{sec:red}, and the three-loop 
master integrals are discussed in Section~\ref{sec:masters}. Explicit 
analytical expressions for them are collected in 
Appendix~\ref{app:mi}.  Our results for the three-loop form factors are 
presented in Section~\ref{sec:ff}, and supplemented by 
Appendix~\ref{app:coeff}. The infrared structure of the QCD form factors up to 
four-loops is analyzed in Section~\ref{sec:ir}. The three-loop hard matching coefficients for Drell-Yan and 
Higgs production in soft-collinear effective theory are determined from the form factors 
in Section~\ref{sec:match}. An outlook on future applications is contained in 
Section~\ref{sec:conc}.

\section{Quark and gluon form factors in perturbative QCD}
\label{sec:renorm}

The form factors are the basic vertex functions of an external off-shell current 
(with virtuality $q^2 = s_{12}$) coupling to a pair of partons
with on-shell momenta $p_1$ and $p_2$. One distinguishes time-like ($s_{12}>0$, i.e.\
with partons both either in the initial or in the final state) and 
space-like ($s_{12}<0$, i.e.\ with one parton in the initial and one in the final state) configurations. 
The form factors are described in terms of scalar functions by contracting the respective vertex functions 
(evaluated in dimensional regularization with $d=4-2\e$ dimensions) with 
projectors.  For massless partons, the full vertex function is described with only a single form factor. 

The quark form factor is obtained from the photon-quark-antiquark vertex $\Gamma^\mu_{q\bar q}$ by
\begin{equation}
{\cal F}^q  = -\frac{1}{4(1-\e)q^2}\, {\mathrm Tr} \left( p_2 \!\!\!\! / \, \Gamma^\mu_{q\bar q} p_1 \!\!\!\! / \, \gamma_\mu\right)\, , \label{eq:projq}
\end{equation}
while the gluon form factor relates to the effective Higgs-gluon-gluon vertex $\Gamma^{\mu \nu}_{gg}$ as
\begin{equation}
{\cal F}^g = \frac{p_1\cdot p_2 \, g_{\mu\nu} - p_{1,\mu} p_{2,\nu} - p_{1,\nu} p_{2,\mu} }{2 (1-\e)} \, 
\Gamma_{gg}^{\mu \nu}\, . \label{eq:projg}
\end{equation}
The form factors are expanded in perturbative QCD in powers of the coupling constant, with each power 
corresponding to a virtual loop. We denote the unrenormalized form factors by ${\cal F}^a$ and the renormalized 
form factors by $F^a$ with $a=q,g$. 

At tree level, the Higgs boson does not couple either to the gluon or to massless 
quarks. In higher orders in perturbation theory, heavy quark loops introduce
a coupling between the Higgs boson and gluons. In the limit of infinitely 
massive quarks, these loops give rise to an effective Lagrangian~\cite{hgg} 
mediating the 
coupling between the scalar Higgs field and the gluon field strength tensor:
\begin{equation}
{\cal L}_{{\rm int}} = -\frac{\lambda}{4} H F_a^{\mu\nu} F_{a,\mu\nu}\ .
\label{eq:lagr}
\end{equation}
The coupling $\lambda$ has inverse mass dimension. It can be computed 
by matching~\cite{kniehl,kniehl2} 
the effective theory to the full standard 
model cross sections~\cite{higgsfull}.

Evaluation of the Feynman diagrams, contributing to the vertex functions at a given loop order yields the bare 
(unrenormalised) form factors,
\begin{eqnarray}
{\cal F}_b^q (\alpha_s^b, s_{12}) &=& 1 + \sum_{n=1}^{\infty} \left( \frac{\alpha_s^b}{4\pi}\right)^n \left(\frac{-s_{12}}{\mu_0^2}\right)^{-n\eps} S_{\eps}^n \, {\cal F}_n^q,\\
{\cal F}_b^g (\alpha_s^b, s_{12}) &=& \lambda^b\left(1 + \sum_{n=1}^{\infty} \left( \frac{\alpha_s^b}{4\pi}\right)^n \left(\frac{-s_{12}}{\mu_0^2}\right)^{-n\eps} 
S_{\eps}^n \,{\cal F}_n^g\right),
\end{eqnarray}
where $\mu_0^2$ is the mass parameter introduced in dimensional regularisation to maintain a dimensionless coupling in the bare Lagrangian density and where
\begin{equation}
S_{\eps} = e^{-\eps \gamma} (4\pi)^{\eps}, \qquad \qquad {\rm with~the~Euler~constant~} \gamma =  0.5772\ldots
\end{equation}

The renormalization of the form factor is carried out by replacing 
the bare coupling $\alpha^{b}$ with the renormalized coupling 
$\alpha_s\equiv \alpha_s(\mu^2)$ evaluated at the renormalization scale $\mu^2$
\begin{equation}
\alpha_s^b \mu_0^{2\epsilon} = Z_{\alpha_s} \mu^{2\epsilon}  \alpha_s(\mu^2).
\end{equation}
For simplicity we set $\mu^2 = |s_{12}|$ so that in the $\overline{{\rm MS}}$ scheme \cite{msbar},
\begin{eqnarray}
Z_{\alpha_s} &=& S_{\eps}^{-1}\Bigg[  
1- \frac{\beta_0}{\e}\left(\frac{\alpha_s}{4\pi}\right) 
+\left(\frac{\beta_0^2}{\e^2}-\frac{\beta_1}{2\e}\right)
\left(\frac{\alpha_s}{4\pi}\right)^2\nonumber \\
&& \hspace{1cm}
-\left(\frac{\beta_0^3}{\e^3}-\frac{7}{6}\frac{\beta_1\beta_0}{\e^2}+\frac{1}{3}\frac{\beta_2}{\e}\right)\left(\frac{\alpha_s}{4\pi}\right)^3+{\cal O}(\alpha_s^4) \Bigg] \; ,
\end{eqnarray}
where $\beta_0$, $\beta_1$ and $\beta_2$ are \cite{beta0,beta1,beta2}
\begin{eqnarray}
\beta_0 &=& 
\frac{11 C_A}{3}-\frac{2 N_F}{3},\\
\beta_1 &=& 
\frac{34 C_A^2}{3}-\frac{10 C_A N_F}{3}-2 C_F N_F,\\
\beta_2 &=& 
\frac{2857 C_A^3}{54}+C_F^2 N_F-\frac{205 C_F C_A N_F}{18}-\frac{1415 C_A^2 N_F}{54}+\frac{11 C_F N_F^2}{9}+\frac{79 C_A N_F^2}{54}.
\end{eqnarray}
The renormalization relation for the effective coupling $\lambda^b$ 
in the $\overline{{\rm MS}}$ scheme
is given by,
\begin{equation}
\lambda^b = Z_{\lambda} \lambda
\end{equation}
with
\begin{eqnarray}
Z_{\lambda}&=& 
1- \frac{\beta_0}{\e}\left(\frac{\alpha_s}{4\pi}\right) 
+\left(\frac{\beta_0^2}{\e^2}-\frac{\beta_1}{\e}\right)
\left(\frac{\alpha_s}{4\pi}\right)^2
\nonumber \\
&& \hspace{1cm}
-\left(\frac{\beta_0^3}{\e^3}-\frac{2\beta_1\beta_0}{\e^2}+\frac{\beta_2}{\e}\right)\left(\frac{\alpha_s}{4\pi}\right)^3+{\cal O}(\alpha_s^4)  \; .
\end{eqnarray}

The $i$-loop contribution to the unrenormalized coefficients is   
${\cal F}_i^a$, while the renormalised coefficient is denoted by
$F_i^a$ where $a=q,g$.    If $s_{12}$ is space-like, the form factors are real, while they acquire imaginary parts for time-like $s_{12}$.  These imaginary parts (and corresponding real parts) arise from the $\epsilon$-expansion of
\begin{equation}
\Delta(s_{12}) = (-{\rm sgn}(s_{12})-i0)^{-\epsilon} 
\end{equation}
so that the renormalized form factors are given by,
\begin{eqnarray}
{F}^q (\alpha_s(\mu^2), s_{12},\mu^2= |s_{12}|) &=& 1 + \sum_{n=1}^{\infty} \left( \frac{\alpha_s (\mu^2)}{4\pi}\right)^n    \, {F}_n^q,\\
{F}^g (\alpha_s(\mu^2), s_{12},\mu^2= |s_{12}|) &=& \lambda \left(1 + \sum_{n=1}^{\infty} \left( \frac{\alpha_s(\mu^2)}{4\pi}\right)^n   
  \,{F}_n^g\right).
\end{eqnarray}

Up to three loops, the renormalized coefficients for the quark form factor (with $\mu^2 = |s_{12}|$) are then obtained as,
\begin{eqnarray}
F_1^q  &=& {\cal F}_1^q \Delta(s_{12}), \nonumber \\
F_2^q   &=& 
{\cal F}_2^q \left(\Delta(s_{12})\right)^2
-\frac{\beta_0}{\e} {\cal F}_1^q \Delta(s_{12}) ,  \nonumber \\
F_3^q  &=& 
{\cal F}_3^q \left(\Delta(s_{12})\right)^3
-\frac{2\beta_0}{\e}
{\cal F}_2^q  \left(\Delta(s_{12})\right)^2
-\left(\frac{\beta_1}{2\e}-\frac{\beta_0^2}{\e^2}\right)
{\cal F}_1^q \Delta(s_{12}),
\label{eq:renq}
\end{eqnarray}
while those for the gluon form factor are given by,
\begin{eqnarray}
F_1^g   &=& 
 {\cal F}_1^g \Delta(s_{12})
-\frac{\beta_0}{\e}   ,  \nonumber \\
F_2^g  &=& 
 {\cal F}_2^g \left(\Delta(s_{12})\right)^2
-\frac{2\beta_0}{\e} 
{\cal F}_1^g  \Delta(s_{12})
-\left(\frac{\beta_1}{\e}-\frac{\beta_0^2}{\e^2}\right),\nonumber \\
F_3^g  &=& 
 {\cal F}_3^g \left(\Delta(s_{12})\right)^3
-\frac{3\beta_0}{\e} 
{\cal F}_2^g  \left(\Delta(s_{12})\right)^2
-\left(\frac{3\beta_1}{2\e}-\frac{3\beta_0^2}{\e^2}\right)
{\cal F}_1^g\Delta(s_{12})
-\left(\frac{\beta_2}{\e}
-\frac{2\beta_1\beta_0}{\e^2}
+\frac{\beta_0^3}{\e^3}\right).\nonumber \\
\label{eq:reng}
\end{eqnarray}
Unless explicitly stated otherwise, the renormalized form factors are given in the space-like case in the 
following sections. 

The one-loop and two-loop form factors were computed in many places in the
literature~\cite{vn,harlander,GHM,MMV1,MMV2}. 
All-order expressions 
in terms of one-loop and two-loop master integrals are given 
in~\cite{GHM}, and are summarized below. 

\subsection{Results at one-loop}

Written in terms of the one-loop bubble integral, which is normalized to 
the factor 
\begin{equation}
 S_{\Gamma} = \frac{(4\pi)^\e}{16\pi^2\Gamma(1-\e)},
\label{eq:sgamma}
\end{equation}
 the unrenormalised
one-loop form factors are given by
\begin{eqnarray}
\label{eq:f1q}
{\cal F}_1^q/S_R =
&&C_F\,B_{2,1}
\left(\frac{4}{(D-4)}+D-3\right)\,,\\
\label{eq:f1g}
{\cal F}_1^g/S_R =
&&C_A\,B_{2,1}\left(
\frac{4}{(D-4)}-\frac{4}{(D-2)}+10-D\right)\,,
\end{eqnarray}
where
\begin{equation}
 S_{R} = \frac{16\pi^2 S_{\Gamma}}{S_{\epsilon}} = \frac{\exp(\epsilon \gamma)}{\Gamma(1-\epsilon)}.
\label{eq:sR}
\end{equation}
Eqs. \eqref{eq:f1q} and \eqref{eq:f1g} agree with eqs.~(8) and (9) of ref.~\cite{GHM} respectively.

Inserting the expansion of the one-loop master integrals and keeping terms through to ${\cal O}(\eps^5)$, we find that
{\allowdisplaybreaks
\begin{eqnarray}
{\cal F}_1^q = C_F\Biggl[
&&-\frac{2}{\eps^2}
-\frac{3}{\eps}
+\left(\zeta_2-8\right)
+\eps\left(\frac{3\zeta_2}{2}+\frac{14\zeta_3}{3}-16\right)
+\eps^2\left(\frac{47\zeta_2^2}{20}+4\zeta_2+7\zeta_3-32\right)\nonumber \\
&&+\eps^3\left(\frac{141\zeta_2^2}{40}-\frac{7\zeta_2\zeta_3}{3}+8\zeta_2+\frac{56\zeta_3}{3}+\frac{62\zeta_5}{5}-64\right)\nonumber \\
&&+\eps^4\left(\frac{949\zeta_2^3}{280}+\frac{47\zeta_2^2}{5}-\frac{7\zeta_2\zeta_3}
{2}-\frac{49\zeta_3^2}{9}+16\zeta_2+\frac{112\zeta_3}{3}+\frac{93\zeta_5}{5}-128\right)\nonumber \\
&&+\eps^5\bigg(
\frac{2847\zeta_2^3}{560}
+\frac{94\zeta_2^2}{5}
-\frac{329\zeta_2^2\zeta_3}{60}
-\frac{28\zeta_2\zeta_3}{3}
-\frac{31\zeta_2\zeta_5}{5}
-\frac{49\zeta_3^2}{6}\nonumber \\
&&\hspace{1cm}
+32\zeta_2
+\frac{224\zeta_3}{3}
+\frac{248\zeta_5}{5}
+\frac{254\zeta_7}{7}
-256\bigg)
\Biggr]\,, \\
{\cal F}_1^g = C_A\Biggl[
&&-\frac{2}{\eps^2}
+\zeta_2
+\eps\left(\frac{14\zeta_3}{3}-2\right)
+\eps^2\left(\frac{47\zeta_2^2}{20}-6\right)\nonumber \\
&&
+\eps^3\left(-\frac{7\zeta_2\zeta_3}{3}+\zeta_2+\frac{62\zeta_5}{5}-14\right)
+\eps^4\left(\frac{949\zeta_2^3}{280}-\frac{49\zeta_3^2}{9}+3\zeta_2+\frac{14\zeta_3}{3}-30\right)\nonumber \\
&&+\eps^5\bigg(
\frac{47\zeta_2^2}{20}
-\frac{329\zeta_2^2\zeta_3}{60}
-\frac{31\zeta_2\zeta_5}{5}
+7\zeta_2
+14\zeta_3 
+\frac{254\zeta_7}{7}
-62\bigg)
\Biggr]
\end{eqnarray}}
where the gluon form factor agrees with eq.~(7) of ref.~\cite{MMV2} through to ${\cal O}(\eps^4)$. Note that at each order in $\epsilon$, the terms of highest harmonic weight are the same for both quark and gluon form-factor.  This is guaranteed by the equivalence of the coefficient of the leading pole in eqs.~\eqref{eq:f1q} and ~\eqref{eq:f1g}. 

\subsection{Results at two-loops}

Written in terms of the two-loop master integrals (listed in the appendix),
  the unrenormalised
two-loop gluon form factor is given by
{\allowdisplaybreaks
\begin{eqnarray}
{\cal F}_2^q/S_R^2 = C_F^2 \Biggl[
&&B_{4,2}
\left(\frac{16}{(D-4)^2}+\frac{8}{(D-4)}+D^2-6D+17\right)\nonumber \\
&&-C_{4,1}
\left(\frac{7D^2}{8}-\frac{983D}{48}-\frac{565}{32(2D-7)}-\frac{20}{9(3D-8)}-\frac{28}{(D-4)}\right.\nonumber \\
&&\left.\qquad-\frac{40}{(D-4)^2}+\frac{10693}{288}\right)\nonumber \\
&&+ B_{3,1} 
\left(\frac{27D^2}{8}-\frac{1293D}{16}+\frac{3955}{32(2D-7)}-\frac{17}{2(D-3)}-\frac{476}{(D-4)}\right.\nonumber \\
&&\left.\qquad-\frac{456}{(D-4)^2}-\frac{288}{(D-4)^3}+\frac{581}{32}\right)\nonumber \\
&&- C_{6,2}
\frac{D^3-20D^2+104D-176}{8(2D-7)} 
\Biggr]\nonumber \\
+C_FC_A\Biggl[
&&-C_{4,1}
\left(\frac{D^2}{16}+\frac{77D}{32}+\frac{565}{64(2D-7)}+\frac{12}{5(3D-8)}+\frac{23}{15(D-1)}\right.\nonumber \\
&&\left.\qquad+\frac{8}{3(D-4)}+\frac{16}{(D-4)^2}+\frac{163}{64}\right)\nonumber \\
&&- B_{3,1} 
\left(\frac{75D^2}{16}-\frac{1837D}{32}+\frac{3955}{64(2D-7)}+\frac{3}{4(D-3)}-\frac{186}{(D-4)}\right.\nonumber \\
&&\left.\qquad-\frac{144}{(D-4)^2}-\frac{96}{(D-4)^3}+\frac{3845}{64}\right)\nonumber \\
&&+C_{6,2}
\frac{D^3-20D^2+104D-176}{16(2D-7)}
\Biggr]\nonumber \\
+C_FN_F\Biggl[
&&-C_{4,1}
\frac{(D-2)(3D^3-31D^2+110D-128)}{(3D-8)(D-4)(D-1)} 
\Biggr] 
\end{eqnarray}
}

{\allowdisplaybreaks
\begin{eqnarray}
{\cal F}_2^g/S_R^2 = C_A^2 \Biggl[
&&B_{4,2}
\left(D^2-20D-\frac{48}{(D-2)}+\frac{32}{(D-4)}+\frac{16}{(D-2)^2}\right.\nonumber \\
&&\left.\qquad+\frac{16}{(D-4)^2}+100\right)\nonumber \\
&&+C_{4,1}
\left(\frac{27D}{2}+\frac{119}{48(2D-5)}+\frac{75}{16(2D-7)}+\frac{10}{3(D-1)}+\frac{80}{(D-2)}\right.\nonumber \\
&&\left.\qquad+\frac{103}{3(D-4)}-\frac{32}{(D-2)^2}+\frac{24}{(D-4)^2}-\frac{609}{8}\right)\nonumber \\
&&+B_{3,1}
\left(24D+\frac{107}{144(2D-5)}+\frac{525}{16(2D-7)}+\frac{116}{9(D-1)}+\frac{96}{(D-2)}\right.\nonumber \\
&&\left.\qquad-\frac{2}{(D-3)}-\frac{1175}{3(D-4)}-\frac{32}{(D-2)^2}-\frac{1388}{3(D-4)^2}-\frac{192}{(D-4)^3}\right.\nonumber \\
&&\left.\qquad-\frac{1955}{8}\right)\nonumber \\
&&+C_{6,2}
\frac{3(3D-8)(D-3)}{4(2D-5)(2D-7)} 
\Biggr]\nonumber \\
+C_AN_F\Biggl[
&& C_{4,1}
\left(\frac{7D}{8}+\frac{119}{12(2D-5)}+\frac{35}{48(2D-7)}+\frac{20}{3(D-1)}-\frac{40}{3(D-2)}\right.\nonumber \\
&&\left.\qquad-\frac{2}{(D-4)}-\frac{45}{16}\right)\nonumber \\
&&-B_{3,1}
\left(\frac{19D}{8}-\frac{107}{36(2D-5)}-\frac{245}{48(2D-7)}-\frac{232}{9(D-1)}+\frac{40}{3(D-2)}\right.\nonumber \\
&&\left.\qquad-\frac{3}{2(D-3)}+\frac{8}{9(D-4)}-\frac{8}{(D-4)^2}-\frac{61}{16}\right)\nonumber \\
&&+C_{6,2}
\frac{(2D^3-25D^2+94D-112)(D-4)}{8(D-2)(2D-5)(2D-7)} 
\Biggr]\nonumber \\
+C_FN_F\Biggl[
&&-C_{4,1}
\frac{(46D^4-545D^3+2395D^2-4606D+3248)(D-6)}{2(2D-7)(2D-5)(D-4)(D-2)}\nonumber \\
&&+B_{3,1}
\left(\frac{35D}{4}-\frac{107}{18(2D-5)}-\frac{245}{24(2D-7)}+\frac{8}{3(D-2)}-\frac{1}{(D-3)}\right.\nonumber \\
&&\left.\qquad-\frac{448}{9(D-4)}-\frac{112}{3(D-4)^2}-\frac{333}{8}\right)\nonumber \\
&&-C_{6,2}
\frac{(2D^3-25D^2+94D-112)(D-4)}{4(D-2)(2D-5)(2D-7)} 
\Biggr]
\end{eqnarray}
}
which, after re-expressing in terms of $N$ and $N_F$ agrees with eqs.~(10) and (11) of ref.~\cite{GHM}.

Inserting the expansion of the two-loop master integrals and keeping terms through to ${\cal O}(\eps^3)$, we find that
{\allowdisplaybreaks
\begin{eqnarray}
{\cal F}_2^q = C_F^2 \Biggl[
&&\frac{2}{\eps^4}
+\frac{6}{\eps^3}
-\frac{1}{\eps^2}\left(2\zeta_2-\frac{41}{2}\right)
-\frac{1}{\eps}\left(\frac{64\zeta_3}{3}-\frac{221}{4}\right)\nonumber \\
&&-\left(13\zeta_2^2-\frac{17\zeta_2}{2}+58\zeta_3-\frac{1151}{8}\right)\nonumber \\
&&-\eps\left(\frac{171\zeta_2^2}{5}-\frac{112\zeta_2\zeta_3}{3}-\frac{213\zeta_2}{4}+\frac{839\zeta_3}{3}+\frac{184\zeta_5}{5}
-\frac{5741}{16}\right)\nonumber \\
&&+\eps^2\left(\frac{223\zeta_2^3}{5}-\frac{3401\zeta_2^2}{20}+54\zeta_2\zeta_3+\frac{2608\zeta_3^2}{9}+\frac{1839\zeta_2}{8}\right.\nonumber \\
&&\left.\qquad-\frac{6989\zeta_3}{6}-\frac{462\zeta_5}{5}+\frac{27911}{32}\right) \nonumber \\
&&+\eps^3\left(\frac{768\zeta_2^3}{7}+\frac{5488\zeta_2^2\zeta_3}{15}-\frac{29157\zeta_2^2}{40}+\frac{757\zeta_2\zeta_3}{3}+\frac{184\zeta_2\zeta_5}{5}
+\frac{2434\zeta_3^2}{3}\right.\nonumber \\
&&\left.\qquad+\frac{13773\zeta_2}{16}-\frac{58283\zeta_3}{12}-\frac{3251\zeta_5}{5}+\frac{8942\zeta_7}{7}+\frac{133781}{64}\right) 
\Biggr]\nonumber \\
+C_FC_A\Biggl[
&&-\frac{11}{6\eps^3}
+\frac{1}{\eps^2}\left(\zeta_2-\frac{83}{9}\right)
-\frac{1}{\eps}\left(\frac{11\zeta_2}{6}-13\zeta_3+\frac{4129}{108}\right)\nonumber \\
&&+\left(\frac{44\zeta_2^2}{5}-\frac{119\zeta_2}{9}+\frac{467\zeta_3}{9}-\frac{89173}{648}\right)\nonumber \\
&&+\eps\left(\frac{1891\zeta_2^2}{60}-\frac{89\zeta_2\zeta_3}{3}-\frac{6505\zeta_2}{108}+\frac{6586\zeta_3}{27}+51\zeta_5-\frac{1775893}{3888}\right)\nonumber \\
&&-\eps^2\left(\frac{809\zeta_2^3}{70}-\frac{2639\zeta_2^2}{18}+\frac{397\zeta_2\zeta_3}{9}+\frac{569\zeta_3^2}{3}+\frac{146197\zeta_2}{648}\right.\nonumber \\
&&\left.\qquad-\frac{159949\zeta_3}{162}-\frac{3491\zeta_5}{15}+\frac{33912061}{23328}\right) \nonumber \\
&&+\eps^3\left(
\frac{3817\zeta_2^3}{140}
-\frac{7103\zeta_2^2\zeta_3}{30}
+\frac{638441\zeta_2^2}{1080}
-\frac{4358\zeta_2\zeta_3}{27}
-\frac{497\zeta_2\zeta_5}{5}
-\frac{16439\zeta_3^2}{27}\right.\nonumber \\
&&\left.\qquad
-\frac{2996725\zeta_2}{3888}
+\frac{3709777\zeta_3}{972}
+\frac{49786\zeta_5}{45}
-372\zeta_7 
-\frac{632412901}{139968}\right) 
\Biggr]\nonumber \\
+C_FN_F\Biggl[
&&\frac{1}{3\eps^3}
+\frac{14}{9\eps^2}
+\frac{1}{\eps}\left(\frac{\zeta_2}{3}+\frac{353}{54}\right)
+\left(\frac{14\zeta_2}{9}-\frac{26\zeta_3}{9}+\frac{7541}{324}\right)\nonumber \\
&&-\eps\left(\frac{41\zeta_2^2}{30}-\frac{353\zeta_2}{54}+\frac{364\zeta_3}{27}-\frac{150125}{1944}\right)\nonumber \\
&&-\eps^2\left(\frac{287\zeta_2^2}{45}+\frac{26\zeta_2\zeta_3}{9}-\frac{7541\zeta_2}{324}+\frac{4589\zeta_3}{81}+\frac{242\zeta_5}{15}-\frac{2877653}{11664}\right) \nonumber \\
&&+\eps^3\left(
-\frac{127\zeta_2^3}{14}
-\frac{14473\zeta_2^2}{540}
-\frac{364\zeta_2\zeta_3}{27}
+\frac{338\zeta_3^2}{27}\right.\nonumber \\
&&\left.\qquad
+\frac{150125\zeta_2}{1944}
-\frac{98033\zeta_3}{486}
-\frac{3388\zeta_5}{45}
+\frac{53933309}{69984}\right) 
\Biggr]\,,
\end{eqnarray}
}which agrees through to ${\cal O}(\eps^2)$ with eq.~(3.6) of ref.~\cite{MMV1} and provides the next term in the expansion.

Similarly we find that the two-loop expansion of the gluon form factor is given by
{\allowdisplaybreaks
\begin{eqnarray}
{\cal F}_2^g = C_A^2 \Biggl[
&&\frac{2}{\eps^4}
-\frac{11}{6\eps^3}
-\frac{1}{\eps^2}\left(\zeta_2+\frac{67}{18}\right)
+\frac{1}{\eps}\left(\frac{11\zeta_2}{2}-\frac{25\zeta_3}{3}+\frac{68}{27}\right)\nonumber \\
&&-\left(\frac{21\zeta_2^2}{5}-\frac{67\zeta_2}{6}-\frac{11\zeta_3}{9}-\frac{5861}{162}\right)\nonumber \\
&&-\eps\left(\frac{77\zeta_2^2}{60}-\frac{23\zeta_2\zeta_3}{3}-\frac{106\zeta_2}{9}+\frac{1139\zeta_3}{27}-\frac{71\zeta_5}{5}
-\frac{158201}{972}\right)\nonumber \\
&&+\eps^2\left(\frac{2313\zeta_2^3}{70}-\frac{1943\zeta_2^2}{60}-\frac{55\zeta_2\zeta_3}{3}+\frac{901\zeta_3^2}{9}+\frac{481\zeta_2}{54}\right.\nonumber \\
&&\left.\qquad-\frac{26218\zeta_3}{81}+\frac{341\zeta_5}{15}+\frac{3484193}{5832}\right) \nonumber \\
&&+\eps^3\left(
\frac{2057\zeta_2^3}{60}
+\frac{1291\zeta_2^2\zeta_3}{10}
-\frac{28826\zeta_2^2}{135}
+\frac{335\zeta_2\zeta_3}{9}
-\frac{313\zeta_2\zeta_5}{5}
+\frac{5137\zeta_3^2}{27}\right.\nonumber \\
&&\left.\qquad
-\frac{4019\zeta_2}{324}
-\frac{397460\zeta_3}{243}
-\frac{5963\zeta_5}{45}
+\frac{6338\zeta_7}{7} 
+\frac{70647113}{34992}\right) 
\Biggr]\nonumber \\
+C_AN_F\Biggl[
&&\frac{1}{3\eps^3}
+\frac{5}{9\eps^2}
-\frac{1}{\eps}\left(\zeta_2+\frac{26}{27}\right)
-\left(\frac{5\zeta_2}{3}+\frac{74\zeta_3}{9}+\frac{808}{81}\right)\nonumber \\
&&-\eps\left(\frac{51\zeta_2^2}{10}+\frac{16\zeta_2}{9}+\frac{604\zeta_3}{27}+\frac{23131}{486}\right)\nonumber \\
&&-\eps^2\left(\frac{257\zeta_2^2}{18}-\frac{50\zeta_2\zeta_3}{3}-\frac{28\zeta_2}{27}+\frac{3962\zeta_3}{81}+\frac{542\zeta_5}{15}
+\frac{540805}{2916}\right) \nonumber \\
&&+\eps^3\left(
-\frac{253\zeta_2^3}{210}
-\frac{103\zeta_2^2}{3}
+\frac{380\zeta_2\zeta_3}{9}
+\frac{2306\zeta_3^2}{27}\right.\nonumber \\
&&\left.\qquad
+\frac{3157\zeta_2}{162}
-\frac{30568\zeta_3}{243}
-\frac{854\zeta_5}{9} 
-\frac{11511241}{17496}\right)
\Biggr]\nonumber \\
+C_FN_F\Biggl[
&&-\frac{1}{\eps}
+\left(8\zeta_3-\frac{67}{6}\right)
+\eps\left(+\frac{16\zeta_2^2}{3}+\frac{7\zeta_2}{3}+\frac{92\zeta_3}{3}-\frac{2027}{36}\right)\nonumber \\
&&+\eps^2\left(\frac{184\zeta_2^2}{9}-\frac{40\zeta_2\zeta_3}{3}+\frac{209\zeta_2}{18}+\frac{1124\zeta_3}{9}+32\zeta_5
-\frac{47491}{216}\right) \nonumber \\
&&+\eps^3\left(
-\frac{176\zeta_2^3}{35}
+\frac{22147\zeta_2^2}{270}
-\frac{460\zeta_2\zeta_3}{9}
-120\zeta_3^2 \right.\nonumber \\
&&\left.\qquad
+\frac{4273\zeta_2}{108}
+\frac{15284\zeta_3}{27}
+\frac{368\zeta_5}{3} 
-\frac{987995}{1296}\right)
\Biggr]\,,
\end{eqnarray}
}which agrees through to ${\cal O}(\eps^2)$ with eq.~(8) of ref.~\cite{MMV2} and provides the next term in the expansion.
Expressions for the renormalized one-loop and two-loop form factors, expanded 
to the appropriate order in $\e$, can be found in~\cite{GHM}. 

\section{Calculation of the three-loop form factors}
\label{sec:red}
To compute the three-loop quark and gluon form factors, we evaluate the relevant three-loop vertex 
functions within dimensional regularisation. 
At this loop order, there are 244 Feynman diagrams contributing to the quark 
form factor, and 1586 diagrams contributing to the gluon form factor. 
We generated these diagrams using QGRAF~\cite{qgraf}. After contraction with 
the projectors~(\ref{eq:projq})--(\ref{eq:projg}), each diagram can be expressed as a linear 
combination of (typically hundreds of) scalar three-loop Feynman integrals. 
The three-loop integrals appearing in the form factors have up to nine
different propagators. The integrands can depend on the three loop momenta, 
and the two on-shell external momenta, such that 12 different scalar
products involving loop momenta can be formed. Consequently, not all 
scalar products can be cancelled against combinations of denominators, and we are left with irreducible scalar 
products in the numerator of the integrand. We denote the  number of different propagators in an 
integral by $t$, the total number of propagators by $r$ and the total number of irreducible scalar products 
by $s$. The topology of each integral is fixed by specifying the set of $t$ different propagators and subtopologies 
are obtained by removing one or more of the propagators. 

Using relations between different integrals based on integration-by-parts (IBP)~\cite{chet1} and Lorentz 
invariance (LI)~\cite{gr}, one can express the large number of different integrals in terms 
of a small number of so-called master integrals. These identities yield large linear systems of
equations, which are solved in an iterative manner using lexicographic ordering~\cite{laporta}.
To carry out the reduction in 
a systematic manner, we introduce so-called auxiliary topologies. Each auxiliary
topology is a set of 12 linearly independent propagators. Within the 
auxiliary topology, the integrand of a three-loop form factor integral 
with $(r,s,t)$ is expressed by $r$ propagators (with exactly $t$ different propagators) in the 
denominator, and $s$ propagators (with at most 12-$t$ different propagators) in the numerator. 
All three-loop form factor integrals can be cast into one of three auxiliary topologies, which are listed in 
Table~\ref{tab:auxtopo}. The first auxiliary topology contains planar integrals only. 
\begin{table}[b]
\begin{center}
\begin{tabular}{lll}
AuxTopo 1\hspace{4mm} & AuxTopo 2\hspace{4mm} & AuxTopo3 \\[2mm]
$k_1^2$      & $k_1^2$ & $k_1^2$ \\
$k_2^2$      & $k_2^2$ & $k_2^2$ \\
$k_3^2$      & $k_3^2$ & $k_3^2$ \\
$(k_1-k_2)^2$ & $(k_1-k_2)^2$ & $(k_1-k_2)^2$ \\
$(k_1-k_3)^2$ & $(k_1-k_3)^2$ & $(k_1-k_3)^2$ \\
$(k_2-k_3)^2$ & $(k_2-k_3)^2$ & $(k_1-k_2-k_3)^2$ \\
$(k_1-p_1)^2$ & $(k_1-k_3-p_2)^2$ & $(k_1-p_1)^2$ \\
$(k_1-p_1-p_2)^2$ & $(k_1-p_1-p_2)^2$ & $(k_1-p_1-p_2)^2$ \\
$(k_2-p_1)^2$ & $(k_2-p_1)^2$ & $(k_2-p_1)^2$ \\
$(k_2-p_1-p_2)^2$ & $(k_1-k_2-p_2)^2$ & $(k_2-p_1-p_2)^2$ \\
$(k_3-p_1)^2$ & $(k_3-p_1)^2$ & $(k_3-p_1)^2$ \\
$(k_3-p_1-p_2)^2$ & $(k_3-p_1-p_2)^2$ & $(k_3-p_1-p_2)^2$ 
\end{tabular}
\end{center}
\caption{Propagators in the three different auxiliary topologies used to represent all three-loop form 
factor integrals.}
\label{tab:auxtopo}
\end{table}

Three-loop integrals with $4\leq t \leq 9$ and $t\leq r \leq 9$ appear in the form factors. These come with 
up to $s=4$ irreducible scalar products for the quark form factor and up to $s=5$ for the gluon form factor.
For a fixed topology and given $(r,s,t)$, there are in total
$$
N_{r,s,t} = \left( {{r-1}\atop {t-1}} \right) \, \left( {{11-t+s}\atop {s}} \right)
$$ 
different integrals. 

 To obtain a reduction, one has to solve very large systems of 
equations. Already for $s\leq 4$, the system for a given auxiliary topology contains $900000$
equations, and its solution is feasible only with dedicated computer algebra tools. For this reduction, we used 
the Mathematica-based package FIRE~\cite{fire} and the C++ package Reduze~\cite{reduze}, which was 
developed most recently by one of us. 

With Reduze, the reduction and its performance are as follows. The topologies with more than 4 propagators are 
reduced after inserting the results of the sub-topologies into the system. 
With increasing $t$ the number of equations decrease as (in general) does the time taken to solve the system 
which is in the range of a few days to less than an hour with the program Reduze on a 
modern desktop computer. 
The total computing time for all the planar diagrams is more than 2 months. However,
the parallelization of topologies with an equal number of propagators 
reduced the overall reduction time to a few weeks.

The three-loop form factors contain in total 22 master integrals, of which 14 are genuine three-loop vertex 
functions, 4 are three-loop propagator integrals and 4 are products of one-loop and two-loop integrals. They are described in detail in the following section.

\section{Three-loop form factor master integrals}
\label{sec:masters}
\begin{figure}[t]
\psfrag{b21}{$B_{2,1}$}
\psfrag{b31}{$B_{3,1}$}
\psfrag{b42}{$B_{4,2}$}
\psfrag{c41}{$C_{4,1}$}
\psfrag{c62}{$C_{6,2}$}
	\begin{center}
		\includegraphics[height=2cm]{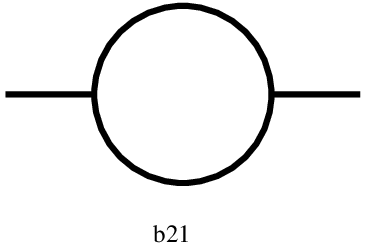}
		\hspace{1cm}
		\includegraphics[height=2cm]{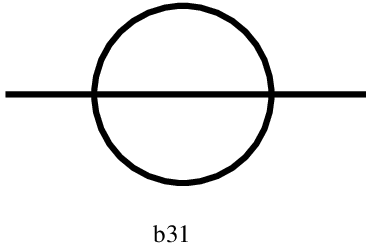}
		\hspace{1cm}
		\includegraphics[height=2cm]{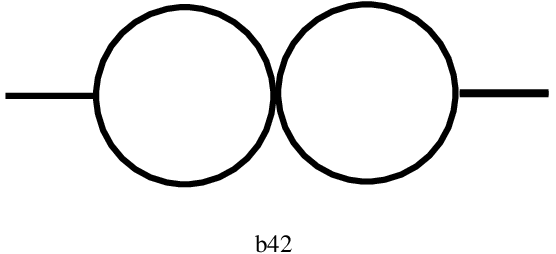}
	\end{center}	 
	\vspace{1cm}
	\begin{center}
		\includegraphics[height=2cm]{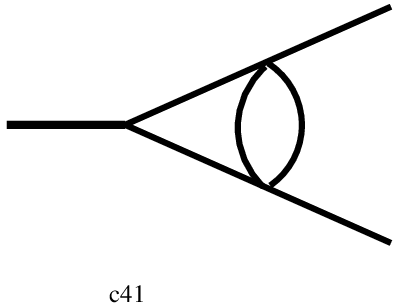}
		\hspace{1cm}
		\includegraphics[height=2cm]{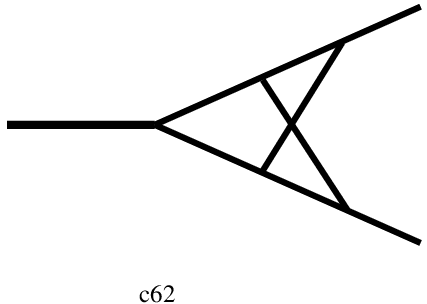}
	\end{center}
\caption{One and two-loop master integrals appearing in the quark and gluon form factors.}
\label{fig:mi1l2l}
\end{figure}
\begin{figure}[t]
\psfrag{b41}{$B_{4,1}$}
\psfrag{b51}{$B_{5,1}$}
\psfrag{b52}{$B_{5,2}$}
\psfrag{b61}{$B_{6,1}$}
\psfrag{b62}{$B_{6,2}$}
\psfrag{b81}{$B_{8,1}$}
\psfrag{c61}{$C_{6,1}$}
\psfrag{c81}{$C_{8,1}$}
	\begin{center}
		\includegraphics[height=2cm]{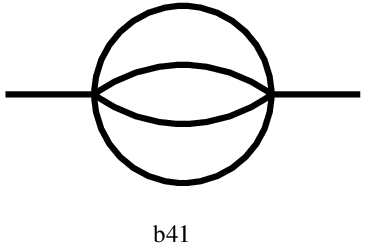}
		\hspace{1cm}
		\includegraphics[height=2cm]{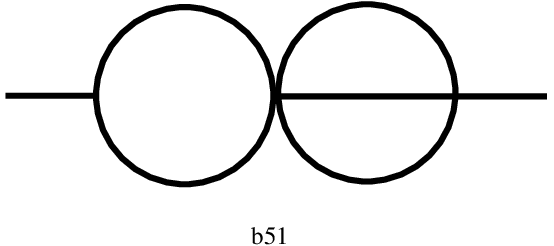}
		\hspace{1cm}
		\includegraphics[height=2cm]{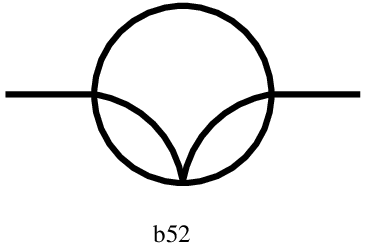}
	\end{center}	 
	\vspace{1cm}
	\begin{center}
		\includegraphics[height=2cm]{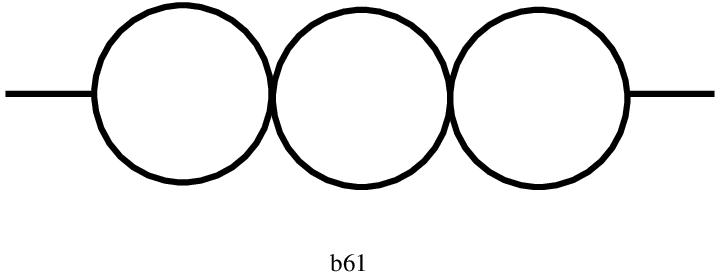}
		\hspace{1cm}
		\includegraphics[height=2cm]{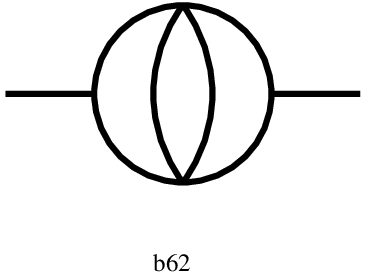}
		\hspace{1cm}
		\includegraphics[height=2cm]{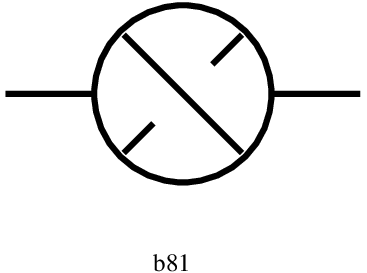}
	\end{center}
	\vspace{1cm}
		 
	\begin{center}
		\includegraphics[height=2cm]{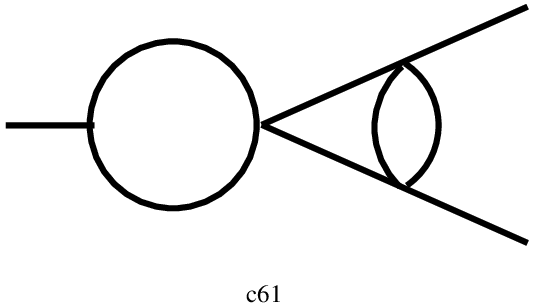}
		\hspace{1cm}
		\includegraphics[height=2cm]{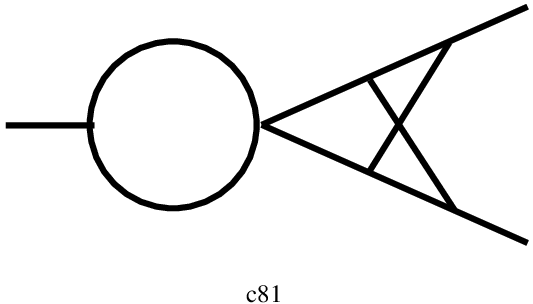}
	\end{center}	 
\caption{Three-loop two-point and factorizable three-point integrals.}
\label{fig:mi3lb}
\end{figure}

Our notation for the master integrals follows~\cite{masterA}, and we distinguish three topological 
types of master integrals: genuine three-loop triangles ($A_{t,i}$-type), 
bubble integrals  ($B_{t,i}$-type) and integrals that contain  two-loop 
triangles   ($C_{t,i}$-type). In this notation, the index $t$ denotes the number of propagators, and $i$ is 
simply enumerating the topologically different integrals with the same number of propagators.

\begin{figure}[t]
\psfrag{a51}{$A_{5,1}$}
\psfrag{a52}{$A_{5,2}$}
\psfrag{a61}{$A_{6,1}$}
\psfrag{a62}{$A_{6,2}$}
\psfrag{a63}{$A_{6,3}$}
\psfrag{a71}{$A_{7,1}$}
\psfrag{a72}{$A_{7,2}$}
\psfrag{a73}{$A_{7,3}$}
\psfrag{a74}{$A_{7,4}$}
\psfrag{a75}{$A_{7,5}$}
\psfrag{a81}{$A_{8,1}$}
\psfrag{a91}{$A_{9,1}$}
\psfrag{a92}{$A_{9,2}$}
\psfrag{a94}{$A_{9,4}$}
	\begin{center}
		\includegraphics[height=2cm]{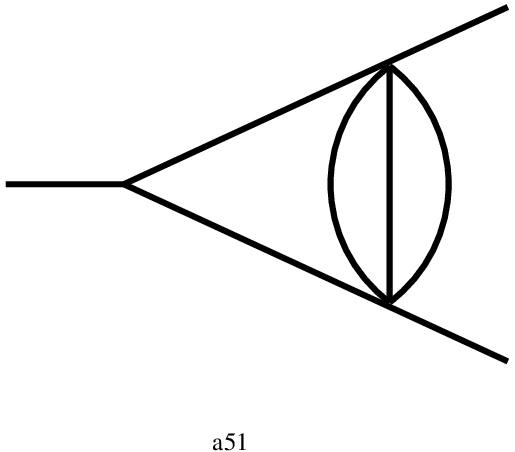}
		\hspace{1cm}
		\includegraphics[height=2cm]{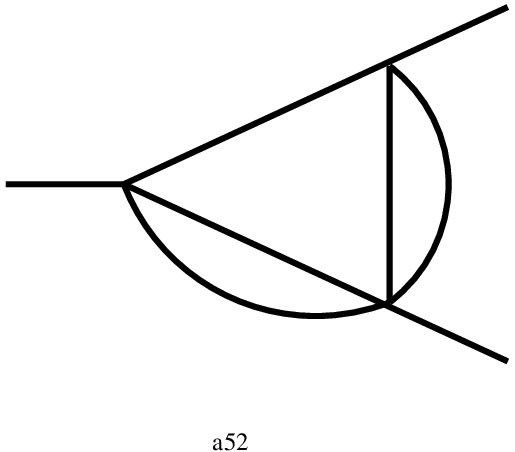}
		\hspace{1cm}
		\includegraphics[height=2cm]{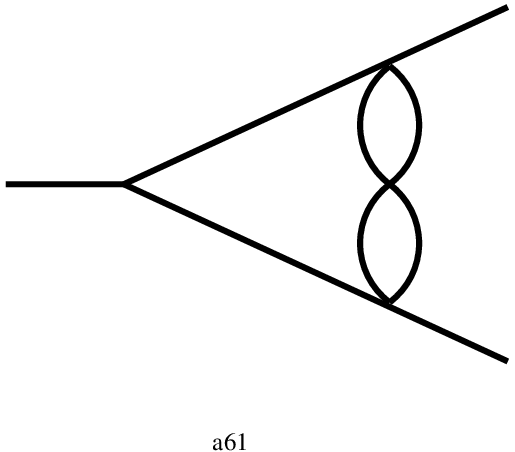}
		\hspace{1cm}
		\includegraphics[height=2cm]{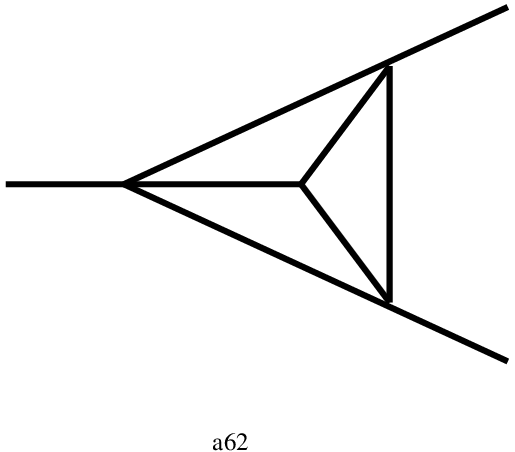}
	\end{center}	 
	\vspace{0.5cm}
	\begin{center}
		\hspace{1cm}
		\includegraphics[height=2.5cm]{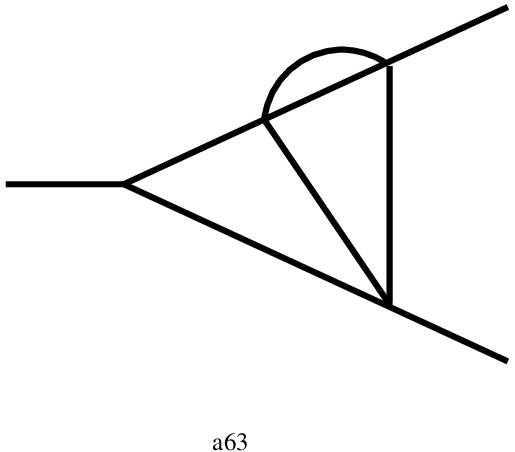}
		\hspace{1cm}
		\includegraphics[height=2.5cm]{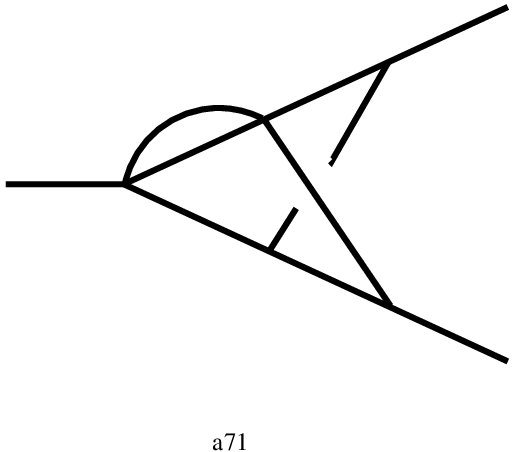}
		\hspace{1cm}
		\includegraphics[height=2.5cm]{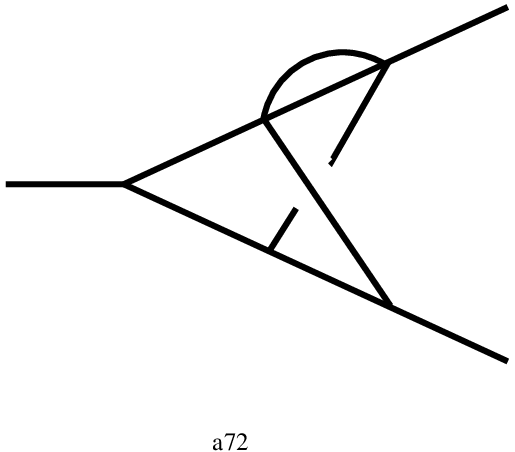}
	\end{center}	 
	\vspace{0.5cm}
	\begin{center}
		\includegraphics[height=2.5cm]{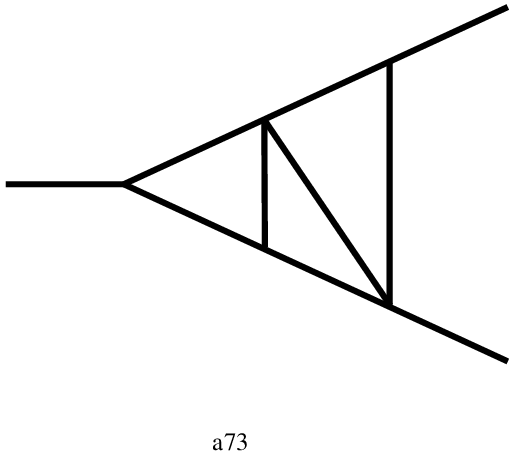}
		\hspace{1cm}
		\includegraphics[height=2.5cm]{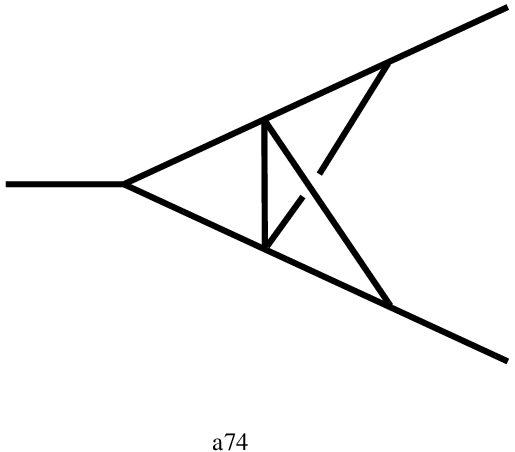}
		\hspace{1cm}
		\includegraphics[height=2.5cm]{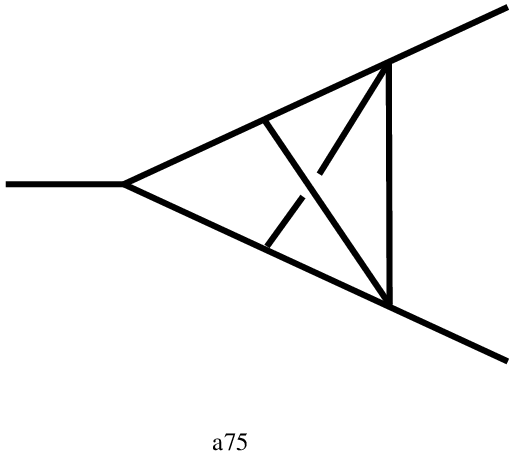}
	\end{center}	 
	\vspace{0.5cm}
	\begin{center}
		\includegraphics[height=2.5cm]{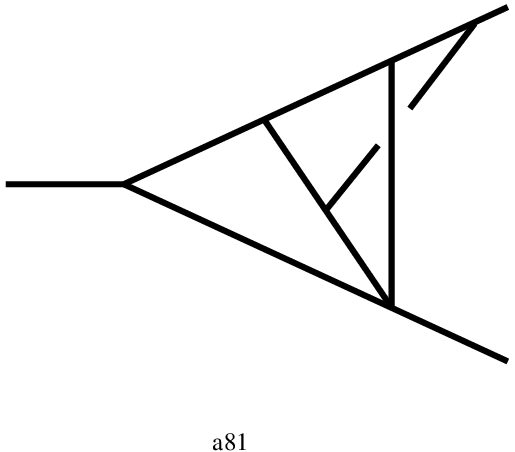}
		\hspace{1cm}
		\includegraphics[height=2.5cm]{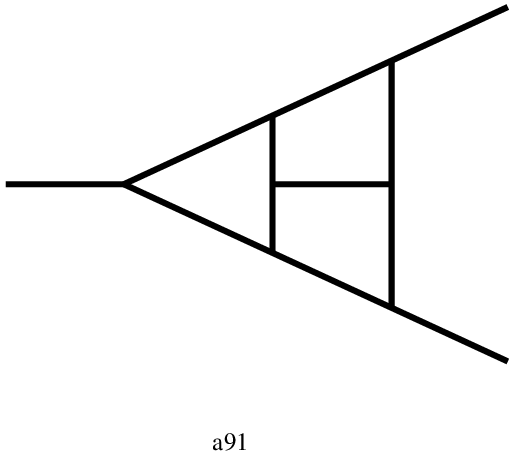}
	\end{center}	 
	\vspace{0.5cm}
	\begin{center}
		\includegraphics[height=2.5cm]{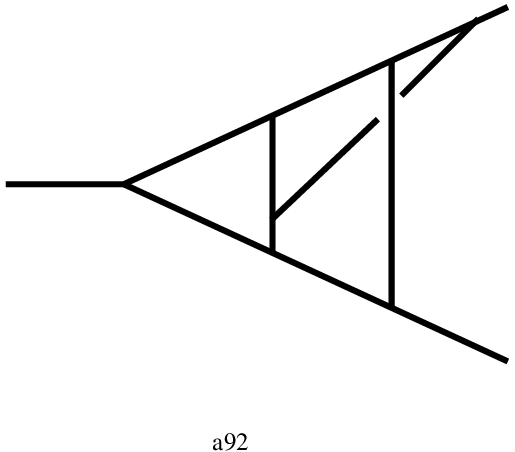}
		\hspace{1cm}
		\includegraphics[height=2.5cm]{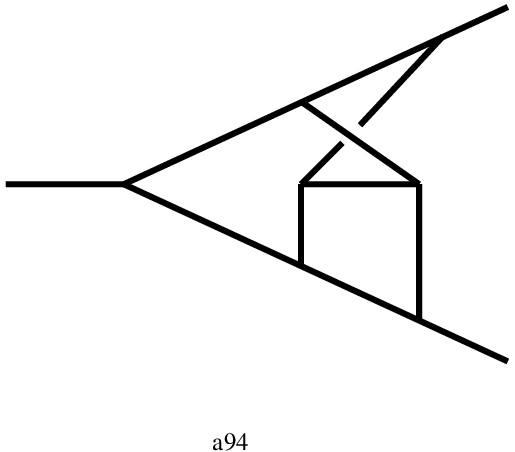}
	\end{center}	 
\caption{Three-point integrals listed in Refs.~\cite{masterA,masterB,masterC}.}
\label{fig:mi3la}
\end{figure}

The one-loop and two-loop master integrals appearing in the form factors at these loop orders 
are displayed in 
Figure~\ref{fig:mi1l2l}. Their expansions to finite order have been known for a long time, all-orders 
expressions were derived in~\cite{GHM}, they can for example be expanded using HypExp~\cite{HypExp}.
$B_{t,i}$-type and $C_{t,i}$-type three-loop integrals are listed in Figure~\ref{fig:mi3lb}. The $B_{t,i}$-type
integrals were computed to finite order in~\cite{chet1,mincer}, and supplemented by the higher order 
terms in~\cite{bekavac}. Finally, the 
genuine three-loop vertex integrals are shown in Figure~\ref{fig:mi3la}, their expansions to finite order 
were derived in~\cite{masterA,masterB,masterC,masterD}.

The calculation of the nine-line three-loop 
integrals was the last missing ingredient to the 
form factor calculation for a long time.  The full result for $A_{9,1}$ and most of the pole parts 
of $A_{9,2}$ and $A_{9,4}$ were computed analytically in~\cite{masterC}. Analytical expressions 
for the remaining pieces 
of the latter two integrals were subsequently obtained in~\cite{masterD}. In~\cite{masterC}, it was pointed out 
that for each of these three integrals one can find an integral from the same topology with an irreducible 
scalar product, which has homogeneous transcendentality. These integrals were named
$A_{9,1n}$, $A_{9,2n}$ and $A_{9,4n}$, and are defined in~\cite{masterC}.
Compared to~\cite{masterC} we increased the numerical precision of the remaining coefficients,
both for $A_{9,2}$ and $A_{9,4}$, by means of conventional packages like {\tt MB.m}~\cite{Czakon:2005rk}.
We reproduce thirteen significant digits of the analytic result of~\cite{masterD} in the case of $A_{9,2}$,
and fourteen in the case of $A_{9,4}$. We also converted our numerical 
results for these two integrals into the corresponding integrals of 
homogeneous transcendentality, $A_{9,2n}$ and $A_{9,4n}$. On the coefficients of 
these integrals, a  PSLQ~\cite{pslq} determination was attempted. For the 
 pole  coefficients, the PSLQ algorithm converged to a unique solution in 
agreement with~\cite{masterD}. For the 
finite coefficients, the numerical precision that we 
obtained is yet insufficient for PSLQ to yield a 
unique solution.  

An analytic result for $A_{9,2}$ and $A_{9,4}$, derived by purely analytic steps and without fitting
rational coefficients to numerical values, is still a desirable task, and remains to be investigated in
the future. This goal is definitely within reach in the case of $A_{9,4}$, whereas the situation is
less clear for $A_{9,2}$.

Expansions of all master integrals to the order 
in $\e$ where transcendentality six first appears 
are listed in the Appendix.

\section{Three-loop form factors}
\label{sec:ff}
The unrenormalised three-loop form factors can be decomposed into different colour 
structures as follows:
\begin{eqnarray}
{\cal F}_3^q/S_R^3 &=&  
C_F^3 ~X^q_{C_F^3}
+C_F^2C_A  ~X^q_{C_F^2C_A} 
+ C_FC_A^2  ~X^q_{C_FC_A^2} 
+ C_F^2 N_F  ~X^q_{C_F^2 N_F}
\nonumber \\
&&
+C_FC_A N_F ~X^q_{C_FC_A N_F}
+C_FN_F^2  ~X^q_{C_FN_F^2} + C_F N_{F,V}\left(\frac{N^2-4}{N}\right)
 ~X^q_{C_FN_{F,V}}
\label{eq:fq3lbare}
\end{eqnarray}
and
\begin{eqnarray}
{\cal F}_3^g/S_R^3 &=&  
  C_A^3 ~X^g_{C_A^3}
+ C_A^2N_F  ~X^g_{C_A^2N_F} 
+ C_AC_FN_F  ~X^g_{C_AC_FN_F} 
+ C_F^2 N_F  ~X^g_{C_F^2 N_F}
\nonumber \\
&&
+ C_A N_F^2 ~X^g_{C_A N_F^2}
+ C_F N_F^2 ~X^g_{C_F N_F^2}\;,
\label{eq:fg3lbare}
\end{eqnarray}
where the last term in the quark form factor 
is generated by graphs where the virtual gauge boson does 
not couple directly to the final-state quarks.  This 
contribution is denoted by $N_{F,V}$ and is proportional to the charge 
weighted sum of the quark flavours.  In the case of purely 
electromagnetic interactions, we find,
\begin{equation}
N_{F,\gamma} = \frac{\sum_q e_q}{e_q}. 
\end{equation}

The coefficient of each colour structure is a linear combination of 
master integrals, resulting from the reduction of the integrals appearing in 
the Feynman diagrams. All coefficients are listed in Appendix~\ref{app:coeff}.

Inserting the expansion of the three-loop master integrals and keeping terms through to ${\cal O}(\eps^0)$, we find that the three-loop coefficients are given by 
{\allowdisplaybreaks
\begin{eqnarray}
{\cal F}_3^q = 
 C_F^3 \Biggl[
&&-\frac{4}{3\eps^6}
-\frac{6}{\eps^5}
+\frac{1}{\eps^4}\left(2\zeta_2-25\right)
+\frac{1}{\eps^3}\left(-3\zeta_2+\frac{100\zeta_3}{3}-83\right)\nonumber \\
&&+\frac{1}{\eps^2}\left(\frac{213\zeta_2^2}{10}-\frac{77\zeta_2}{2}+138\zeta_3-\frac{515}{2}\right)\nonumber \\
&&+\frac{1}{\eps}\left(\frac{1461\zeta_2^2}{20}-\frac{214\zeta_2\zeta_3}{3}-\frac{467\zeta_2}{2}+\frac{2119\zeta_3}{3}+\frac{644\zeta_5}{5}-\frac{9073}{12}\right)\nonumber \\
&&+\left(
          -\frac{ 53675}{24}
          - \frac{13001\zeta_2}{12}
          + \frac{12743\zeta_2^2}{40}
          - \frac{9095\zeta_2^3}{252}
          + 2669 \zeta_3
          + 61\zeta_3\zeta_2 \right. \nonumber \\
&&    \left.    \qquad  - \frac{1826\zeta_3^2}{3}
          + \frac{4238\zeta_5}{5}
\right)
\Biggr]\nonumber \\
+C_F^2C_A \Biggl[
&&\frac{11}{3\eps^5}
+\frac{1}{\eps^4}\left(-2\zeta_2+\frac{431}{18}\right)
+\frac{1}{\eps^3}\left(-\frac{7\zeta_2}{6}-26\zeta_3+\frac{6415}{54}\right)\nonumber \\
&&+\frac{1}{\eps^2}\left(-\frac{83\zeta_2^2}{5}+\frac{1487\zeta_2}{36}-210\zeta_3+\frac{79277}{162}\right)\nonumber \\
&&+\frac{1}{\eps}\left(-\frac{9839\zeta_2^2}{72}+\frac{215\zeta_2\zeta_3}{3}+\frac{38623\zeta_2}{108}-\frac{6703\zeta_3}{6}-142\zeta_5+\frac{1773839}{972}\right)\nonumber \\
&&+\left(
           \frac{37684115}{5832}
          + \frac{664325\zeta_2}{324}
          - \frac{1265467\zeta_2^2}{2160}
          - \frac{18619\zeta_2^3}{1260} \right. \nonumber \\
&& \left.          \qquad
          - \frac{96715\zeta_3}{18}
          + \frac{46\zeta_2\zeta_3}{9}
          + \frac{1616\zeta_3^2}{3}
          - \frac{46594\zeta_5}{45}
\right) \Biggr]\nonumber \\
+C_FC_A^2 \Biggl[
&&-\frac{242}{81\eps^4}
+\frac{1}{\eps^3}\left(\frac{88\zeta_2}{27}-\frac{6521}{243}\right)
+\frac{1}{\eps^2}\left(-\frac{88\zeta_2^2}{45}-\frac{553\zeta_2}{81}+\frac{1672\zeta_3}{27}-\frac{40289}{243}\right)\nonumber \\
&&+\frac{1}{\eps}\left(\frac{802\zeta_2^2}{15}-\frac{88\zeta_2\zeta_3}{9}-\frac{68497\zeta_2}{486}+\frac{12106\zeta_3}{27}-\frac{136\zeta_5}{3}
-\frac{1870564}{2187}\right)\nonumber \\
&& +\left(
          - \frac{52268375}{13122}
          - \frac{767320\zeta_2}{729}
          + \frac{152059\zeta_2^2}{540}
          - \frac{6152\zeta_2^3}{189}\right. \nonumber \\
&& \left.        \qquad  + \frac{1341553\zeta_3}{486}
          - \frac{710\zeta_2\zeta_3}{9}
          - \frac{1136\zeta_3^2}{9}
          + \frac{2932\zeta_5}{9}
\right) 
\Biggr]\nonumber \\
+C_F^2N_F \Biggl[
&&-\frac{2}{3\eps^5}-\frac{37}{9\eps^4}+\frac{1}{\eps^3}\left(-\frac{\zeta_2}{3}-\frac{545}{27}\right)+\frac{1}{\eps^2}\left(-\frac{133\zeta_2}{18}+\frac{146\zeta_3}{9}-\frac{6499}{81}\right)\nonumber \\
&&+\frac{1}{\eps}\left(\frac{337\zeta_2^2}{36}-\frac{2849\zeta_2}{54}+\frac{2557\zeta_3}{27}-\frac{138865}{486}\right)\nonumber \\
&&+\left(\frac{8149\zeta_2^2}{216}-\frac{343\zeta_2\zeta_3}{9}-\frac{45235\zeta_2}{162}+\frac{51005\zeta_3}{81}+\frac{278\zeta_5}{45}-\frac{2732173}{2916}\right) 
\Biggr]\nonumber \\
+C_FC_AN_F \Biggl[
&&\frac{88}{81\eps^4}+\frac{1}{\eps^3}\left(-\frac{16\zeta_2}{27}+\frac{2254}{243}\right)+\frac{1}{\eps^2}\left(\frac{316\zeta_2}{81}-\frac{256\zeta_3}{27}+\frac{13679}{243}\right)\nonumber \\
&&+\frac{1}{\eps}\left(-\frac{44\zeta_2^2}{5}+\frac{11027\zeta_2}{243}-\frac{6436\zeta_3}{81}+\frac{623987}{2187}\right)\nonumber \\
&&+\left(-\frac{1093\zeta_2^2}{27}+\frac{368\zeta_2\zeta_3}{9}+\frac{442961\zeta_2}{1458}-\frac{45074\zeta_3}{81}-\frac{208\zeta_5}{3}+\frac{8560052}{6561}\right) 
\Biggr]\nonumber \\
+C_FN_F^2 \Biggl[
&&-\frac{8}{81\eps^4}-\frac{188}{243\eps^3}+\frac{1}{\eps^2}\left(-\frac{4\zeta_2}{9}-\frac{124}{27}\right)+\frac{1}{\eps}\left(-\frac{94\zeta_2}{27}+\frac{136\zeta_3}{81}-\frac{49900}{2187}\right)\nonumber \\
&&+\left(-\frac{83\zeta_2^2}{135}-\frac{62\zeta_2}{3}+\frac{3196\zeta_3}{243}-\frac{677716}{6561}\right) 
\Biggr]\nonumber \\
+ C_FN_{F,V}&&\left(\frac{N^2-4}{N}\right) 
\Biggl[
4-\frac{2\zeta_2^2}{5}+10\zeta_2+\frac{14\zeta_3}{3}-\frac{80\zeta_5}{3}
\Biggr] \;.
\end{eqnarray}

The pole contributions of ${\cal F}_3^q$ are given in eq.~(3.7) of ref.~\cite{MMV1} while the finite parts of the $N_F^2$, $C_AN_F$ and $C_FN_F$ contributions are given in eq.~(6) of ref.~\cite{MMV2}.  The finite $N_{F,V}$ contribution can be obtained from the $\delta(1-x)$ contribution to the $d^{abc}d_{abc}$ colour factor in eq.~(6.6) of ref.~\cite{MMV3}.
The remaining finite contributions are given in eqs.~(8) and (9) of ref.~\cite{BCSSS}.

Similarly, the expansion of the gluon form factor at three-loops is given by
\begin{eqnarray}
{\cal F}_3^g = 
 C_A^3 \Biggl[
&&-\frac{4}{3\eps^6}
+\frac{11}{3\eps^5}
+\frac{361}{81\eps^4}
+\frac{1}{\eps^3}\left(-\frac{517\zeta_2}{54}+\frac{22\zeta_3}{3}-\frac{3506}{243}\right)\nonumber \\
&&+\frac{1}{\eps^2}\left(\frac{247\zeta_2^2}{90}+\frac{481\zeta_2}{162}-\frac{209\zeta_3}{27}-\frac{17741}{243}\right)\nonumber \\
&&+\frac{1}{\eps}\left(-\frac{3751\zeta_2^2}{360}-\frac{85\zeta_2\zeta_3}{9}+\frac{20329\zeta_2}{243}+\frac{241\zeta_3}{9}-\frac{878\zeta_5}{15}
-\frac{145219}{2187}\right)\nonumber \\
&&+\left(
            \frac{14474131}{13122}
          + \frac{307057\zeta_2}{1458}
          + \frac{8459\zeta_2^2}{1080}
          - \frac{22523\zeta_2^3}{270}\right. \nonumber \\
&& \qquad \left.
          - \frac{68590\zeta_3}{243}
          + \frac{77\zeta_2\zeta_3}{18}
          - \frac{1766\zeta_3^2}{9}
          + \frac{20911\zeta_5}{45}
          \right) 
\Biggr]\nonumber \\
+C_A^2N_F\Biggl[
&&-\frac{2}{3\eps^5}
-\frac{2}{81\eps^4}
+\frac{1}{\eps^3}\left(\frac{47\zeta_2}{27}+\frac{1534}{243}\right)
+\frac{1}{\eps^2}\left(-\frac{425\zeta_2}{81}+\frac{518\zeta_3}{27}+\frac{4280}{243}\right)\nonumber \\
&&+\frac{1}{\eps}\left(\frac{2453\zeta_2^2}{180}-\frac{7561\zeta_2}{243}+\frac{1022\zeta_3}{81}-\frac{92449}{2187}\right)\nonumber \\
&&+\left(\frac{437\zeta_2^2}{60}-\frac{439\zeta_2\zeta_3}{9}-\frac{37868\zeta_2}{729}-\frac{754\zeta_3}{27}+\frac{3238\zeta_5}{45}
-\frac{10021313}{13122}\right) 
\Biggr]\nonumber \\
+C_AC_FN_F\Biggl[
&&\frac{20}{9\eps^3}
+\frac{1}{\eps^2}\left(-\frac{160\zeta_3}{9}+\frac{526}{27}\right)
+\frac{1}{\eps}\left(-\frac{176\zeta_2^2}{15}-\frac{22\zeta_2}{3}-\frac{224\zeta_3}{27}+\frac{2783}{81}\right)\nonumber \\
&&+\left(-\frac{16\zeta_2^2}{5}+48\zeta_2\zeta_3-\frac{41\zeta_2}{3}+\frac{11792\zeta_3}{81}+\frac{32\zeta_5}{9}
-\frac{155629}{486}\right) 
\Biggr]\nonumber \\
+C_F^2N_F\Biggl[
&&\frac{2}{3\eps}
+\left(\frac{296\zeta_3}{3}-160\zeta_5+\frac{304}{9}\right) 
\Biggr]\nonumber \\
+C_AN_F^2\Biggl[
&&-\frac{8}{81\eps^4}
-\frac{80}{243\eps^3}
+\frac{1}{\eps^2}\left(\frac{20\zeta_2}{27}+\frac{8}{9}\right)
+\frac{1}{\eps}\left(\frac{200\zeta_2}{81}+\frac{664\zeta_3}{81}+\frac{34097}{2187}\right)\nonumber \\
&&+\left(\frac{797\zeta_2^2}{135}+\frac{76\zeta_2}{27}+\frac{11824\zeta_3}{243}+\frac{1479109}{13122}\right) 
\Biggr]\nonumber \\
+C_FN_F^2\Biggl[
&&\frac{8}{9\eps^2}
+\frac{1}{\eps}\left(-\frac{32\zeta_3}{3}+\frac{424}{27}\right)
+\left(-\frac{112\zeta_2^2}{15}-\frac{16\zeta_2}{3}-\frac{704\zeta_3}{9}+\frac{10562}{81}\right) 
\Biggr]\,.
\end{eqnarray}

The divergent parts agree with eq.~(8) of ref.~\cite{MMV2} while the finite contributions agree with eq.~(10) of ref.~\cite{BCSSS}.

Using our knowledge of the three-loop form factors, we can also write down the ${\cal O}(\eps)$ contributions to the $N_F$ parts of the quark and gluon form factors.
For the quark form-factor we find that,
\begin{eqnarray}
{\cal F}_3^q|_{N_F} &=& 
         C_FN_F^2\eps \Biggl(  
       - \frac{2913928}{6561} 
       + \frac{2248}{135}\zeta_5 
       + \frac{2108}{27}\zeta_3 
       - \frac{24950}{243}\zeta_2 
       + \frac{68}{9}\zeta_2\zeta_3 
       - \frac{3901}{810}\zeta_2^2 
       \Biggr)\nonumber \\
   &&   + C_FC_AN_F\eps \Biggl ( 
      \frac{24570881}{4374} 
      - \frac{28156}{45}\zeta_5 
      - \frac{2418896}{729}\zeta_3 
      + \frac{10816}{27}\zeta_3^2 
      + \frac{7137385}{4374}\zeta_2
     \nonumber \\
     &&  \hspace{3cm}
      + \frac{2674}{27}\zeta_2\zeta_3 
      - \frac{352559}{1620}\zeta_2^2 
      + \frac{17324}{945}\zeta_2^3 
      \Biggr)\nonumber \\
   &&    + C_F^2N_F\eps \Biggl (  
       - \frac{50187205}{17496} 
       + \frac{5863}{135}\zeta_5 
       + \frac{929587}{243}\zeta_3 
       - \frac{5771}{9}\zeta_3^2 
       - \frac{1263505}{972}\zeta_2 
     \nonumber \\
     &&  \hspace{3cm}
       - \frac{8515}{54}\zeta_2\zeta_3 
       + \frac{821749}{3240}\zeta_2^2 
       - \frac{875381}{7560}\zeta_2^3 
       \Biggr)\nonumber \\
   &&    + C_FN_{F,V}\left(\frac{N^2-4}{N}\right)\eps \Biggl ( 
       \frac{170}{3} 
       + \frac{752}{9}\zeta_5 
       + \frac{94}{9}\zeta_3 
       - \frac{344}{3}\zeta_3^2 
       + \frac{260}{3}\zeta_2 
     \nonumber \\
     &&  \hspace{3cm}
       + 30\zeta_2\zeta_3 
       - \frac{196}{15}\zeta_2^2 
       - \frac{9728}{315}\zeta_2^3 
       \Biggr)\;,
\end{eqnarray}
and for the gluon form factor
\begin{eqnarray}
{\cal F}_3^g|_{N_F} &=& 
        C_AN_F^2\eps \Biggl( 
      \frac{16823771}{26244} 
      + \frac{9368}{135}\zeta_5 
      + \frac{5440}{27}\zeta_3 
      - \frac{30283}{1458}\zeta_2 
      - \frac{988}{27}\zeta_2\zeta_3 
      + \frac{14018}{405}\zeta_2^2 
      \Biggr)\nonumber \\
      &&+ C_A^2N_F\eps \Biggl (  
      - \frac{48658741}{8748} 
      - \frac{10066}{45}\zeta_5 
      + \frac{349918}{729}\zeta_3 
      - \frac{11657}{27}\zeta_3^2 
      + \frac{904045}{4374}\zeta_2 
     \nonumber \\
     &&  \hspace{3cm}
      + \frac{791}{9}\zeta_2\zeta_3 
      - \frac{34931}{1620}\zeta_2^2 
      - \frac{52283}{1080}\zeta_2^3
      \Biggr)\nonumber \\
      &&+ C_FN_F^2\eps \Biggl ( 
      \frac{196900}{243} 
      - \frac{800}{9}\zeta_5 
      - \frac{4208}{9}\zeta_3 
      - 54\zeta_2 
      + \frac{112}{3}\zeta_2\zeta_3 
      - \frac{2464}{45}\zeta_2^2 
      \Biggr)\nonumber \\
      &&+ C_FC_AN_F\eps \Biggl (  
      - \frac{10508593}{2916} 
      + \frac{17092}{27}\zeta_5 
      + \frac{240934}{243}\zeta_3 
      + \frac{4064}{9}\zeta_3^2 
      + \frac{8869}{54}\zeta_2 
     \nonumber \\
     && \hspace{3cm} 
      + \frac{640}{9}\zeta_2\zeta_3 
      + \frac{28823}{270}\zeta_2^2 
      + \frac{23624}{315}\zeta_2^3 \Biggr)\nonumber \\
      &&+ C_F^2N_F\eps \Biggl ( 
      \frac{18613}{54} 
      - \frac{3080}{3}\zeta_5 
      + \frac{10552}{9}\zeta_3 
      - 272\zeta_3^2 
      - \frac{74}{3}\zeta_2 
     \nonumber \\
     &&  \hspace{3cm}
      - 16\zeta_2\zeta_3 
      + \frac{328}{5}\zeta_2^2 
      - \frac{35648}{315}\zeta_2^3 
      \Biggr )
\end{eqnarray}

The UV-renormalization of the form factors is derived in Section~\ref{sec:renorm} above. 
Applying (\ref{eq:renq}) and (\ref{eq:reng}) yields the expansion coefficients of the renormalized form factors. 
These are in the space-like kinematics:
{\allowdisplaybreaks
\begin{eqnarray}
{F}_3^q = 
 C_F^3 \Biggl[
&&-\frac{4}{3\eps^6}
-\frac{6}{\eps^5}
+\frac{1}{\eps^4}\left(2\zeta_2-25\right)
-\frac{1}{\eps^3}\left(3\zeta_2-\frac{100\zeta_3}{3}+83\right)\nonumber \\
&&+\frac{1}{\eps^2}\left(\frac{213\zeta_2^2}{10}-\frac{77\zeta_2}{2}+138\zeta_3-\frac{515}{2}\right)\nonumber \\
&&+\frac{1}{\eps}\left(\frac{1461\zeta_2^2}{20}-\frac{214\zeta_2\zeta_3}{3}-\frac{467\zeta_2}{2}+\frac{2119\zeta_3}{3}+\frac{644\zeta_5}{5}-\frac{9073}{12}\right)\nonumber \\
&&+\left(
          -\frac{ 53675}{24}
          - \frac{13001\zeta_2}{12}
          + \frac{12743\zeta_2^2}{40}
          - \frac{9095\zeta_2^3}{252}
          + 2669 \zeta_3
          + 61\zeta_3\zeta_2 \right. \nonumber \\
&&    \left.    \qquad  - \frac{1826\zeta_3^2}{3}
          + \frac{4238\zeta_5}{5}
\right)
\Biggr]\nonumber \\
+C_F^2C_A \Biggl[
&& - \frac{11}{\eps^5}
-\frac{1}{\eps^4}\left(
           \frac{361}{18}
          + 2 \zeta_2
\right)
+ \frac{1}{\eps^3}\left(
          - \frac{1703}{54}
          - 26 \zeta_3
          + \frac{27\zeta_2}{2}
\right)\nonumber \\
&&+\frac{1}{\eps^2}\left(
            \frac{6820}{81}
          - \frac{482\zeta_3}{9}
          + \frac{1487\zeta_2}{36}
          - \frac{83\zeta_2^2}{5}
\right)\nonumber \\
&&+\frac{1}{\eps}\left(
            \frac{374149}{486}
          - 142\zeta_5
          + \frac{215\zeta_3\zeta_2}{3}
          - \frac{4151\zeta_3}{6}
          + \frac{31891\zeta_2}{108}
          - \frac{2975\zeta_2^2}{72}
\right)\nonumber \\
&&+\left(
            \frac{11169211}{2916}
          - \frac{6890\zeta_5}{9}
          - \frac{806\zeta_3\zeta_2}{3}
          - \frac{19933\zeta_3}{6}
 \right. \nonumber \\
&& \left.          \qquad
          + \frac{1616\zeta_3^2}{3}
          + \frac{537803\zeta_2}{324}
          - \frac{723739\zeta_2^2}{2160}
          - \frac{18619\zeta_2^3}{1260}
\right) \Biggr]\nonumber \\
+C_FC_A^2 \Biggl[
&&-\frac{1331}{81\eps^4}
+\frac{1}{\eps^3}\left(
            \frac{2866}{243}
          - \frac{110\zeta_2}{27}
\right)
+\frac{1}{\eps^2}\left(
            \frac{11669}{486}
          - \frac{902\zeta_3}{27}
          + \frac{1625\zeta_2}{81}
          - \frac{88\zeta_2^2}{45}
\right)\nonumber \\
&&+\frac{1}{\eps}\left(
          - \frac{139345}{8748}
          - \frac{136\zeta_5}{3}
          - \frac{88\zeta_3\zeta_2}{9}
          + \frac{3526\zeta_3}{27}
          - \frac{7163\zeta_2}{243}
          - \frac{166\zeta_2^2}{15}
\right)\nonumber \\
&& +\left(
         - \frac{51082685}{52488}
          - \frac{434\zeta_5}{9}
          + \frac{416\zeta_3\zeta_2}{3}
          + \frac{505087\zeta_3}{486}
\right. \nonumber \\
&& \left.        \qquad  
          - \frac{1136\zeta_3^2}{9}
          - \frac{412315\zeta_2}{729}
          + \frac{22157\zeta_2^2}{270}
          - \frac{6152\zeta_2^3}{189}
\right) 
\Biggr]\nonumber \\
+C_F^2N_F \Biggl[
&& \frac{2}{\eps^5}+\frac{35}{9\eps^4}
+\frac{1}{\eps^3}\left(
            \frac{139}{27}
          - 3\zeta_2
\right)+\frac{1}{\eps^2}\left(
          - \frac{775}{81}
          - \frac{110\zeta_3}{9}
          - \frac{133\zeta_2}{18}
\right)\nonumber \\
&&+\frac{1}{\eps}\left(
          - \frac{24761}{243}
          + \frac{469\zeta_3}{27}
          - \frac{2183\zeta_2}{54}
          - \frac{287\zeta_2^2}{36}
\right)\nonumber \\
&&+\left(
          - \frac{691883}{1458}
          - \frac{386\zeta_5}{9}
          + \frac{35\zeta_3\zeta_2}{3}
          + \frac{21179\zeta_3}{81}
          - \frac{16745\zeta_2}{81}
          - \frac{8503\zeta_2^2}{1080}
\right) 
\Biggr]\nonumber \\
+C_FC_AN_F \Biggl[
&& \frac{484}{81\eps^4}+\frac{1}{\eps^3}\left(
          - \frac{752}{243}
          + \frac{20\zeta_2}{27}
\right)+\frac{1}{\eps^2}\left(
          - \frac{2068}{243}
          + \frac{212\zeta_3}{27}
          - \frac{476\zeta_2}{81}
\right)\nonumber \\
&&+\frac{1}{\eps}\left(
          - \frac{8659}{2187}
          - \frac{964\zeta_3}{81}
          + \frac{2594\zeta_2}{243}
          + \frac{44\zeta_2^2}{15}
\right)\nonumber \\
&&+\left(
            \frac{1700171}{6561}
          - \frac{4\zeta_5}{3}
          + \frac{4\zeta_3\zeta_2}{3}
          - \frac{4288\zeta_3}{27}
          + \frac{115555\zeta_2}{729}
          + \frac{2\zeta_2^2}{27}
\right) 
\Biggr]\nonumber \\
+C_FN_F^2 \Biggl[
&&-\frac{44}{81\eps^4}-\frac{8}{243\eps^3}+\frac{1}{\eps^2}\left(
            \frac{46}{81}
          + \frac{4}{9}\zeta_2
\right)+\frac{1}{\eps}\left(
            \frac{2417}{2187}
          - \frac{8}{81}\zeta_3
          - \frac{20}{27}\zeta_2
\right)\nonumber \\
&&+\left(
           -\frac{190931}{13122}
          - \frac{416}{243}\zeta_3
          - \frac{824}{81}\zeta_2
          - \frac{188}{135}\zeta_2^2
\right) 
\Biggr]\nonumber \\
+C_FN_{F,V}&&\left(\frac{N^2-4}{N}\right) 
\Biggl[
4-\frac{2\zeta_2^2}{5}+10\zeta_2+\frac{14\zeta_3}{3}-\frac{80\zeta_5}{3}
\Biggr]\;, \label{eq:f3qr} \\
{F}_3^g = 
 C_A^3 \Biggl[
&&-\frac{4}{3\eps^6}
-\frac{55}{3\eps^5}
-\frac{9079}{162\eps^4}
+\frac{1}{\eps^3}\left(
            \frac{5453}{486}
          + \frac{22\zeta_3}{3}
          + \frac{77\zeta_2}{54}
\right)\nonumber \\
&&+\frac{1}{\eps^2}\left(
          - \frac{4277}{243}
          + \frac{2266\zeta_3}{27}
          - \frac{1393\zeta_2}{81}
          + \frac{247\zeta_2^2}{90}
\right)\nonumber \\
&&+\frac{1}{\eps}\left(
          - \frac{1307704}{2187}
          - \frac{878\zeta_5}{15} 
          - \frac{85\zeta_3\zeta_2}{9}
          + \frac{1814\zeta_3}{9}
          - \frac{27301\zeta_2}{486}
          + \frac{12881\zeta_2^2}{360}
\right)\nonumber \\
&&+\left(
          - \frac{23496187}{26244}
          + \frac{13882\zeta_5}{45}
          - \frac{1441\zeta_3\zeta_2}{18}
          + \frac{24893\zeta_3}{243}
              \right. \nonumber \\
&& \qquad \left.
          - \frac{1766\zeta_3^2}{9}
          + \frac{118165\zeta_2}{1458}
          + \frac{126071\zeta_2^2}{1080}
          - \frac{22523\zeta_2^3}{270}
          \right) 
\Biggr]\nonumber \\
+C_A^2N_F\Biggl[
&&\frac{10}{3\eps^5}
+\frac{1780}{81\eps^4}
+\frac{1}{\eps^3}\left(
          \frac{2344}{243}
          - \frac{7\zeta_2}{27}
\right)
+\frac{1}{\eps^2}\left(
          - \frac{1534}{243}
          + \frac{68\zeta_3}{27}
          + \frac{169\zeta_2}{81}
\right)\nonumber \\
&&+\frac{1}{\eps}\left(
            \frac{854467}{4374}
          + \frac{3002\zeta_3}{81}
          + \frac{3536\zeta_2}{243}
          + \frac{941\zeta_2^2}{180}
\right)\nonumber \\
&&+\left(
            \frac{2143537}{13122}
          + \frac{4516\zeta_5}{45}
          - \frac{301\zeta_3\zeta_2}{9}
          + \frac{1414\zeta_3}{9}
          - \frac{6440\zeta_2}{729}
          + \frac{527\zeta_2^2}{20}
\right) 
\Biggr]\nonumber \\
+C_AC_FN_F\Biggl[
&&-\frac{34}{9\eps^3}
+\frac{1}{\eps^2}\left(
            \frac{427}{27}
          - \frac{160\zeta_3}{9}
\right)
+\frac{1}{\eps}\left(
            \frac{13655}{81}
          - \frac{2600\zeta_3}{27}
          - \frac{13\zeta_2}{3}
          - \frac{176\zeta_2^2}{15}
\right)\nonumber \\
&&+\left(
            \frac{284929}{972}
          + \frac{32\zeta_5}{9}
          + 48\zeta_3\zeta_2
          - \frac{14398\zeta_3}{81}
          - \frac{118\zeta_2}{3}
          - \frac{928\zeta_2^2}{15}
\right) 
\Biggr]\nonumber \\
+C_F^2N_F\Biggl[
&&-\frac{1}{3\eps}
+\left(
          \frac{304}{9}
          - 160\zeta_5
          + \frac{296\zeta_3}{3}
\right) 
\Biggr]\nonumber \\
+C_AN_F^2\Biggl[
&&-\frac{170}{81\eps^4}
-\frac{998}{243\eps^3}
+\frac{1}{\eps^2}\left(
            \frac{92}{27}
          + \frac{2\zeta_2}{27}
\right)
+\frac{1}{\eps}\left(
         - \frac{37133}{4374}
          - \frac{164\zeta_3}{81}
          - \frac{70\zeta_2}{81}
\right)\nonumber \\
&&+\left(
            \frac{125059}{13122}
          + \frac{952\zeta_3}{243}
          - \frac{20\zeta_2}{27}
          - \frac{157\zeta_2^2}{135}
\right) 
\Biggr]\nonumber \\
+C_FN_F^2\Biggl[
&& \frac{14}{9\eps^2}
+\frac{1}{\eps}\left(
         - \frac{212}{27}
          + \frac{16\zeta_3}{3}
\right)
+\left(
            \frac{2881}{162}
          - \frac{152\zeta_3}{9}
          - \frac{2\zeta_2}{3}
          + \frac{16\zeta_2^2}{5}
\right) 
\Biggr] \nonumber \\
+N_F^3 \Biggl[&& \frac{8}{27\e^3} \Biggr] \label{eq:f3gr}
\end{eqnarray}}

The  ${\cal O}(\eps)$ contributions to the $N_F$ parts of the UV-renormalized space-like 
quark and gluon form factors are given by,
\begin{eqnarray}
F_3^q|_{N_F} &=& 
       + C_FN_F^2\eps \Biggl(  
       - \frac{3769249}{26244} 
       + \frac{88}{135}\zeta_5 
       + \frac{2632}{243}\zeta_3 
       - \frac{5515}{81}\zeta_2 
       + \frac{8}{3}\zeta_2\zeta_3 
       - \frac{952}{81}\zeta_2^2 
       \Biggr)\nonumber \\
   &&   + C_FC_AN_F\eps \Biggl ( 
      \frac{1552436}{729} 
      - \frac{11596}{45}\zeta_5 
      - \frac{1214351}{729}\zeta_3 
      + \frac{3988}{27}\zeta_3^2 
      + \frac{4933141}{4374}\zeta_2
     \nonumber \\
     &&  \hspace{3cm}
      + \frac{1966}{27}\zeta_2\zeta_3 
      + \frac{4579}{405}\zeta_2^2 
      + \frac{2762}{945}\zeta_2^3 
      \Biggr)\nonumber \\
   &&    + C_F^2N_F\eps \Biggl (  
       - \frac{15199979}{8748} 
       - \frac{10769}{135}\zeta_5 
       + \frac{553882}{243}\zeta_3 
       - \frac{6881}{27}\zeta_3^2 
       - \frac{961699}{972}\zeta_2 
     \nonumber \\
     &&  \hspace{3cm}
       - \frac{4627}{54}\zeta_2\zeta_3 
       + \frac{94747}{3240}\zeta_2^2 
       - \frac{425813}{7560}\zeta_2^3 
       \Biggr)\nonumber \\
   &&    + C_FN_{F,V}\left(\frac{N^2-4}{N}\right)\eps \Biggl ( 
       \frac{170}{3} 
       + \frac{752}{9}\zeta_5 
       + \frac{94}{9}\zeta_3 
       - \frac{344}{3}\zeta_3^2 
       + \frac{260}{3}\zeta_2 
     \nonumber \\
     &&  \hspace{3cm}
       + 30\zeta_2\zeta_3 
       - \frac{196}{15}\zeta_2^2 
       - \frac{9728}{315}\zeta_2^3 
       \Biggr)\;,
\end{eqnarray}
and for the gluon form factor
\begin{eqnarray}
F_3^g|_{N_F} &=& 
      + C_AN_F^2\eps \Biggl( 
      \frac{6599393}{26244} 
      + \frac{1844}{135}\zeta_5 
      + \frac{8396}{81}\zeta_3 
      - \frac{25315}{1458}\zeta_2 
      - \frac{172}{27}\zeta_2\zeta_3 
      + \frac{2453}{405}\zeta_2^2 
      \Biggr)\nonumber \\
      &&+ C_A^2N_F\eps \Biggl (  
      - \frac{18825781}{8748} 
      + \frac{1682}{45}\zeta_5 
      + \frac{270232}{729}\zeta_3 
      - \frac{6251}{27}\zeta_3^2 
      + \frac{867919}{4374}\zeta_2 
     \nonumber \\
     &&  \hspace{3cm}
      - \frac{881}{9}\zeta_2\zeta_3 
      + \frac{33403}{405}\zeta_2^2 
      + \frac{133627}{7560}\zeta_2^3
      \Biggr)\nonumber \\
      &&+ C_FN_F^2\eps \Biggl ( 
      \frac{360181}{972} 
      - \frac{224}{9}\zeta_5 
      - \frac{1960}{9}\zeta_3 
      - \frac{277}{9}\zeta_2 
      + \frac{32}{3}\zeta_2\zeta_3 
      - \frac{208}{15}\zeta_2^2 
      \Biggr)\nonumber \\
      &&+ C_FC_AN_F\eps \Biggl (  
      - \frac{7017335}{5832} 
      + \frac{7588}{27}\zeta_5 
      - \frac{92894}{243}\zeta_3 
      + \frac{4064}{9}\zeta_3^2 
      + \frac{986}{27}\zeta_2 
     \nonumber \\
     && \hspace{3cm} 
      + \frac{1960}{9}\zeta_2\zeta_3 
      - \frac{59987}{540}\zeta_2^2 
      + \frac{23624}{315}\zeta_2^3 \Biggr)\nonumber \\
      &&+ C_F^2N_F\eps \Biggl ( 
      \frac{18613}{54} 
      - \frac{3080}{3}\zeta_5 
      + \frac{10552}{9}\zeta_3 
      - 272\zeta_3^2 
      - \frac{74}{3}\zeta_2 
     \nonumber \\
     &&  \hspace{3cm}
      - 16\zeta_2\zeta_3 
      + \frac{328}{5}\zeta_2^2 
      - \frac{35648}{315}\zeta_2^3 
      \Biggr ).
\end{eqnarray}
}

\section{Infrared pole structure}
\label{sec:ir}

According to ref.~\cite{cusp1,Becher:2009qa}, the general infrared pole structure of a renormalised 
QCD amplitude is related to the ultraviolet behaviour of an effective operator in 
 soft-collinear effective theory.  
These poles can therefore be subtracted by means of a multiplicative 
renormalization factor ${\bf Z}$.
This means that the finite remainders of a scattering amplitude ${\bf M^F}$ 
is obtained from the full amplitude ${\bf M}$ via the relation,
\begin{equation}
\label{eq:mfinite}
{\bf M^F} = {\bf Z}^{-1} {\bf M}.
\end{equation}
In general, the scattering amplitude ${\bf M}$ and ${\bf Z}$ are matrices in colour space. However, 
in the context of the quark and gluon form factors, the colour matrix is trivial.  
The UV renormalised amplitudes ${M}$ and ${M_F}$ have perturbative expansions,
\begin{eqnarray}
{ M} &=& 1 + \sum_{i=1} \left(\frac{\alpha_s(\mu^2)}{4\pi}\right)^i { M}_i,\\
{ M^F} &=& 1 + \sum_{i=1} \left(\frac{\alpha_s(\mu^2)}{4\pi}\right)^i { M_i^F},
\end{eqnarray}
while 
\begin{equation}
\log({ Z}) = \sum_{i=1} \left(\frac{\alpha_s(\mu^2)}{4\pi}\right)^i { Z}_i.
\end{equation}

We can now solve eq.~\eqref{eq:mfinite} order by order in the strong coupling,
\begin{eqnarray}
\Poles{(M_1)} &=& Z_1,\\
\Poles{(M_2)} &=& Z_2+\frac{M_1^2}{2},\\
\Poles{(M_3)} &=& Z_3-\frac{M_1^3}{3}+M_2M_1,\\
\Poles{(M_4)} &=& Z_4+\frac{M_1^4}{4}-M_1^2M_2+M_1M_3+\frac{M_2^2}{2},\\
\Poles{(M_5)} &=& Z_5-\frac{M_1^5}{5}+M_1^3M_2-M_1^2M_3-M_1M_2^2+M_1M_4+M_2M_3.
\end{eqnarray}
The deepest infrared pole for the $i$-loop amplitude is $\eps^{-2i}$.  However, the deepest pole in the $Z_i$-factor is $\eps^{-i-1}$.  All of the deepest poles are obtained directly from the lower loop amplitudes - which must be known to an appropriately high order in $\eps$.  For example, to obtain the correct pole structure for $M_i$, one needs knowledge of $M_1$ through to ${\cal O}(\eps^{2i-3})$.

We find that the infrared pole structure of the renormalised  form factors  is given by ($i=q,g$ and $C_q = C_F$, 
$C_g=C_A$ for the cusp anomalous dimension):
\begin{eqnarray}
\label{polef1q}
\Poles{(F_1^i)} &=& 
-\frac{C_i \gamma^{\rm cusp}_0}{2 \eps^2}+\frac{\gamma^i_0}{\eps}\,,\\
\label{polef2q}
\Poles{(F_2^i)} &=& 
\frac{3 C_i \gamma^{\rm cusp}_0 \beta_0}{8 \eps^3}+\frac{1}{\eps^{2}}\biggl(-\frac{\beta_0 \gamma^i_0}{2}-\frac{C_i 
\gamma^{\rm cusp}_1}{8}\biggr)
+\frac{\gamma^i_1}{2 \eps} +\frac{\left(F_1^i\right)^2}{2}\,,\\
\label{polef3q}
\Poles{(F_3^i)} &=& 
-\frac{11 \beta_0^2 C_i \gamma^{\rm cusp}_0}{36 \eps^4}+\frac{1}{\eps^{3}}
\biggl( \frac{5 \beta_0 C_i \gamma^{\rm cusp}_1}{36}+\frac{\beta_0^2 \gamma^i_0}{3}+\frac{2 C_i \gamma^{\rm cusp}_0 \beta_1}{9}\biggr)\nonumber \\ & &
+\frac{1}{\eps^{2}}\biggl(-\frac{\beta_0 \gamma^i_1}{3}-\frac{C_i \gamma^{\rm cusp}_2}{18}-\frac{\beta_1 \gamma^i_0}{3}\biggr)
+\frac{\gamma^i_2}{3 \eps}
-\frac{\left(F_1^i\right)^3}{3}+F_2^iF_1^i\,.
\end{eqnarray}
Note that the full (all-orders) expressions for $F^i_n$ are recycled on the right-hand-side.
The coefficients of the cusp soft anomalous dimension $\gamma^{\rm cusp}_i$ are known to three-loop order~\cite{MMV1} and are given by:
\begin{eqnarray}
\gamma^{\rm cusp}_0 &=& 4\,,\\
\gamma^{\rm cusp}_1 &=& 
C_A\bigg(\frac{268}{9}-\frac{4\pi^2}{3}\bigg) 
-\frac{40N_F}{9}\,,\\
\gamma^{\rm cusp}_2 &=& 
C_A^2\bigg(\frac{490}{3}-\frac{536\pi^2}{27}+\frac{44\pi^4}{45}+\frac{88\zeta_3}{3}\bigg) 
+C_AN_F\bigg(-\frac{836}{27}+\frac{80\pi^2}{27}-\frac{112\zeta_3}{3}\bigg)\nonumber \\ & &
+C_FN_F\bigg(-\frac{110}{3}+32\zeta_3\bigg) 
-\frac{16N_F^2}{27}\,.
\end{eqnarray}
while the quark and gluon collinear anomalous dimensions $\gamma^{q}_i$  and $\gamma^{g}_i$ in the conventional dimensional regularisation scheme are also known to three-loop order~\cite{Becher:2006mr,Becher:2009qa} and are given by:
{\allowdisplaybreaks
\begin{eqnarray}
\gamma^q_0 &=& -3C_F\,,\\
\gamma^q_1 &=& 
C_F^2\biggl(-\frac{3}{2}+2\pi^2-24\zeta_3\bigg) 
+C_FC_A\biggl(-\frac{961}{54}-\frac{11\pi^2}{6}+26\zeta_3\bigg)\nonumber \\ & &
+C_FN_F\biggl(\frac{65}{27}+\frac{\pi^2}{3}\bigg)\,,\\
\gamma^q_2 &=& 
C_F^2N_F\biggl(\frac{2953}{54}-\frac{13\pi^2}{9}-\frac{14\pi^4}{27}+\frac{256\zeta_3}{9}\bigg) 
+C_FN_F^2\biggl(\frac{2417}{729}-\frac{10\pi^2}{27}-\frac{8\zeta_3}{27}\bigg)\nonumber \\ & &
+C_FC_AN_F\biggl(-\frac{8659}{729}+\frac{1297\pi^2}{243}+\frac{11\pi^4}{45}-\frac{964\zeta_3}{27}\bigg) \nonumber \\ & &
+C_F^3\biggl(-\frac{29}{2}-3\pi^2-\frac{8\pi^4}{5}-68\zeta_3+\frac{16\pi^2\zeta_3}{3}+240\zeta_5\bigg)\nonumber \\ & &
+C_AC_F^2\biggl(-\frac{151}{4}+\frac{205\pi^2}{9}+\frac{247\pi^4}{135}-\frac{844\zeta_3}{3}-\frac{8\pi^2\zeta_3}{3}-120\zeta_5\bigg)\nonumber \\ & &
+C_A^2C_F\biggl(-\frac{139345}{2916}-\frac{7163\pi^2}{486}-\frac{83\pi^4}{90}+\frac{3526\zeta_3}{9}-\frac{44\pi^2\zeta_3}{9}-136\zeta_5\bigg)\,,\\
\gamma^g_0 &=& 
-\frac{11C_A}{3}+\frac{2N_F}{3}\,,\\
\gamma^g_1 &=& 
C_A^2\biggl(-\frac{692}{27}+\frac{11\pi^2}{18}+2\zeta_3\bigg) 
+C_AN_F\biggl( \frac{128}{27}-\frac{\pi^2}{9}\bigg) 
+2C_FN_F\\
\gamma^g_2 &=& 
C_A^3\biggl(-\frac{97186}{729}+\frac{6109\pi^2}{486}-\frac{319\pi^4}{270}+\frac{122\zeta_3}{3}-\frac{20\pi^2\zeta_3}{9}-16\zeta_5\bigg)\nonumber\, \\ & &
+C_A^2N_F\biggl(\frac{30715}{1458}-\frac{599\pi^2}{243}+\frac{41\pi^4}{135}+\frac{356\zeta_3}{27}\bigg)\nonumber \\ & &
+C_FC_AN_F\biggl(\frac{1217}{27}-\frac{\pi^2}{3}-\frac{4\pi^4}{45}-\frac{152\zeta_3}{9}\bigg)-C_F^2N_F\nonumber \\ & &
+C_AN_F^2\biggl(-\frac{269}{1458}+\frac{10\pi^2}{81}-\frac{56\zeta_3}{27}\bigg)-\frac{11C_FN_F^2}{9}\;.
\end{eqnarray}
}

Taking this one step further, we find that the pole structure of the renormalised four-loop quark form factor is given by
\begin{eqnarray}
\label{polef4q}
\Poles{(F_4^i)} &=& 
\frac{25 \beta_0^3 C_i \gamma^{\rm cusp}_0}{96 \eps^5}-\frac{\beta_0 (24 \beta_0^2 \gamma^i_0+13 \beta_0 C_i \gamma^{\rm cusp}_1+40 C_i \gamma^{\rm cusp}_0 \beta_1)}{96 \eps^4}\nonumber \\ & &+\frac{1}{\eps^{3}}\biggl(\frac{7 \beta_0 C_i \gamma^{\rm cusp}_2}{96}+\frac{3 \beta_1 C_i \gamma^{\rm cusp}_1}{32}+\frac{\beta_0^2 \gamma^i_1}{4}+\frac{\beta_1 \beta_0 \gamma^i_0}{2}+\frac{5 C_i \gamma^{\rm cusp}_0 \beta_2}{32}\biggr)\nonumber \\ & &
+\frac{1}{\eps^{2}}\biggl(-\frac{\beta_1 \gamma^i_1}{4}-\frac{C_i \gamma^{\rm cusp}_3}{32}-\frac{\beta_0 \gamma^i_2}{4}-\frac{\beta_2 \gamma^i_0}{4}\biggr)
+\frac{\gamma^i_3}{4 \eps}\nonumber \\ & &
+ \frac{\left(F_1^i\right)^4}{4}
- \left(F_1^i\right)^2 F_2^i+\frac{\left(F_2^i\right)^2}{2}+F_1^i F_3^i  \,.
\end{eqnarray}
In this expression, we assume Casimir scaling of the cusp anomalous dimension to hold at four 
loops~\cite{casimir,Becher:2009qa}, such that only a universal $\gamma^{\rm cusp}_3$ appears. If, contrary to 
expectations, Casimir scaling should be violated at this order, different $\gamma^{\rm cusp}_3$ would 
appear in the double pole terms of the quark and gluon form factors at four loops. 

Eq.~\eqref{polef4q} shows that 
in order to make use of a calculation of the pole parts of the four-loop form factors to extract the cusp and 
collinear anomalous dimensions, one requires the finite parts of the three-loop form factor for $\gamma^{\rm cusp}_3$,
and of the subleading ${\cal O}(\e)$ parts for $\gamma_3^{q,g}$. For all colour-factor 
contributions proportional to $N_F$, these are provided in the previous section. The required subleading terms 
in higher orders in $\e$ from the one-loop and two-loop form factors were summarized in Section~\ref{sec:renorm}
above.

\section{Effective Theory Matching Coefficients}
\label{sec:match}

It is well known that fixed-order perturbation theory is not 
necessarily reliable for physical quantities involving several disparate scales. In such cases, higher-order corrections are enhanced by large logarithms of scale ratios. 
Experimentally relevant examples are the Drell-Yan and Higgs production processes in hadron-hadron colliders. 
When the phase space for soft gluon emission is constrained, large logarithmic threshold corrections appear 
of the form
\begin{equation}
\label{eq:dlog}
\alpha_s^k \left[ \frac{\ln^{m-1}(1-z)}{(1-z)}\right]_+, \qquad (m \leq 2k),
\end{equation}
where $(1-z)$ is the fraction of centre-of-mass energy of the initial partons available for soft gluon radiation.
These can spoil the 
convergence of the perturbative series.
The resummation of these so-called Sudakov-logarithms has been accomplished to fourth logarithmic order~\cite{resum1}, 
using the exponentiation properties of the coefficient functions in moment space~\cite{sterman2}. 

An alternative resummation framework is provided by soft-collinear effective field theory (SCET), which is 
based on the idea to 
 split the calculation into a series of single-scale problems by successively integrating out the physics associated with the
largest remaining scale. The SCET framework~\cite{scet1} originated in the study of heavy quarks, and has been subsequently generalized 
to massless collider processes~\cite{scet2}. 
The infrared poles in the high energy theory (QCD)
get transformed into ultraviolet poles in the effective theory~\cite{cusp2,cusp1} and can then be resummed by renormalization-group (RG) evolution from the larger scales to the smaller ones. Of course the SCET must match precisely onto the high energy theory, and this is achieved by computing matrix elements in both the SCET and QCD and adjusting the Wilson coefficients so that they agree.  If the matching is performed on-shell, then the matching coefficients relevant for Drell Yan and Higgs production can be obtained from the quark and gluon form factors respectively.   Therefore, we can utilise the results presented in the previous sections to compute the matching conditions through to three-loops. Results up to two loops were obtained previously 
in~\cite{Idilbi:2005ky,Idilbi:2006dg,Ahrens:2008nc}.

The renormalised form-factors are infrared divergent.   In the effective field theory, these infrared divergences are transformed into ultraviolet poles.   The matching coefficient $C^i$ $(i=q,g)$ is obtained by extracting the poles using a renormalisation factor such that,
\begin{equation}
C^i(\alpha_s(\mu^2),s_{12},\mu^2) = \lim_{\epsilon \to 0} Z_i^{-1}(\epsilon,s_{12},\mu) F^i(\epsilon,s_{12},\mu^2).
\end{equation}
The matching coefficients have the perturbative expansion,
\begin{eqnarray}
{C}^i (\alpha_s(\mu^2), s_{12},\mu^2) &=& 1 + \sum_{n=1}^{\infty} \left( \frac{\alpha_s (\mu^2)}{4\pi}\right)^n  \, {C}_n^i (s_{12},\mu^2).
\end{eqnarray}
They are are known to two loop order for Drell-Yan~\cite{Idilbi:2005ky,Idilbi:2006dg} and Higgs~\cite{Ahrens:2008nc} production,  
{\allowdisplaybreaks
\begin{eqnarray}
C^q_1 &=& C_F \bigg(- L^2 + 3 L - 8 + \zeta_2  \bigg) , \\
C^q_2 &=&
C_F^2 \bigg(
\frac{1}{2}L^4-3L^3+\bigg(\frac{25}{2}-\zeta_2\bigg)L^2+\bigg(-\frac{45}{2}+24\zeta_3-9\zeta_2\bigg)L\nonumber \\ &&\hspace{2.5cm}+\frac{255}{8}-30\zeta_3+21\zeta_2-\frac{83}{10}\zeta_2^2 \bigg)\nonumber \\ &&
+C_FC_A \bigg(
\frac{11}{9}L^3+\bigg(-\frac{233}{18}+2\zeta_2\bigg)L^2+\bigg(\frac{2545}{54}-26\zeta_3+\frac{22}{3}\zeta_2\bigg)L\nonumber \\ &&\hspace{2.5cm} -\frac{51157}{648}+\frac{313}{9}\zeta_3-\frac{337}{18}\zeta_2+\frac{44}{5}\zeta_2^2 \bigg)\nonumber \\ &&
+C_FN_F \bigg(
-\frac{2}{9}L^3+\frac{19}{9}L^2+\bigg(-\frac{209}{27}-\frac{4}{3}\zeta_2\bigg)L+\frac{4085}{324}+\frac{2}{9}\zeta_3+\frac{23}{9}\zeta_2 \bigg) , \\ 
C^g_1 &=& C_A \bigg(- L^2 + \zeta_2  \bigg),  \\
C^g_2 &=&
C_A^2 \bigg(
\frac{1}{2}L^4+\frac{11}{9}L^3+\bigg(-\frac{67}{9}+\zeta_2\bigg)L^2+\bigg(\frac{80}{27}-2\zeta_3-\frac{22}{3}\zeta_2\bigg)L\nonumber \\ &&\hspace{2.5cm}+\frac{5105}{162}-\frac{143}{9}\zeta_3+\frac{67}{6}\zeta_2+\frac{1}{2}\zeta_2^2 \bigg)\nonumber \\ &&
+C_AN_F \bigg(
-\frac{2}{9}L^3+\frac{10}{9}L^2+\bigg(\frac{52}{27}+\frac{4}{3}\zeta_2\bigg)L-\frac{916}{81}-\frac{46}{9}\zeta_3-\frac{5}{3}\zeta_2 \bigg)\nonumber \\ &&
+C_FN_F \bigg(
2L -\frac{67}{6}+8\zeta_3 \bigg)  
\end{eqnarray}
}
where $L=\log(-s_{12}/\mu^2)$.

Exploiting the expressions for the renormalised quark and gluon form factors 
given in eqs.~\eqref{eq:f3qr} and \eqref{eq:f3gr} respectively, we find that
the three-loop matching coefficients are
{\allowdisplaybreaks

\begin{eqnarray}
C^q_3 &=&
C_F^3 \bigg(
-\frac{1}{6}L^6+\frac{3}{2}L^5+\bigg(-\frac{17}{2}+\frac{1}{2}\zeta_2\bigg)L^4+\bigg(9\zeta_2+27-24\zeta_3\bigg)L^3\nonumber \\ &&\hspace{2cm}
+\bigg(102\zeta_3-\frac{507}{8}-\frac{105}{2}\zeta_2+\frac{83}{10}\zeta_2^2\bigg)L^2\nonumber \\ &&\hspace{2cm}+\bigg(-214\zeta_3-240\zeta_5-8\zeta_2\zeta_3+\frac{357}{2}\zeta_2+\frac{207}{10}\zeta_2^2+\frac{785}{8}\bigg)L\nonumber \\ &&\hspace{2cm}-\frac{413}{5}\zeta_2^2+664\zeta_5-\frac{6451}{24}\zeta_2+\frac{37729}{630}\zeta_2^3-470\zeta_3+250\zeta_2\zeta_3-\frac{2539}{12}+16\zeta_3^2 \bigg)\nonumber \\ && 
+C_F^2C_A \bigg(
-\frac{11}{9}L^5+\bigg(\frac{299}{18}-2\zeta_2\bigg)L^4+\bigg(-\frac{2585}{27}+26\zeta_3-\frac{1}{9}\zeta_2\bigg)L^3\nonumber \\ &&\hspace{2cm}+\bigg(\frac{206317}{648}-\frac{1807}{9}\zeta_3+\frac{502}{9}\zeta_2-\frac{34}{5}\zeta_2^2\bigg)L^2\nonumber \\ &&\hspace{2cm}+\bigg(-\frac{13805}{24}+120\zeta_5+\frac{2441}{3}\zeta_3-\frac{11260}{27}\zeta_2-10\zeta_2\zeta_3+\frac{162}{5}\zeta_2^2\bigg)L\nonumber \\ &&\hspace{2cm}+\frac{415025}{648}-\frac{2756}{9}\zeta_5-\frac{18770}{27}\zeta_3+\frac{296}{3}\zeta_3^2+\frac{538835}{648}\zeta_2-\frac{3751}{9}\zeta_2\zeta_3\nonumber \\ &&\hspace{2cm}-\frac{4943}{270}\zeta_2^2-\frac{12676}{315}\zeta_2^3 \bigg)\nonumber \\ &&
+C_F^2N_F \bigg(
\frac{2}{9}L^5-\frac{25}{9}L^4+\bigg(\frac{410}{27}+\frac{10}{9}\zeta_2\bigg)L^3+\bigg(-\frac{12815}{324}+\frac{70}{9}\zeta_3-\frac{112}{9}\zeta_2\bigg)L^2\nonumber \\ &&\hspace{2cm}+\bigg(\frac{3121}{108}-\frac{610}{9}\zeta_3+\frac{1618}{27}\zeta_2+\frac{28}{5}\zeta_2^2\bigg)L\nonumber \\ &&\hspace{2cm}+\frac{41077}{972}-\frac{416}{9}\zeta_5+\frac{13184}{81}\zeta_3-\frac{31729}{324}\zeta_2-\frac{38}{9}\zeta_2\zeta_3-\frac{331}{27}\zeta_2^2 \bigg)\nonumber \\ &&
+C_FC_A^2 \bigg(
-\frac{121}{54}L^4+\bigg(\frac{2869}{81}-\frac{44}{9}\zeta_2\bigg)L^3+\bigg(-\frac{18682}{81}+88\zeta_3+\frac{26}{9}\zeta_2-\frac{44}{5}\zeta_2^2\bigg)L^2\nonumber \\ &&\hspace{2cm}+\bigg(\frac{1045955}{1458}+136\zeta_5-\frac{17464}{27}\zeta_3+\frac{17366}{81}\zeta_2+\frac{88}{3}\zeta_2\zeta_3-\frac{94}{3}\zeta_2^2\bigg)L\nonumber \\ &&\hspace{2cm}-\frac{51082685}{52488}-\frac{434}{9}\zeta_5+\frac{505087}{486}\zeta_3-\frac{1136}{9}\zeta_3^2-\frac{412315}{729}\zeta_2+\frac{416}{3}\zeta_2\zeta_3\nonumber \\ &&\hspace{2cm}+\frac{22157}{270}\zeta_2^2-\frac{6152}{189}\zeta_2^3 \bigg)\nonumber \\ &&
+C_FC_AN_F \bigg(
\frac{22}{27}L^4+\bigg(-\frac{974}{81}+\frac{8}{9}\zeta_2\bigg)L^3+\bigg(\frac{5876}{81}-8\zeta_3+\frac{16}{3}\zeta_2\bigg)L^2\nonumber \\ &&\hspace{2cm}+\bigg(-\frac{154919}{729}+\frac{724}{9}\zeta_3-\frac{5864}{81}\zeta_2+\frac{44}{15}\zeta_2^2\bigg)L\nonumber \\ &&\hspace{2cm}+\frac{1700171}{6561}-\frac{4}{3}\zeta_5-\frac{4288}{27}\zeta_3+\frac{115555}{729}\zeta_2+\frac{4}{3}\zeta_2\zeta_3+\frac{2}{27}\zeta_2^2 \bigg)\nonumber \\ &&
+C_FN_F^2 \bigg(
-\frac{2}{27}L^4+\frac{76}{81}L^3+\bigg(-\frac{406}{81}-\frac{8}{9}\zeta_2\bigg)L^2+\bigg(\frac{9838}{729}+\frac{16}{27}\zeta_3+\frac{152}{27}\zeta_2\bigg)L\nonumber \\ &&\hspace{2cm}-\frac{190931}{13122}-\frac{416}{243}\zeta_3-\frac{824}{81}\zeta_2-\frac{188}{135}\zeta_2^2 \bigg)\nonumber \\ &&
+C_FN_{F,V}\left(\frac{N^2-4}{N}\right) \bigg(
4-\frac{80}{3}\zeta_5+\frac{14}{3}\zeta_3+10\zeta_2-\frac{2}{5}\zeta_2^2 \bigg)  
\end{eqnarray}
}
and,
{\allowdisplaybreaks
\begin{eqnarray}
C^g_3 &=&
C_A^3 \bigg(
-\frac{1}{6}L^6-\frac{11}{9}L^5+\bigg(\frac{281}{54}-\frac{3}{2}\zeta_2\bigg)L^4+\bigg(\frac{11}{3}\zeta_2+\frac{1540}{81}+2\zeta_3\bigg)L^3\nonumber \\ &&\hspace{2cm}+\bigg(\frac{143}{9}\zeta_3-\frac{6740}{81}+\frac{685}{18}\zeta_2-\frac{73}{10}\zeta_2^2\bigg)L^2\nonumber \\ &&\hspace{2cm}+\bigg(\frac{2048}{27}\zeta_3+16\zeta_5+\frac{34}{3}\zeta_2\zeta_3-\frac{13420}{81}\zeta_2+\frac{176}{5}\zeta_2^2-\frac{373975}{1458}\bigg)L\nonumber \\ &&\hspace{2cm}-\frac{1939}{270}\zeta_2^2+\frac{2222}{9}\zeta_5+\frac{105617}{729}\zeta_2-\frac{24389}{1890}\zeta_2^3-\frac{152716}{243}\zeta_3-\frac{605}{9}\zeta_2\zeta_3\nonumber \\ &&\hspace{2cm}+\frac{29639273}{26244}-\frac{104}{9}\zeta_3^2 \bigg)\nonumber \\ &&
+C_A^2N_F \bigg(
\frac{2}{9}L^5-\frac{8}{27}L^4+\bigg(-\frac{734}{81}-\frac{2}{3}\zeta_2\bigg)L^3+\bigg(\frac{377}{27}+\frac{118}{9}\zeta_3-\frac{103}{9}\zeta_2\bigg)L^2\nonumber \\ &&\hspace{2cm}+\bigg(\frac{133036}{729}+\frac{28}{9}\zeta_3+\frac{3820}{81}\zeta_2-\frac{48}{5}\zeta_2^2\bigg)L\nonumber \\ &&\hspace{2cm}-\frac{3765007}{6561}+\frac{428}{9}\zeta_5-\frac{460}{81}\zeta_3-\frac{14189}{729}\zeta_2-\frac{82}{9}\zeta_2\zeta_3+\frac{73}{45}\zeta_2^2 \bigg)\nonumber \\ &&
+C_AN_F^2 \bigg(
-\frac{2}{27}L^4+\frac{40}{81}L^3+\bigg(\frac{116}{81}+\frac{8}{9}\zeta_2\bigg)L^2+\bigg(-\frac{14057}{729}-\frac{128}{27}\zeta_3-\frac{80}{27}\zeta_2\bigg)L\nonumber \\ &&\hspace{2cm}+\frac{611401}{13122}+\frac{4576}{243}\zeta_3+\frac{4}{9}\zeta_2+\frac{4}{27}\zeta_2^2 \bigg)\nonumber \\ &&
+C_FN_F^2 \bigg(
\frac{4}{3}L^2+\bigg(-\frac{52}{3}+\frac{32}{3}\zeta_3\bigg)L+\frac{4481}{81}-\frac{112}{3}\zeta_3-\frac{20}{9}\zeta_2-\frac{16}{45}\zeta_2^2 \bigg)\nonumber \\ &&
+C_FC_AN_F \bigg(
-\frac{8}{3}L^3+\bigg(13-16\zeta_3\bigg)L^2+\bigg(\frac{3833}{54}-\frac{376}{9}\zeta_3+6\zeta_2+\frac{16}{5}\zeta_2^2\bigg)L\nonumber \\ &&\hspace{2cm}-\frac{341219}{972}+\frac{608}{9}\zeta_5+\frac{14564}{81}\zeta_3-\frac{68}{9}\zeta_2+\frac{64}{3}\zeta_2\zeta_3-\frac{64}{45}\zeta_2^2 \bigg)\nonumber \\ &&
+C_F^2N_F \bigg(
-2L + \frac{304}{9}-160\zeta_5+\frac{296}{3}\zeta_3 \bigg).
\end{eqnarray}
}
These matching coefficients allow to perform the three-loop matching of the SCET-based resummation onto the full 
QCD calculation.

\section{Conclusions}
\label{sec:conc}

In this paper, we described the calculation of the three-loop quark and gluon form factors in detail. Our 
results confirm earlier expressions obtained by Baikov et al.~\cite{BCSSS}, which we extended by subleading 
terms in the fermionic corrections. 

The form factors are the simplest QCD objects with non-trivial infrared structure. Recent 
findings on the relation between massless on-shell QCD amplitudes and operators in soft-collinear effective 
theory~\cite{Becher:2009cu}, combined with constraints from factorization, has led to the 
conjecture~\cite{Becher:2009qa} that their pole terms at a given loop level contain all information 
needed to predict the pole structure of massless on-shell multi-leg amplitudes at the same loop order. In 
particular, the cusp anomalous dimension can be extracted from the double pole, and the 
collinear anomalous dimension from the single pole. At a given loop order, finite and subleading terms from 
lower loop orders are also required. In this respect, the finite terms presented here 
will be instrumental for the extraction of the four-loop cusp anomalous dimension, while the subleading 
terms contribute to the four-loop quark and gluon collinear anomalous dimension. 

The three-loop form factors are key ingredients for the fourth order (N$^3$LO) corrections to the 
inclusive Drell-Yan and Higgs boson production cross sections. The calculation of these, at least in 
an improvement to the soft approximation~\cite{mv,Idilbi:2006dg}, could be envisaged in future work. In view of 
this application, we derived the hard matching coefficients of the SCET operators to this order. 
Inclusion of these corrections will 
lead to a further stabilization of the perturbative prediction under scale variations, and 
are thus important for precision physics at hadron colliders.

\section*{Acknowledgements}
This research was supported in part by
the Swiss National Science Foundation (SNF) under contract 200020-126691, by the 
Forschungskredit der Universit\"at Z\"urich,  the UK Science and Technology Facilities
Council, by the European Commission's Marie-Curie Research Training Network
under contract MRTN-CT-2006-035505 `Tools and Precision Calculations for Physics
Discoveries at Colliders', and by the Helmholtz Alliance ``Physics at the Terascale''.
EWNG gratefully acknowledges the support of the
Wolfson Foundation and the Royal Society.

\begin{appendix}
\section{Master Integrals for three-loop form factors}
\label{app:mi}

In this appendix, we summarize the $\e$-expansions of all master integrals 
needed for the three-loop form 
factors. Our notation for the integrals follows~\cite{masterA}, using 
a Minkowskian loop integration measure 
$\d^d k/(2\pi)^d$. All master 
integrals are defined in Section~\ref{sec:masters} above. 

With the normalization  $S_\Gamma$ defined in (\ref{eq:sgamma}),
all $M$-loop integrals have an overall factor of 
\begin{equation}
\label{eq:normalisation}
s_{12}^n\, \left(iS_\Gamma \left(-s_{12}-i0\right)^{-\eps}\right)^M 
\end{equation}
where $n$ is fixed by dimensional arguments.   Unlike Refs.~\cite{masterA,masterB,masterC}, 
there is no $(-1)^n$ factor. We expand to the 
required order for the three-loop form factors (which is typically the order 
where transcendentality 6 first appears). 

The one-loop master integral is:
{\allowdisplaybreaks
\begin{eqnarray}
B_{2,1}=
&& \frac{1}{\eps}
+2
+4\eps
-\eps^2\left( 2\zeta_3-8\right)
-\eps^3\left( \frac{6\zeta_2^2}{5}+4\zeta_3-16\right)\nonumber \\
&&
-\eps^4\left( \frac{12\zeta_2^2}{5}+8\zeta_3+6\zeta_5-32\right)\nonumber \\
&&-\eps^5\left( \frac{16\zeta_2^3}{7}+\frac{24\zeta_2^2}{5}-2\zeta_3^2+16\zeta_3+12\zeta_5\
-64\right) \nonumber \\
&&+\eps^6\left(128 - 18\zeta_7 - 24\zeta_5 - 32\zeta_3 + 4\zeta_3^2 - \frac{48}{5}
         \zeta_2^2 + \frac{12}{5}\zeta_2^2\zeta_3 - \frac{32}{7}\zeta_2^3\right)
+{\cal O} (\eps^7)\,. \nonumber \\
\end{eqnarray}}

The two-loop two-point and three-point master integrals are:
{\allowdisplaybreaks
\begin{eqnarray}
B_{3,1}=
&&-\frac{1}{4\eps}
-\frac{13}{8}
-\frac{115\eps}{16}
+\eps^2\left( \frac{5\zeta_3}{2}-\frac{865}{32}\right)
+\eps^3\left( \frac{3\zeta_2^2}{2}+\frac{65\zeta_3}{4}-\frac{5971}{64}\right)\nonumber \\
&&+\eps^4\left( \frac{39\zeta_2^2}{4}+\frac{575\zeta_3}{8}+\frac{27\zeta_5}{2}-\frac{39193}{128}\right)\nonumber \\
&&+\eps^5\left( \frac{44\zeta_2^3}{7}+\frac{345\zeta_2^2}{8}-\frac{25\zeta_3^2}{2}+\frac{4325\zeta_3}{16}+\frac{351\zeta_5}{4}
-\frac{249355}{256}\right)\nonumber \\
&&+\eps^6\left(- \frac{1555105}{512} + \frac{165}{2}\zeta_7 + \frac{3105}{8}\zeta_5 + \frac{29855}{32}\zeta_3
          - \frac{325}{4}\zeta_3^2 + \frac{2595}{16}\zeta_2^2 - 15\zeta_2^2\zeta_3 + \frac{286}{7}\zeta_2^3
\right)\nonumber \\
&&
+ {\cal O} (\eps^7)\,, \\
B_{4,2}=
&&+\frac{1}{\eps^2}
+\frac{4}{\eps}
+12-\eps\left( 4\zeta_3-32\right)
-\eps^2\left( \frac{12\zeta_2^2}{5}+16\zeta_3-80\right)\nonumber\\
&&
-\eps^3\left( \frac{48\zeta_2^2}{5}+48\zeta_3+12\zeta_5-192\right)\nonumber \\
&&-\eps^4\left( \frac{32\zeta_2^3}{7}+\frac{144\zeta_2^2}{5}-8\zeta_3^2+128\zeta_3+48\zeta_5
-448\right)\nonumber \\
&&+\eps^5\left(1024 - 36\zeta_7 - 144\zeta_5 - 320\zeta_3 + 32\zeta_3^2 - \frac{384}{5}
         \zeta_2^2 + \frac{48}{5}\zeta_2^2\zeta_3 - \frac{128}{7}\zeta_2^3\right)
\nonumber \\ &&
+ {\cal O} (\eps^6)\,,\\
C_{4,1}=
&&+\frac{1}{2\eps^2}
+\frac{5}{2\eps}
+\left( \zeta_2+\frac{19}{2}\right)
+\eps\left( 5\zeta_2-4\zeta_3+\frac{65}{2}\right)\nonumber \\
&&
-\eps^2\left( \frac{6\zeta_2^2}{5}-19\zeta_2+20\zeta_3-\frac{211}{2}\right)\nonumber \\
&&-\eps^3\left( 6\zeta_2^2+8\zeta_2\zeta_3-65\zeta_2+76\zeta_3+24\zeta_5
-\frac{665}{2}\right)\nonumber \\
&&-\eps^4\left( \frac{528\zeta_2^3}{35}+\frac{114\zeta_2^2}{5}+40\zeta_2\zeta_3-16\zeta_3^2-211\zeta_2
+260\zeta_3+120\zeta_5-\frac{2059}{2}\right)\nonumber \\
&&+\eps^5\bigg(\frac{6305}{2} - 156\zeta_7 - 456\zeta_5 - 844\zeta_3 + 80\zeta_3^2 + 665
         \zeta_2 - 48\zeta_2\zeta_5 - 152\zeta_2\zeta_3 - 78\zeta_2^2 \nonumber \\
	 &&+ \frac{48}{5}\zeta_2^2
         \zeta_3 - \frac{528}{7}\zeta_2^3\bigg)
+ {\cal O} (\eps^6)\,,\\
C_{6,2}=
&&+\frac{1}{\eps^4}
-\frac{5\zeta_2}{\eps^2}
-\frac{27\zeta_3}{\eps}
-23\zeta_2^2
+\eps\left( 48\zeta_2\zeta_3-117\zeta_5\right)
-\eps^2\left( \frac{456\zeta_2^3}{35}-267\zeta_3^2\right)
\nonumber \\ && +\eps^3\left(6\zeta_7 + 240\zeta_2\zeta_5 + \frac{1962}{5}\zeta_2^2\zeta_3 \right) + {\cal O} (\eps^4)\,.
\end{eqnarray}}
The $B_{t,i}$-type and $C_{t,i}$-type 
master integrals read at three loops:
{\allowdisplaybreaks
\begin{eqnarray}
B_{4,1} &=& 
\frac{1}{36\eps}+\frac{71}{216}+\frac{3115\eps}{1296}+\eps^2\biggl(-\frac{7\zeta_3}{9}+\frac{109403}{7776}\biggr) 
+\eps^3\biggl(-\frac{497\zeta_3}{54}-\frac{7\pi^4}{540}+\frac{3386467}{46656}\biggr)\nonumber \\ & &
+\eps^4\biggl(-\frac{21805\zeta_3}{324}-7\zeta_5-\frac{497\pi^4}{3240}+\frac{96885467}{279936}\biggr)\nonumber \\ & &
+\eps^5\biggl(-\frac{765821\zeta_3}{1944}-\frac{497\zeta_5}{6}-\frac{4361\pi^4}{3888}-\frac{4\pi^6}{243}+\frac{98\zeta_3^2}{9}+\frac{2631913075}{1679616}\biggr)
\nonumber \\
&&+ {\cal O} (\eps^6)\,, \\
B_{5,1} &=& 
-\frac{1}{4\eps^2}-\frac{17}{8\eps}-\frac{183}{16}+\eps\biggl(3\zeta_3-\frac{1597}{32}\biggr) 
+\eps^2\biggl(\frac{51\zeta_3}{2}+\frac{\pi^4}{20}-\frac{12359}{64}\biggr)\nonumber \\ & &
+\eps^3\biggl(\frac{549\zeta_3}{4}+15\zeta_5+\frac{17\pi^4}{40}-\frac{88629}{128}\biggr)\nonumber \\ & &
+\eps^4\biggl(\frac{4791\zeta_3}{8}+\frac{255\zeta_5}{2}+\frac{183\pi^4}{80}+\frac{2\pi^6}{63}-18\zeta_3^2-\frac{603871}{256}\biggr) 
+ {\cal O} (\eps^5)\,,\\
B_{5,2} &=& 
-\frac{1}{3\eps^2}-\frac{10}{3\eps}-\frac{64}{3}+\eps\biggl(-112+\frac{22\zeta_3}{3}\biggr) 
+\eps^2\biggl(-528+\frac{220\zeta_3}{3}+\frac{11\pi^4}{90}\biggr)\nonumber \\ & &
+\eps^3\biggl(-2336+\frac{1408\zeta_3}{3}+70\zeta_5+\frac{11\pi^4}{9}\biggr)\nonumber \\ & &
+\eps^4\biggl(\frac{352\pi^4}{45}+2464\zeta_3+700\zeta_5-\frac{29824}{3}+\frac{94\pi^6}{567}-\frac{242\zeta_3^2}{3}\biggr) 
+{\cal O}(\eps^5)\,, \\
B_{6,1} &=& 
\frac{1}{\eps^3}+\frac{6}{\eps^2}+\frac{24}{\eps}+\biggl(80-6\zeta_3\biggr)
+\eps\biggl(240-36\zeta_3-\frac{\pi^4}{10}\biggr)\nonumber \\ & &
+\eps^2\biggl(672-144\zeta_3-18\zeta_5-\frac{3\pi^4}{5}\biggr)\nonumber \\ & &
+\eps^3\biggl(1792-480\zeta_3-108\zeta_5-\frac{12\pi^4}{5}-\frac{2\pi^6}{63}+18\zeta_3^2\biggr) + {\cal O} (\eps^4)\,, \\
B_{6,2} &=& 
\frac{1}{3\eps^3}+\frac{7}{3\eps^2}+\frac{31}{3\eps}+\biggl(\frac{8\zeta_3}{3}+\frac{103}{3}\biggr)
+\eps\biggl(\frac{235}{3}+\frac{56\zeta_3}{3}+\frac{2\pi^4}{45}\biggr)\nonumber \\ & &
+\eps^2\biggl(\frac{19}{3}+120\zeta_5+\frac{320\zeta_3}{3}+\frac{14\pi^4}{45}\biggr)\nonumber \\ & &
+\eps^3\biggl(-\frac{3953}{3}+840\zeta_5+\frac{1832\zeta_3}{3}+\frac{16\pi^4}{9}+\frac{176\pi^6}{567}-\frac{292\zeta_3^2}{3}\biggr)
+ {\cal O} (\eps^4)\,,\end{eqnarray}
\begin{eqnarray}
B_{8,1} &=& 
20\zeta_5+\eps\biggl(68\zeta_3^2+40\zeta_5+\frac{10\pi^6}{189}\biggr)
\nonumber \\
&&+\eps^2\biggl(136\zeta_3^2+\frac{34\pi^4\zeta_3}{15}+80\zeta_5+\frac{20\pi^6}{189}+450\zeta_7\biggr)\nonumber \\ & &
+ {\cal O} (\eps^3) \,,\\
C_{6,1} &=& 
\frac{1}{2\eps^3}+\frac{7}{2\eps^2}+\frac{1}{\eps}\biggl(\frac{\pi^2}{6}+\frac{33}{2}\biggr) 
+\biggl(\frac{7\pi^2}{6}-5\zeta_3+\frac{131}{2}\biggr)\nonumber \\ & & 
+\eps\biggl(\frac{11\pi^2}{2}-35\zeta_3-\frac{\pi^4}{20}+\frac{473}{2}\biggr)\nonumber \\ & &
+\eps^2\biggl(\frac{131\pi^2}{6}-\frac{5\pi^2\zeta_3}{3}-165\zeta_3-27\zeta_5-\frac{7\pi^4}{20}+\frac{1611}{2}\biggr)\nonumber \\ & &
+\eps^3\biggl(\frac{473\pi^2}{6}-\frac{35\pi^2\zeta_3}{3}-655\zeta_3-189\zeta_5-\frac{33\pi^4}{20}-\frac{61\pi^6}{756}+25\zeta_3^2+\frac{5281}{2}\biggr) 
\nonumber \\
&&+ {\cal O} (\eps^4) \,, \\
C_{8,1} &=& 
\frac{1}{\eps^5}+\frac{2}{\eps^4}+\frac{1}{\eps^{3}}\biggl(-\frac{5\pi^2}{6}+4\biggr)
+\frac{1}{\eps^{2}}\biggl(8-\frac{5\pi^2}{3}-29\zeta_3\biggr) \nonumber \\ & &
+\frac{1}{\eps}\biggl(16-\frac{10\pi^2}{3}-58\zeta_3-\frac{121\pi^4}{180}\biggr)\nonumber \\ & &
+\biggl(32-\frac{20\pi^2}{3}+\frac{29\pi^2\zeta_3}{3}-116\zeta_3-123\zeta_5-\frac{121\pi^4}{90}\biggr)\nonumber \\ & &
+\eps\biggl(\frac{58\pi^2\zeta_3}{3}-232\zeta_3-246\zeta_5-\frac{40\pi^2}{3}+323\zeta_3^2-\frac{121\pi^4}{45}+64-\frac{163\pi^6}{3780}\biggr) 
\nonumber \\ & &+ {\cal O} (\eps^2) \,.
\end{eqnarray}
}
The genuine three-loop vertex integrals are:
{\allowdisplaybreaks
\begin{eqnarray}
A_{5,1} &=& 
\frac{1}{24\eps^2}+\frac{19}{48\eps}+\biggl(\frac{233}{96}+\frac{\pi^2}{24}\biggr) 
+\eps\biggl(\frac{2363}{192}+\frac{19\pi^2}{48}-\frac{11\zeta_3}{12}\biggr)\nonumber \\ & &
+\eps^2\biggl(\frac{7227}{128}+\frac{233\pi^2}{96}+\frac{\pi^4}{80}-\frac{209\zeta_3}{24}\biggr)\nonumber \\ & &
+\eps^3\biggl(\frac{62641}{256}+\frac{2363\pi^2}{192}+\frac{19\pi^4}{160}-\frac{2563\zeta_3}{48}-\frac{11\pi^2\zeta_3}{12}-\frac{35\zeta_5}{4}\biggr)\nonumber \\ & &
+\eps^4\biggl(\frac{1575481}{1536}+\frac{7227\pi^2}{128}+\frac{233\pi^4}{320}-\frac{919\pi^6}{45360}-\frac{25993\zeta_3}{96}\nonumber \\ & &
-\frac{209\pi^2\zeta_3}{24}+\frac{121\zeta_3^2}{12}-\frac{665\zeta_5}{8}\biggr)\nonumber \\ & &
+ {\cal O} (\eps^5)\,,\\
A_{5,2} &=& 
-\frac{1}{6\eps^2}-\frac{5}{3\eps}+\biggl(-\frac{32}{3}-\frac{\pi^2}{12}\biggr)
+\eps\biggl(-56-\frac{5\pi^2}{6}+\frac{11\zeta_3}{3}\biggr)\nonumber \\ & &
+\eps^2\biggl(-264-\frac{16\pi^2}{3}+\frac{19\pi^4}{720}+\frac{110\zeta_3}{3}\biggr)\nonumber \\ & &
+\eps^3\biggl(-1168-28\pi^2+\frac{19\pi^4}{72}+\frac{704\zeta_3}{3}+\frac{11\pi^2\zeta_3}{6}+35\zeta_5\biggr)\nonumber \\ & &
+\eps^4\biggl(-\frac{14912}{3}-132\pi^2+\frac{76\pi^4}{45}+\frac{9011\pi^6}{90720}+1232\zeta_3+\frac{55\pi^2\zeta_3}{3}-\frac{121\zeta_3^2}{3}+350\zeta_5\biggr)\nonumber \\ & &
+ {\cal O} (\eps^5)\,, \\
A_{6,1} &=& 
\frac{1}{3\eps^3}+\frac{8}{3\eps^2}+\frac{1}{\eps}\biggl(\frac{44}{3}+\frac{\pi^2}{3}\biggr) 
+\biggl(\frac{8\pi^2}{3}+\frac{208}{3}-\frac{16\zeta_3}{3}\biggr)\nonumber \\ & &
+\eps\biggl(304+\frac{44\pi^2}{3}+\frac{2\pi^4}{15}-\frac{128\zeta_3}{3}\biggr)\nonumber \\ & &
+\eps^2\biggl(1280+\frac{208\pi^2}{3}+\frac{16\pi^4}{15}-\frac{704\zeta_3}{3}-\frac{16\pi^2\zeta_3}{3}-56\zeta_5\biggr)\nonumber \\ & &
+\eps^3\biggl(\frac{15808}{3}+304\pi^2+\frac{88\pi^4}{15}-\frac{55\pi^6}{567}-\frac{3328\zeta_3}{3}-\frac{128\pi^2\zeta_3}{3}+\frac{128\zeta_3^2}{3}-448\zeta_5\biggr)\nonumber \\ & &
+ {\cal O} (\eps^4)\,, \\
A_{6,2} &=& 
\frac{2\zeta_3}{\eps}+\biggl(\frac{7\pi^4}{180}+18\zeta_3\biggr)\nonumber \\ & &
+\eps\biggl(\frac{7\pi^4}{20}+122\zeta_3-\frac{2\pi^2\zeta_3}{3}+10\zeta_5\biggr) \nonumber \\ & &
+\eps^2\biggl(\frac{427\pi^4}{180}-\frac{163\pi^6}{7560}+738\zeta_3-6\pi^2\zeta_3-76\zeta_3^2+90\zeta_5\biggr)
+ {\cal O} (\eps^3) \,, \\
A_{6,3} &=& 
\frac{1}{6\eps^3}+\frac{3}{2\eps^2}+\frac{1}{\eps}\biggl(\frac{55}{6}+\frac{\pi^2}{6}\biggr) 
+\biggl(\frac{3\pi^2}{2}+\frac{95}{2}-\frac{17\zeta_3}{3}\biggr)\nonumber \\ & &
+\eps\biggl(\frac{1351}{6}+\frac{55\pi^2}{6}+\frac{\pi^4}{90}-51\zeta_3\biggr)\nonumber \\ & &
+\eps^2\biggl(\frac{2023}{2}+\frac{95\pi^2}{2}+\frac{\pi^4}{10}-\frac{935\zeta_3}{3}-\frac{10\pi^2\zeta_3}{3}-65\zeta_5\biggr)\nonumber \\ & &
+\eps^3\biggl(\frac{26335}{6}+\frac{1351\pi^2}{6}+\frac{11\pi^4}{18}-\frac{7\pi^6}{54}-1615\zeta_3-30\pi^2\zeta_3+\frac{268\zeta_3^2}{3}-585\zeta_5\biggr)
\nonumber \\ & &+ {\cal O} (\eps^4)\,, \\
A_{7,1} &=& 
\frac{1}{4\eps^5}+\frac{1}{2\eps^4}+\frac{1}{\eps^{3}}\biggl(1-\frac{\pi^2}{6}\biggr) 
+\frac{1}{\eps^{2}}\biggl(2-\frac{\pi^2}{3}-10\zeta_3\biggr)\nonumber \\ & &
+\frac{1}{\eps}\biggl(4-\frac{2\pi^2}{3}-\frac{11\pi^4}{45}-20\zeta_3\biggr)\nonumber \\ & &
+\biggl(-\frac{22\pi^4}{45}-\frac{4\pi^2}{3}+\frac{14\pi^2\zeta_3}{3}+8-40\zeta_3-88\zeta_5\biggr)\nonumber \\ & &
+\eps\biggl(16-\frac{8\pi^2}{3}-\frac{44\pi^4}{45}-\frac{943\pi^6}{7560}-80\zeta_3+\frac{28\pi^2\zeta_3}{3}+196\zeta_3^2-176\zeta_5\biggr) 
\nonumber \\ & &
+ {\cal O} (\eps^2) \,, \\
A_{7,2} &=& 
\frac{\pi^2}{12\eps^3}+\frac{1}{\eps^{2}}\biggl(\frac{\pi^2}{6}+2\zeta_3\biggr) 
+\frac{1}{\eps}\biggl(\frac{\pi^2}{3}+\frac{83\pi^4}{720}+4\zeta_3\biggr)\nonumber \\ & &
+\biggl(\frac{2\pi^2}{3}+\frac{83\pi^4}{360}+8\zeta_3-\frac{5\pi^2\zeta_3}{3}+15\zeta_5\biggr)\nonumber \\ & &
+\eps\biggl(\frac{4\pi^2}{3}+\frac{83\pi^4}{180}+\frac{2741\pi^6}{90720}+16\zeta_3-\frac{10\pi^2\zeta_3}{3}-73\zeta_3^2+30\zeta_5\biggr) 
\nonumber \\ & &+ {\cal O} (\eps^2)\,, \\
A_{7,3} &=& 
+\frac{1}{\eps}\biggl(-\frac{\pi^2\zeta_3}{6}-10\zeta_5\biggr) 
+\biggl(-\frac{119\pi^6}{2160}-\frac{31\zeta_3^2}{2}\biggr) 
+ {\cal O} (\eps )\,, \\
A_{7,4} &=& 
\frac{6\zeta_3}{\eps^2}+\frac{1}{\eps}\biggl(\frac{11\pi^4}{90}+36\zeta_3\biggr)
+\biggl(216\zeta_3-2\pi^2\zeta_3+\frac{11\pi^4}{15}+46\zeta_5\biggr)\nonumber \\ & &
+\eps\biggl(\frac{22\pi^4}{5}-\frac{19\pi^6}{270}+1296\zeta_3-12\pi^2\zeta_3-282\zeta_3^2+276\zeta_5\biggr) 
+ {\cal O} (\eps^2)\,, \\
A_{7,5} &=& 
+\biggl(2\pi^2\zeta_3+10\zeta_5\biggr) 
+\eps\biggl(\frac{11\pi^6}{162}+12\pi^2\zeta_3+18\zeta_3^2+60\zeta_5\biggr) 
+ {\cal O} (\eps^2)\,, \\
A_{8,1} &=& 
-\frac{8\zeta_3}{3\eps^2}+\frac{1}{\eps}\biggl(-\frac{5\pi^4}{27}+8\zeta_3\biggr) 
+\biggl(\frac{5\pi^4}{9}-24\zeta_3+\frac{52\pi^2\zeta_3}{9}-\frac{352\zeta_5}{3}\biggr)\nonumber \\ & &
+\eps\biggl(-\frac{5\pi^4}{3}-\frac{1709\pi^6}{8505}+72\zeta_3-\frac{52\pi^2\zeta_3}{3}+\frac{332\zeta_3^2}{3}+352\zeta_5\biggr) 
+ {\cal O} (\eps^2)\,.
\end{eqnarray}} 
The most complicated three-loop vertex integrals 
are the nine-line master integrals~\cite{masterD}:
{\allowdisplaybreaks
\begin{eqnarray}
A_{9,1} &=& 
-\frac{1}{18\eps^5}+\frac{1}{2\eps^4}+\frac{1}{\eps^{3}}\biggl(-\frac{53}{18}-\frac{4\pi^2}{27}\biggr)\nonumber \\ & &
+\frac{1}{\eps^{2}}\biggl(\frac{29}{2}+\frac{22\pi^2}{27}-2\zeta_3\biggr)\nonumber \\ & &
+\frac{1}{\eps}\biggl(-\frac{8\pi^2}{3}+\frac{158\zeta_3}{9}-\frac{20\pi^4}{81}-\frac{129}{2}\biggr)\nonumber \\ & &
+\biggl(\frac{322\pi^4}{405}+6\pi^2-\frac{14\pi^2\zeta_3}{3}+\frac{537}{2}-\frac{578\zeta_3}{9}-\frac{238\zeta_5}{3}\biggr)\nonumber \\ & &
+\eps\biggl(-\frac{2133}{2}-4\pi^2-\frac{302\pi^4}{135}-\frac{2398\pi^6}{5103}+158\zeta_3-\frac{26\pi^2\zeta_3}{3}-\frac{466\zeta_3^2}{3}+\frac{826\zeta_5}{3}\biggr) 
 \nonumber\\ & &
+ {\cal O} (\eps^2) \,, \\
A_{9,2} & =&
\frac{2}{9\eps^6}+\frac{5}{6\eps^5}+\frac{1}{\eps^{4}}\biggl(-\frac{20}{9}-\frac{7\pi^2}{27}\biggr)\nonumber \\ & &
+\frac{1}{\eps^{3}}\biggl(\frac{50}{9}-\frac{17\pi^2}{27}-\frac{91\zeta_3}{9}\biggr)\nonumber \\ & &
+\frac{1}{\eps^{2}}\biggl(\frac{4\pi^2}{3}-\frac{166\zeta_3}{9}-\frac{373\pi^4}{1080}-\frac{110}{9}\biggr)\nonumber \\ & &
+\frac{1}{\eps}\biggl(\frac{494\zeta_3}{9}+\frac{179\pi^2\zeta_3}{27}-167\zeta_5-\frac{16\pi^2}{9}-\frac{187\pi^4}{540}+\frac{170}{9}\biggr) 
\nonumber \\&&
+\biggl(\frac{130}{9}-\frac{32\pi^2}{9}-\frac{1466\zeta_3}{9}+
\frac{679\pi^4}{540}+\frac{682\pi^2\zeta_3}{27}-\frac{1390\zeta_5}{3}
-\frac{59797\pi^6}{136080}+\frac{169\zeta_3^2}{9}\biggr)\nonumber \\ & &
+ {\cal O} (\eps) \,, \\
A_{9,4} & =& 
\frac{1}{9\eps^6}+\frac{8}{9\eps^5}+\frac{1}{\eps^{4}}\biggl(-1-\frac{10\pi^2}{27}\biggr)\nonumber \\ & &
+\frac{1}{\eps^{3}}\biggl(-\frac{14}{9}-\frac{47\pi^2}{27}-12\zeta_3\biggr)\nonumber \\ & &
+\frac{1}{\eps^{2}}\biggl(17+\frac{71\pi^2}{27}-\frac{200\zeta_3}{3}-\frac{47\pi^4}{810}\biggr)\nonumber \\ & &
+ \frac{1}{\eps} \biggl(
-84-\pi^2+\frac{940\zeta_3}{9}-\frac{671\pi^4}{540}
+\frac{652\pi^2\zeta_3}{27}-\frac{692\zeta_5}{9}
 \biggr)\nonumber \\ & &
+ \biggl(
339-15\pi^2-\frac{448\zeta_3}{9}+\frac{2689\pi^4}{1620}
+\frac{2188\pi^2\zeta_3}{27}-524\zeta_5+\frac{53563\pi^6}{102060}
+\frac{4564\zeta_3^2}{9}
\biggr)\nonumber \\ & &
+ {\cal O} (\eps) \,,
\end{eqnarray} }
where the analytic expressions for 
$A_{9,1}$ and for the pole parts of $A_{9,2}$ and $A_{9,4}$ were obtained 
independently in~\cite{masterC}.
  For the corresponding integrals with homogeneous transcendentality, one finds:
{\allowdisplaybreaks 
\begin{eqnarray}
A_{9,1n} &=&\frac{1}{36
    \e^6}+\frac{\pi^2}{18\e^4}+\frac{14\zeta_3}{9\e^3}+\frac{47\pi^4}{405\e^2}
\nonumber \\ &&
  \hspace{15mm}
  +\left(\frac{85}{27}\pi^2\zeta_3+20\zeta_5\right)\frac{1}{\e}
  +\frac{1160\pi^6}{5103}+\frac{137}{3}\zeta_3^2 +{\cal{O}}(\e) \\
A_{9,2n} &=&\frac{2}{9 \e^6}-\frac{7\pi^2}{27\e^4}-\frac{91\zeta_3}{9\e^3}-\frac{373\pi^4}{1080\e^2}\nonumber\\
&& \hspace{15mm}
+\left(\frac{179}{27}\pi^2\zeta_3-167\zeta_5\right)\frac{1}{\e}
-\frac{59797}{136080}\pi^6 + \frac{169}{9}\zeta_3^2 +{\cal{O}}(\e) \\
A_{9,4n} &=& - \frac{1}{9 \e^6}+\frac{10\pi^2}{27\e^4}+\frac{12\zeta_3}{\e^3}+\frac{47\pi^4}{810\e^2}\nonumber\\
&& \hspace{15mm}
+\left(-\frac{652}{27}\pi^2\zeta_3+\frac{692}{9}\zeta_5\right)\frac{1}{\e}
-\frac{53563}{102060}\pi^6 - \frac{4564}{9}\zeta_3^2 +{\cal{O}}(\e)
\end{eqnarray}
}

\section{Form factors in terms of master integrals}
\label{app:coeff}

The unrenormalised three-loop form factors can be expressed as a linear combination of master integrals. 
In the colour factor decomposition as defined in (\ref{eq:fq3lbare}) and (\ref{eq:fg3lbare}), these coefficients
read:
{\allowdisplaybreaks
\begin{eqnarray*}
\lefteqn{X^q_{C_F^3}=}\nonumber \\
&&{-}B_{4,1} 
\left(+\frac{489406D^3}{625}-\frac{43304589D^2}{3125}+\frac{615952127D}{7500}+\frac{34015}{4(2D-7)}-\frac{109222498}{75(2D-9)}\right.\nonumber \\
&&\left.\qquad+\frac{50720}{9(3D-10)}+\frac{6816654}{11(3D-14)}+\frac{89728}{25(D-2)}-\frac{12581}{12(D-3)}+\frac{6489724}{15(D-4)}\right.\nonumber \\
&&\left.\qquad+\frac{19326056092}{7734375(5D-16)}-\frac{7186019918}{78125(5D-18)}+\frac{643118017984}{703125(5D-22)}-\frac{1024}{3(D-2)^2}-\frac{779}{12(D-3)^2}\right.\nonumber \\
&&\left.\qquad+\frac{884312}{5(D-4)^2}+\frac{1187096}{15(D-4)^3}+\frac{745376}{15(D-4)^4}+\frac{91648}{5(D-4)^5}-\frac{53258146831}{562500}\right)\nonumber \\
&&{-}A_{5,1} 
\left(+\frac{54568D^3}{625}-\frac{16060301D^2}{9375}+\frac{135964099D}{11250}-\frac{7315}{2(2D-7)}-\frac{59657807}{1300(2D-9)}\right.\nonumber \\
&&\left.\qquad-\frac{36208}{27(3D-10)}+\frac{142784}{75(D-2)}-\frac{106}{3(D-3)}+\frac{770008}{15(D-4)}+\frac{3481535536}{3046875(5D-16)}\right.\nonumber \\
&&\left.\qquad+\frac{2887120096}{78125(5D-18)}-\frac{32265012416}{234375(5D-22)}+\frac{83104}{3(D-4)^2}+\frac{12800}{(D-4)^3}+\frac{26112}{5(D-4)^4}\right.\nonumber \\
&&\left.\qquad-\frac{35332079719}{1687500}\right)\nonumber \\
&&{+}A_{5,2} 
\left(+\frac{87316D^3}{1875}-\frac{8532244D^2}{9375}+\frac{15436454D}{3125}+\frac{418}{(2D-7)}+\frac{3273151}{325(2D-9)}\right.\nonumber \\
&&\left.\qquad+\frac{9920}{27(3D-10)}+\frac{2913100}{99(3D-14)}+\frac{400}{27(3D-8)}+\frac{152056}{225(D-2)}-\frac{70952}{15(D-4)}\right.\nonumber \\
&&\left.\qquad-\frac{9365062376}{33515625(5D-16)}-\frac{368980436}{78125(5D-18)}-\frac{37325556352}{703125(5D-22)}-\frac{512}{3(D-2)^2}-\frac{97088}{15(D-4)^2}\right.\nonumber \\
&&\left.\qquad-\frac{19392}{5(D-4)^3}-\frac{22016}{15(D-4)^4}-\frac{3578943149}{421875}\right)\nonumber \\
&&{-}B_{5,1}
\left(+\frac{46827D^3}{5000}-\frac{7169631D^2}{50000}+\frac{221676243D}{100000}+\frac{177975}{64(2D-7)}-\frac{2274503}{130(2D-9)}\right.\nonumber \\
&&\left.\qquad+\frac{9728}{15(D-2)}-\frac{151}{2(D-3)}+\frac{44476}{5(D-4)}-\frac{50689072}{3046875(5D-16)}+\frac{648026848}{78125(5D-18)}\right.\nonumber \\
&&\left.\qquad+\frac{1055401984}{78125(5D-22)}+\frac{53792}{5(D-4)^2}+\frac{33824}{5(D-4)^3}+\frac{19072}{5(D-4)^4}-\frac{3774996391}{1000000}\right)\nonumber \\
&&{+}B_{5,2} 
\left(+\frac{105167D^3}{1875}-\frac{33225224D^2}{28125}+\frac{792891607D}{84375}-\frac{1654184}{65(2D-9)}-\frac{4912}{81(3D-10)}\right.\nonumber \\
&&\left.\qquad+\frac{29696}{15(D-2)}+\frac{150868}{15(D-4)}+\frac{500415488}{3046875(5D-16)}+\frac{470084528}{78125(5D-18)}+\frac{3580901184}{78125(5D-22)}\right.\nonumber \\
&&\left.\qquad+\frac{29936}{3(D-4)^2}+\frac{22112}{3(D-4)^3}+\frac{46592}{15(D-4)^4}-\frac{21148347004}{1265625}\right)\nonumber \\
&&{-}A_{6,1}
\left(+\frac{2207D^3}{375}-\frac{7837D^2}{50}+\frac{62616143D}{56250}+\frac{291310}{297(3D-14)}+\frac{2252}{9(D-2)}\right.\nonumber \\
&&\left.\qquad+\frac{23}{(D-3)}+\frac{3136}{15(D-4)}+\frac{1553908}{61875(5D-16)}-\frac{5323758}{15625(5D-18)}-\frac{74762464}{46875(5D-22)}\right.\nonumber \\
&&\left.\qquad+\frac{496}{5(D-4)^2}+\frac{192}{5(D-4)^3}-\frac{183504334}{84375}\right)\nonumber \\
&&{+}A_{6,2}
\left(+\frac{39857D^3}{3000}-\frac{4628009D^2}{15000}+\frac{22107268D}{9375}-\frac{627}{16(2D-7)}+\frac{235409}{600(2D-9)}\right.\nonumber \\
&&\left.\qquad+\frac{98768}{225(D-2)}-\frac{4993}{280(D-3)}+\frac{1024}{15(D-4)}+\frac{36333448}{140625(5D-16)}+\frac{446887648}{984375(5D-22)}\right.\nonumber \\
&&\left.\qquad+\frac{1024}{3(D-2)^2}+\frac{58}{3(D-3)^2}+\frac{256}{15(D-4)^2}-\frac{817543919}{150000}\right)\nonumber \\
&&{-}A_{6,3}
\left(+\frac{3422D^3}{125}-\frac{2017249D^2}{3750}+\frac{835107683D}{225000}+\frac{1045}{8(2D-7)}+\frac{4880379}{5200(2D-9)}\right.\nonumber \\
&&\left.\qquad+\frac{80}{27(3D-8)}+\frac{28736}{25(D-2)}-\frac{161}{4(D-3)}+\frac{4336}{5(D-4)}+\frac{30753004}{203125(5D-16)}\right.\nonumber \\
&&\left.\qquad-\frac{1526704}{3125(5D-18)}-\frac{85442816}{15625(5D-22)}+\frac{576}{(D-4)^2}+\frac{3392}{15(D-4)^3}-\frac{10891722217}{1350000}\right)\nonumber \\
&&{+}B_{6,1}
\frac{(D^2-7D+16)^3}{(D-4)^3}\nonumber \\
&&{-}B_{6,2}
\left(+\frac{7D^3}{8}-\frac{1109D^2}{48}+\frac{29395D}{288}+\frac{8475}{64(2D-7)}+\frac{200}{27(3D-8)}\right.\nonumber \\
&&\left.\qquad-\frac{264}{(D-4)}-\frac{152}{(D-4)^2}-\frac{160}{(D-4)^3}-\frac{374753}{1728}\right)\nonumber \\
&&{-}C_{6,1}
\left(+\frac{7D^3}{8}-\frac{1109D^2}{48}+\frac{29395D}{288}+\frac{8475}{64(2D-7)}+\frac{200}{27(3D-8)}\right.\nonumber \\
&&\left.\qquad-\frac{264}{(D-4)}-\frac{152}{(D-4)^2}-\frac{160}{(D-4)^3}-\frac{374753}{1728}\right)\nonumber \\
&&{-}A_{7,1}
\left(+\frac{21D^3}{50}-\frac{5907D^2}{500}+\frac{523857D}{5000}-\frac{1213}{12(2D-7)}+\frac{29539}{624(2D-9)}\right.\nonumber \\
&&\left.\qquad+\frac{64}{3(D-2)}+\frac{80}{(D-4)}-\frac{655856}{121875(5D-16)}-\frac{388064}{9375(5D-18)}-\frac{2151447}{10000}\right)\nonumber \\
&&{+}A_{7,2}
\left(+\frac{15D^3}{16}-\frac{733D^2}{32}+\frac{228267D}{1600}+\frac{42745}{288(2D-7)}+\frac{232399}{8320(2D-9)}\right.\nonumber \\
&&\left.\qquad+\frac{488}{45(D-2)}-\frac{128}{(D-4)}-\frac{2821088}{14625(5D-16)}+\frac{47928}{125(5D-18)}-\frac{4633049}{16000}\right)\nonumber \\
&&{+}A_{7,3}
\left(+\frac{601D^3}{1250}-\frac{60199D^2}{6250}+\frac{760189D}{6250}-\frac{1213}{12(2D-7)}+\frac{29539}{2340(2D-9)}\right.\nonumber \\
&&\left.\qquad+\frac{496}{(D-2)}+\frac{8329}{210(D-3)}-\frac{909167104}{3046875(5D-16)}+\frac{101225984}{703125(5D-18)}-\frac{15661504}{546875(5D-22)}\right.\nonumber \\
&&\left.\qquad+\frac{21}{(D-3)^2}-\frac{8803773}{15625}\right)\nonumber \\
&&{-}A_{7,4}
\left(+\frac{2489D^3}{5000}-\frac{686707D^2}{50000}+\frac{7042751D}{60000}+\frac{865}{72(2D-7)}+\frac{235409}{2880(2D-9)}\right.\nonumber \\
&&\left.\qquad+\frac{556}{45(D-2)}-\frac{4397}{420(D-3)}-\frac{16}{5(D-4)}+\frac{93131696}{703125(5D-16)}-\frac{9430916}{703125(5D-18)}\right.\nonumber \\
&&\left.\qquad-\frac{365471216}{4921875(5D-22)}+\frac{1}{3(D-3)^2}-\frac{277480707}{1000000}\right)\nonumber \\
&&{-}A_{7,5}
\left(+\frac{2489D^3}{5000}-\frac{686707D^2}{50000}+\frac{7042751D}{60000}+\frac{865}{72(2D-7)}+\frac{235409}{2880(2D-9)}\right.\nonumber \\
&&\left.\qquad+\frac{556}{45(D-2)}-\frac{4397}{420(D-3)}-\frac{16}{5(D-4)}+\frac{93131696}{703125(5D-16)}-\frac{9430916}{703125(5D-18)}\right.\nonumber \\
&&\left.\qquad-\frac{365471216}{4921875(5D-22)}+\frac{1}{3(D-3)^2}-\frac{277480707}{1000000}\right)\nonumber \\
&&{+}A_{8,1}
\left(+\frac{3411D^3}{80000}-\frac{758793D^2}{800000}+\frac{3243781D}{320000}+\frac{33573}{1024(2D-7)}+\frac{32}{(D-2)}\right.\nonumber \\
&&\left.\qquad-\frac{3015}{448(D-3)}+\frac{4}{5(D-4)}+\frac{3411716}{78125(5D-16)}-\frac{8269536}{78125(5D-18)}-\frac{1425936}{546875(5D-22)}\right.\nonumber \\
&&\left.\qquad+\frac{4389}{256(2D-7)^2}-\frac{663954073}{16000000}\right)\nonumber \\
&&{-}B_{8,1}
\frac{(D^3-20D^2+104D-176)(D^2-7D+16)}{8(2D-7)(D-4)}\nonumber \\
&&{-}C_{8,1}
\frac{(D^3-20D^2+104D-176)(D^2-7D+16)}{8(2D-7)(D-4)}\nonumber \\
&&{+}A_{9,1}
\left(+\frac{243D^3}{1250}-\frac{14661D^2}{3125}+\frac{257769D}{6250}+\frac{256}{5(D-2)}-\frac{225}{14(D-3)}\right.\nonumber \\
&&\left.\qquad-\frac{48}{5(D-4)}+\frac{4083992}{78125(5D-16)}+\frac{6463488}{78125(5D-18)}+\frac{2127008}{546875(5D-22)}-\frac{4086513}{31250}\right)\nonumber \\
&&{-}A_{9,2}
\frac{3(3D-14)(D^6-41D^5+661D^4-4992D^3+19276D^2-37104D+28288)}{10(5D-16)(5D-18)(5D-22)(D-3)}\nonumber \\
&&{-}A_{9,4}
\left(+\frac{567D^3}{80000}-\frac{125091D^2}{800000}+\frac{1808937D}{1600000}+\frac{4067}{2304(2D-7)}+\frac{232399}{998400(2D-9)}\right.\nonumber \\
&&\left.\qquad-\frac{16}{75(D-2)}-\frac{225}{448(D-3)}+\frac{8388688}{3046875(5D-16)}-\frac{574016}{234375(5D-18)}-\frac{7557808}{4921875(5D-22)}\right.\nonumber \\
&&\left.\qquad-\frac{38866491}{16000000}\right) 
\end{eqnarray*}
}

{\allowdisplaybreaks
\begin{eqnarray*}
\lefteqn{X^q_{C_F^2C_A}=}\nonumber \\
&&{+}B_{4,1} 
\left(+\frac{225717D^3}{250}-\frac{9995657D^2}{625}+\frac{893831341D}{9375}+\frac{45685}{24(2D-7)}-\frac{990312631}{975(2D-9)}\right.\nonumber \\
&&\left.\qquad+\frac{116080}{21(3D-10)}+\frac{707967}{(3D-14)}-\frac{371482}{6237(D-1)}+\frac{125032}{25(D-2)}-\frac{875}{4(D-3)}\right.\nonumber \\
&&\left.\qquad+\frac{172908907}{405(D-4)}-\frac{5622111978}{2234375(5D-16)}-\frac{345702357}{3125(5D-18)}-\frac{5032168544}{46875(5D-22)}-\frac{1280}{3(D-2)^2}\right.\nonumber \\
&&\left.\qquad+\frac{9}{4(D-3)^2}+\frac{22959056}{135(D-4)^2}+\frac{3875224}{45(D-4)^3}+\frac{2622656}{45(D-4)^4}+\frac{340096}{15(D-4)^5}\right.\nonumber \\
&&\left.\qquad-\frac{2728978211}{25000}\right)\nonumber \\
&&{+}A_{5,1} 
\left(+\frac{71304D^3}{625}-\frac{12785517D^2}{6250}+\frac{156484099D}{12500}-\frac{17815}{3(2D-7)}-\frac{97578701}{2600(2D-9)}\right.\nonumber \\
&&\left.\qquad-\frac{73232}{63(3D-10)}+\frac{2467480}{27027(D-1)}+\frac{108736}{75(D-2)}-\frac{1}{(D-3)}+\frac{6422264}{135(D-4)}\right.\nonumber \\
&&\left.\qquad+\frac{9290021216}{11171875(5D-16)}+\frac{26729196704}{1015625(5D-18)}-\frac{10782693376}{78125(5D-22)}+\frac{179840}{9(D-4)^2}+\frac{130816}{15(D-4)^3}\right.\nonumber \\
&&\left.\qquad+\frac{50176}{15(D-4)^4}-\frac{23118311953}{1125000}\right)\nonumber \\
&&{-}A_{5,2} 
\left(+\frac{25782D^3}{625}-\frac{24183202D^2}{28125}+\frac{437583827D}{84375}+\frac{2036}{3(2D-7)}+\frac{7969573}{650(2D-9)}\right.\nonumber \\
&&\left.\qquad+\frac{190048}{567(3D-10)}+\frac{100850}{3(3D-14)}+\frac{16}{(3D-8)}-\frac{2178524}{81081(D-1)}+\frac{71828}{75(D-2)}\right.\nonumber \\
&&\left.\qquad-\frac{1092208}{405(D-4)}-\frac{9583273136}{33515625(5D-16)}-\frac{4168502038}{1015625(5D-18)}-\frac{14910736064}{234375(5D-22)}-\frac{640}{3(D-2)^2}\right.\nonumber \\
&&\left.\qquad-\frac{523712}{135(D-4)^2}-\frac{90176}{45(D-4)^3}-\frac{35072}{45(D-4)^4}-\frac{21872512759}{2531250}\right)\nonumber \\
&&{+}B_{5,1}
\left(+\frac{114279D^3}{10000}-\frac{22958287D^2}{100000}+\frac{448177891D}{200000}+\frac{177975}{128(2D-7)}-\frac{326249}{26(2D-9)}\right.\nonumber \\
&&\left.\qquad+\frac{6128}{15(D-2)}-\frac{2}{(D-3)}+\frac{35294}{5(D-4)}-\frac{25589872}{3046875(5D-16)}+\frac{831628448}{78125(5D-18)}\right.\nonumber \\
&&\left.\qquad-\frac{130364416}{78125(5D-22)}+\frac{7272}{(D-4)^2}+\frac{17696}{5(D-4)^3}+\frac{10624}{5(D-4)^4}-\frac{7313613927}{2000000}\right)\nonumber \\
&&{-}B_{5,2} 
\left(+\frac{44339D^3}{625}-\frac{4405736D^2}{3125}+\frac{287753909D}{28125}-\frac{237272}{13(2D-9)}+\frac{16}{27(3D-10)}\right.\nonumber \\
&&\left.\qquad+\frac{25456}{15(D-2)}+\frac{154624}{15(D-4)}-\frac{76140512}{3046875(5D-16)}+\frac{650239528}{78125(5D-18)}+\frac{1665432384}{78125(5D-22)}\right.\nonumber \\
&&\left.\qquad+\frac{47624}{5(D-4)^2}+\frac{32416}{5(D-4)^3}+\frac{41984}{15(D-4)^4}-\frac{7361965928}{421875}\right)\nonumber \\
&&{+}A_{6,1}
\left(+\frac{4678D^3}{625}-\frac{609667D^2}{3125}+\frac{25888633D}{18750}+\frac{10085}{9(3D-14)}+\frac{545}{54(D-1)}\right.\nonumber \\
&&\left.\qquad+\frac{5366}{15(D-2)}+\frac{18}{(D-3)}+\frac{31492}{135(D-4)}+\frac{5042024}{78125(5D-16)}-\frac{23129849}{78125(5D-18)}\right.\nonumber \\
&&\left.\qquad-\frac{142238992}{78125(5D-22)}+\frac{1264}{9(D-4)^2}+\frac{512}{15(D-4)^3}-\frac{777739429}{281250}\right)\nonumber \\
&&{-}A_{6,2}
\left(+\frac{9671D^3}{1200}-\frac{1150061D^2}{6000}+\frac{27382277D}{18000}-\frac{509}{8(2D-7)}+\frac{5481191}{15600(2D-9)}\right.\nonumber \\
&&\left.\qquad+\frac{728}{25(D-2)}-\frac{257261}{1680(D-3)}+\frac{3164}{15(D-4)}+\frac{92725544}{121875(5D-16)}+\frac{20493136}{65625(5D-22)}\right.\nonumber \\
&&\left.\qquad+\frac{1280}{3(D-2)^2}+\frac{1}{2(D-3)^2}+\frac{3952}{45(D-4)^2}-\frac{379243601}{112500}\right)\nonumber \\
&&{+}A_{6,3}
\left(+\frac{60679D^3}{2500}-\frac{33322501D^2}{75000}+\frac{436696447D}{150000}+\frac{2545}{12(2D-7)}+\frac{14641137}{10400(2D-9)}\right.\nonumber \\
&&\left.\qquad+\frac{16}{5(3D-8)}-\frac{250462}{19305(D-1)}+\frac{37352}{75(D-2)}-\frac{1}{4(D-3)}+\frac{106784}{135(D-4)}\right.\nonumber \\
&&\left.\qquad+\frac{3533209912}{33515625(5D-16)}-\frac{433927224}{1015625(5D-18)}-\frac{487676544}{78125(5D-22)}+\frac{27568}{45(D-4)^2}+\frac{640}{3(D-4)^3}\right.\nonumber \\
&&\left.\qquad-\frac{2875843347}{500000}\right)\nonumber \\
&&{-}B_{6,2}
\left(+\frac{D^3}{16}+\frac{71D^2}{32}-\frac{283D}{64}-\frac{8475}{128(2D-7)}-\frac{8}{(3D-8)}\right.\nonumber \\
&&\left.\qquad-\frac{46}{9(D-1)}+\frac{1000}{9(D-4)}+\frac{80}{3(D-4)^2}+\frac{64}{(D-4)^3}+\frac{1587}{128}\right)\nonumber \\
&&{-}C_{6,1}
\left(+\frac{D^3}{16}+\frac{71D^2}{32}-\frac{283D}{64}-\frac{8475}{128(2D-7)}-\frac{8}{(3D-8)}\right.\nonumber \\
&&\left.\qquad-\frac{46}{9(D-1)}+\frac{1000}{9(D-4)}+\frac{80}{3(D-4)^2}+\frac{64}{(D-4)^3}+\frac{1587}{128}\right)\nonumber \\
&&{+}A_{7,1}
\left(+\frac{27D^3}{50}-\frac{7009D^2}{500}+\frac{584559D}{5000}-\frac{301}{6(2D-7)}+\frac{21185}{624(2D-9)}\right.\nonumber \\
&&\left.\qquad+\frac{32}{(D-2)}+\frac{80}{(D-4)}-\frac{6069872}{121875(5D-16)}-\frac{827368}{9375(5D-18)}-\frac{2492249}{10000}\right)\nonumber \\
&&{-}A_{7,2}
\left(+\frac{517D^3}{800}-\frac{142459D^2}{8000}+\frac{9023129D}{80000}+\frac{42745}{192(2D-7)}+\frac{697197}{16640(2D-9)}\right.\nonumber \\
&&\left.\qquad-\frac{1124}{143(D-1)}-\frac{76}{15(D-2)}-\frac{112}{(D-4)}-\frac{106444064}{446875(5D-16)}+\frac{9644612}{40625(5D-18)}\right.\nonumber \\
&&\left.\qquad-\frac{33226167}{160000}\right)\nonumber \\
&&{-}A_{7,3}
\left(+\frac{601D^3}{1250}-\frac{29899D^2}{6250}+\frac{152417D}{3125}-\frac{301}{6(2D-7)}+\frac{4237}{468(2D-9)}\right.\nonumber \\
&&\left.\qquad+\frac{544}{3(D-2)}+\frac{3683}{210(D-3)}+\frac{10424832}{1015625(5D-16)}+\frac{54045184}{703125(5D-18)}-\frac{12403904}{546875(5D-22)}\right.\nonumber \\
&&\left.\qquad-\frac{3}{(D-3)^2}-\frac{15077947}{62500}\right)\nonumber \\
&&{+}A_{7,4}
\left(+\frac{19D^3}{80}-\frac{458683D^2}{60000}+\frac{40603349D}{600000}-\frac{235}{32(2D-7)}+\frac{5481191}{74880(2D-9)}\right.\nonumber \\
&&\left.\qquad+\frac{118}{15(D-2)}-\frac{26393}{840(D-3)}-\frac{24}{5(D-4)}+\frac{46026288}{203125(5D-16)}-\frac{501158}{140625(5D-18)}\right.\nonumber \\
&&\left.\qquad-\frac{21760904}{328125(5D-22)}-\frac{62067409}{400000}\right)\nonumber \\
&&{+}A_{7,5}
\left(+\frac{19D^3}{80}-\frac{458683D^2}{60000}+\frac{40603349D}{600000}-\frac{235}{32(2D-7)}+\frac{5481191}{74880(2D-9)}\right.\nonumber \\
&&\left.\qquad+\frac{118}{15(D-2)}-\frac{26393}{840(D-3)}-\frac{24}{5(D-4)}+\frac{46026288}{203125(5D-16)}-\frac{501158}{140625(5D-18)}\right.\nonumber \\
&&\left.\qquad-\frac{21760904}{328125(5D-22)}-\frac{62067409}{400000}\right)\nonumber \\
&&{-}A_{8,1}
\left(+\frac{1197D^3}{160000}-\frac{979611D^2}{1600000}+\frac{8338443D}{640000}+\frac{29027}{2048(2D-7)}+\frac{160}{3(D-2)}\right.\nonumber \\
&&\left.\qquad-\frac{13665}{896(D-3)}-\frac{34}{5(D-4)}+\frac{23109548}{234375(5D-16)}-\frac{3147936}{78125(5D-18)}-\frac{1018736}{546875(5D-22)}\right.\nonumber \\
&&\left.\qquad+\frac{3563}{128(2D-7)^2}-\frac{2065843091}{32000000}\right)\nonumber \\
&&{+}B_{8,1}
\frac{(D^3-20D^2+104D-176)(D^2-7D+16)}{16(2D-7)(D-4)}\nonumber \\
&&{+}C_{8,1}
\frac{(D^3-20D^2+104D-176)(D^2-7D+16)}{16(2D-7)(D-4)}\nonumber \\
&&{-}A_{9,1}
\left(+\frac{243D^3}{1250}-\frac{14661D^2}{3125}+\frac{257769D}{6250}+\frac{256}{5(D-2)}-\frac{225}{14(D-3)}\right.\nonumber \\
&&\left.\qquad-\frac{48}{5(D-4)}+\frac{4083992}{78125(5D-16)}+\frac{6463488}{78125(5D-18)}+\frac{2127008}{546875(5D-22)}-\frac{4086513}{31250}\right)\nonumber \\
&&{+}A_{9,2}
\frac{(3D-14)(3D^6-108D^5+1586D^4-11304D^3+41928D^2-78208D+57984)}{10(5D-16)(5D-18)(5D-22)(D-3)}\nonumber \\
&&{+}A_{9,4}
\left(+\frac{1701D^3}{160000}-\frac{375273D^2}{1600000}+\frac{5426811D}{3200000}+\frac{4067}{1536(2D-7)}+\frac{232399}{665600(2D-9)}\right.\nonumber \\
&&\left.\qquad-\frac{8}{25(D-2)}-\frac{675}{896(D-3)}+\frac{4194344}{1015625(5D-16)}-\frac{287008}{78125(5D-18)}-\frac{3778904}{1640625(5D-22)}\right.\nonumber \\
&&\left.\qquad-\frac{116599473}{32000000}\right) 
\end{eqnarray*}
}

{\allowdisplaybreaks
\begin{eqnarray*}
\lefteqn{X^q_{C_FC_A^2}=}\nonumber \\
&&{-}B_{4,1} 
\left(+\frac{153701D^3}{625}-\frac{45111262D^2}{9375}+\frac{3307905503D}{112500}-\frac{7045}{6(2D-7)}-\frac{140183197}{975(2D-9)}\right.\nonumber \\
&&\left.\qquad-\frac{2060}{27(3D-10)}+\frac{9383166}{55(3D-14)}-\frac{165455}{2673(D-1)}+\frac{44164}{25(D-2)}+\frac{659}{12(D-3)}\right.\nonumber \\
&&\left.\qquad+\frac{416301857}{4860(D-4)}-\frac{138263099401}{402187500(5D-16)}-\frac{5093619454}{234375(5D-18)}-\frac{144904142656}{703125(5D-22)}-\frac{128}{(D-2)^2}\right.\nonumber \\
&&\left.\qquad+\frac{143}{12(D-3)^2}+\frac{8793673}{405(D-4)^2}+\frac{915068}{135(D-4)^3}+\frac{345128}{45(D-4)^4}+\frac{20864}{5(D-4)^5}\right.\nonumber \\
&&\left.\qquad-\frac{31855488829}{843750}\right)\nonumber \\
&&{-}A_{5,1} 
\left(+\frac{4277D^3}{125}-\frac{243647D^2}{375}+\frac{106547887D}{28125}-\frac{49315}{24(2D-7)}-\frac{18960447}{2600(2D-9)}\right.\nonumber \\
&&\left.\qquad+\frac{357584}{1323(3D-10)}+\frac{7618840}{567567(D-1)}+\frac{10064}{25(D-2)}+\frac{10}{3(D-3)}+\frac{3703898}{405(D-4)}\right.\nonumber \\
&&\left.\qquad+\frac{332977894}{1340625(5D-16)}+\frac{563608248}{203125(5D-18)}-\frac{1415791832}{46875(5D-22)}+\frac{3488}{189(D-1)^2}+\frac{81212}{45(D-4)^2}\right.\nonumber \\
&&\left.\qquad+\frac{11584}{45(D-4)^3}+\frac{256}{3(D-4)^4}-\frac{452116231}{67500}\right)\nonumber \\
&&{+}A_{5,2} 
\left(+\frac{20594D^3}{1875}-\frac{6630308D^2}{28125}+\frac{25700042D}{16875}+\frac{1409}{6(2D-7)}+\frac{2348211}{650(2D-9)}\right.\nonumber \\
&&\left.\qquad-\frac{11944}{567(3D-10)}+\frac{801980}{99(3D-14)}-\frac{10360820}{243243(D-1)}+\frac{75128}{225(D-2)}-\frac{559813}{1215(D-4)}\right.\nonumber \\
&&\left.\qquad-\frac{1022764469}{33515625(5D-16)}-\frac{1052157372}{1015625(5D-18)}-\frac{11705152928}{703125(5D-22)}-\frac{64}{(D-2)^2}-\frac{306428}{405(D-4)^2}\right.\nonumber \\
&&\left.\qquad-\frac{43696}{135(D-4)^3}-\frac{1408}{15(D-4)^4}-\frac{3344023858}{1265625}\right)\nonumber \\
&&{-}B_{5,1}
\left(+\frac{7497D^3}{1250}-\frac{394839D^2}{3125}+\frac{10983001D}{12500}-\frac{75999}{40(2D-9)}+\frac{1424}{15(D-2)}\right.\nonumber \\
&&\left.\qquad-\frac{1}{(D-3)}+\frac{1055}{(D-4)}-\frac{712201}{234375(5D-16)}+\frac{200430972}{78125(5D-18)}-\frac{146129984}{78125(5D-22)}\right.\nonumber \\
&&\left.\qquad+\frac{4176}{5(D-4)^2}+\frac{296}{(D-4)^3}+\frac{1152}{5(D-4)^4}-\frac{189456643}{125000}\right)\nonumber \\
&&{+}B_{5,2} 
\left(+\frac{13666D^3}{625}-\frac{4406452D^2}{9375}+\frac{92206756D}{28125}-\frac{13818}{5(2D-9)}+\frac{320}{27(3D-10)}\right.\nonumber \\
&&\left.\qquad+\frac{2536}{5(D-2)}+\frac{10325}{6(D-4)}-\frac{3748279}{156250(5D-16)}+\frac{169244232}{78125(5D-18)}+\frac{122152296}{78125(5D-22)}\right.\nonumber \\
&&\left.\qquad+\frac{22774}{15(D-4)^2}+\frac{15736}{15(D-4)^3}+\frac{7552}{15(D-4)^4}-\frac{2504444962}{421875}\right)\nonumber \\
&&{-}A_{6,1}
\left(+\frac{17033D^3}{7500}-\frac{1493603D^2}{25000}+\frac{95481487D}{225000}+\frac{80198}{297(3D-14)}+\frac{1711}{288(D-1)}\right.\nonumber \\
&&\left.\qquad+\frac{1090}{9(D-2)}+\frac{113}{32(D-3)}+\frac{479}{18(D-4)}+\frac{294544271}{15468750(5D-16)}-\frac{3011532}{78125(5D-18)}\right.\nonumber \\
&&\left.\qquad-\frac{104077288}{234375(5D-22)}+\frac{1853}{144(D-1)^2}+\frac{136}{9(D-4)^2}-\frac{64}{15(D-4)^3}-\frac{1505097247}{1687500}\right)\nonumber \\
&&{+}A_{6,2}
\left(+\frac{4249D^3}{6000}-\frac{322949D^2}{15000}+\frac{61126331D}{300000}-\frac{1409}{64(2D-7)}+\frac{403479}{5200(2D-9)}\right.\nonumber \\
&&\left.\qquad-\frac{872}{297(D-1)}+\frac{5584}{225(D-2)}-\frac{117431}{1680(D-3)}+\frac{9704}{135(D-4)}+\frac{5192329489}{20109375(5D-16)}\right.\nonumber \\
&&\left.\qquad+\frac{41976608}{984375(5D-22)}+\frac{128}{(D-2)^2}-\frac{25}{12(D-3)^2}+\frac{176}{5(D-4)^2}-\frac{846754451}{1800000}\right)\nonumber \\
&&{-}A_{6,3}
\left(+\frac{3073D^3}{625}-\frac{831416D^2}{9375}+\frac{41933917D}{75000}+\frac{7045}{96(2D-7)}+\frac{4880379}{10400(2D-9)}\right.\nonumber \\
&&\left.\qquad-\frac{13867}{429(D-1)}+\frac{9004}{75(D-2)}+\frac{13}{4(D-3)}+\frac{293}{30(D-4)}+\frac{1849361647}{67031250(5D-16)}\right.\nonumber \\
&&\left.\qquad-\frac{49983292}{1015625(5D-18)}-\frac{118945472}{78125(5D-22)}+\frac{236}{5(D-4)^2}+\frac{96}{5(D-4)^3}-\frac{51691069}{46875}\right)\nonumber \\
&&{-}A_{7,1}
\left(+\frac{33D^3}{200}-\frac{8111D^2}{2000}+\frac{645261D}{20000}+\frac{3}{16(2D-7)}+\frac{329}{64(2D-9)}\right.\nonumber \\
&&\left.\qquad+\frac{32}{3(D-2)}+\frac{20}{(D-4)}-\frac{220844}{9375(5D-16)}-\frac{105556}{3125(5D-18)}-\frac{2833051}{40000}\right)\nonumber \\
&&{+}A_{7,2}
\left(+\frac{71D^3}{800}-\frac{25417D^2}{8000}+\frac{1658227D}{80000}+\frac{42745}{576(2D-7)}+\frac{232399}{16640(2D-9)}\right.\nonumber \\
&&\left.\qquad-\frac{562}{143(D-1)}-\frac{236}{45(D-2)}-\frac{24}{(D-4)}-\frac{285048488}{4021875(5D-16)}+\frac{928156}{40625(5D-18)}\right.\nonumber \\
&&\left.\qquad-\frac{5030461}{160000}\right)\nonumber \\
&&{-}A_{7,3}
\left(+\frac{3D^3}{625}-\frac{2023D^2}{12500}-\frac{41819D}{12500}-\frac{3}{16(2D-7)}-\frac{329}{240(2D-9)}\right.\nonumber \\
&&\left.\qquad-\frac{44}{3(D-2)}-\frac{263}{105(D-3)}-\frac{2204608}{234375(5D-16)}-\frac{33404}{78125(5D-18)}+\frac{2035336}{546875(5D-22)}\right.\nonumber \\
&&\left.\qquad+\frac{7}{4(D-3)^2}+\frac{421802}{15625}\right)\nonumber \\
&&{+}A_{7,4}
\left(+\frac{57D^3}{10000}+\frac{116647D^2}{300000}-\frac{2694797D}{600000}+\frac{3845}{576(2D-7)}-\frac{134493}{8320(2D-9)}\right.\nonumber \\
&&\left.\qquad-\frac{38}{45(D-2)}+\frac{1833}{140(D-3)}+\frac{8}{5(D-4)}-\frac{732913468}{9140625(5D-16)}-\frac{368278}{234375(5D-18)}\right.\nonumber \\
&&\left.\qquad+\frac{71838976}{4921875(5D-22)}+\frac{1}{12(D-3)^2}+\frac{16428169}{2000000}\right)\nonumber \\
&&{+}A_{7,5}
\left(+\frac{57D^3}{10000}+\frac{116647D^2}{300000}-\frac{2694797D}{600000}+\frac{3845}{576(2D-7)}-\frac{134493}{8320(2D-9)}\right.\nonumber \\
&&\left.\qquad-\frac{38}{45(D-2)}+\frac{1833}{140(D-3)}+\frac{8}{5(D-4)}-\frac{732913468}{9140625(5D-16)}-\frac{368278}{234375(5D-18)}\right.\nonumber \\
&&\left.\qquad+\frac{71838976}{4921875(5D-22)}+\frac{1}{12(D-3)^2}+\frac{16428169}{2000000}\right)\nonumber \\
&&{-}A_{8,1}
\left(+\frac{1107D^3}{160000}+\frac{110409D^2}{1600000}-\frac{2547331D}{640000}+\frac{2273}{2048(2D-7)}-\frac{56}{3(D-2)}\right.\nonumber \\
&&\left.\qquad+\frac{5325}{896(D-3)}+\frac{18}{5(D-4)}-\frac{8995987}{234375(5D-16)}-\frac{493416}{78125(5D-18)}+\frac{152884}{546875(5D-22)}\right.\nonumber \\
&&\left.\qquad-\frac{9863}{1024(2D-7)^2}+\frac{700944509}{32000000}\right)\nonumber \\
&&{+}A_{9,1}
\left(+\frac{243D^3}{5000}-\frac{62289D^2}{50000}+\frac{283527D}{25000}+\frac{88}{5(D-2)}-\frac{285}{56(D-3)}\right.\nonumber \\
&&\left.\qquad-\frac{2}{(D-4)}+\frac{1549163}{78125(5D-16)}+\frac{1250592}{78125(5D-18)}+\frac{406132}{546875(5D-22)}-\frac{4692843}{125000}\right)\nonumber \\
&&{-}A_{9,2}
\frac{(3D-14)(3D^6-93D^5+1189D^4-7632D^3+26028D^2-45104D+31104)}{40(5D-16)(5D-18)(5D-22)(D-3)}\nonumber \\
&&{-}A_{9,4}
\left(+\frac{567D^3}{160000}-\frac{125091D^2}{1600000}+\frac{1808937D}{3200000}+\frac{4067}{4608(2D-7)}+\frac{232399}{1996800(2D-9)}\right.\nonumber \\
&&\left.\qquad-\frac{8}{75(D-2)}-\frac{225}{896(D-3)}+\frac{4194344}{3046875(5D-16)}-\frac{287008}{234375(5D-18)}-\frac{3778904}{4921875(5D-22)}\right.\nonumber \\
&&\left.\qquad-\frac{38866491}{32000000}\right)
\end{eqnarray*}
}

{\allowdisplaybreaks
\begin{eqnarray*}
\lefteqn{X^q_{C_F^2 N_F}=}\nonumber \\
&&{+}B_{4,1} 
\left(+\frac{72D^2}{5}-\frac{11524D}{75}-\frac{37120}{63(3D-10)}-\frac{742964}{6237(D-1)}+\frac{4}{(D-3)}\right.\nonumber \\
&&\left.\qquad+\frac{10576}{81(D-4)}+\frac{319872}{1375(5D-16)}-\frac{3088}{27(D-4)^2}+\frac{1024}{9(D-4)^3}+\frac{256}{3(D-4)^4}\right.\nonumber \\
&&\left.\qquad+\frac{412948}{1125}\right)\nonumber \\
&&{-}A_{5,1} 
\left(+\frac{1216D^2}{75}-\frac{83936D}{375}-\frac{512}{3(3D-10)}-\frac{2293120}{11583(D-1)}-\frac{19648}{81(D-4)}\right.\nonumber \\
&&\left.\qquad+\frac{297024}{6875(5D-16)}+\frac{3698688}{8125(5D-18)}-\frac{6272}{27(D-4)^2}-\frac{1024}{9(D-4)^3}+\frac{983968}{1875}\right)\nonumber \\
&&{-}A_{5,2} 
\left(+\frac{4D^2}{75}-\frac{78712D}{1125}-\frac{6656}{189(3D-10)}-\frac{32}{9(3D-8)}-\frac{4357048}{81081(D-1)}\right.\nonumber \\
&&\left.\qquad-\frac{20464}{81(D-4)}+\frac{131376}{6875(5D-16)}+\frac{1585152}{8125(5D-18)}-\frac{5312}{27(D-4)^2}-\frac{1024}{9(D-4)^3}\right.\nonumber \\
&&\left.\qquad+\frac{2009608}{16875}\right)\nonumber \\
&&{-}A_{6,1}
\frac{(D^2-7D+16)(6D^3-65D^2+238D-288)(D-2)}{2(D-3)(D-1)(D-4)^2}\nonumber \\
&&{+}A_{6,3}
\left(+\frac{28D^2}{25}-\frac{2868D}{125}-\frac{32}{45(3D-8)}-\frac{500924}{19305(D-1)}-\frac{400}{27(D-4)}\right.\nonumber \\
&&\left.\qquad-\frac{39984}{6875(5D-16)}-\frac{176128}{8125(5D-18)}-\frac{128}{9(D-4)^2}+\frac{334516}{5625}\right)\nonumber \\
&&{-}B_{6,2}
\frac{(D^2-7D+16)(3D^3-31D^2+110D-128)(D-2)}{(D-1)(3D-8)(D-4)^2}\nonumber \\
&&{-}C_{6,1}
\frac{(D^2-7D+16)(3D^3-31D^2+110D-128)(D-2)}{(D-1)(3D-8)(D-4)^2}\nonumber \\
&&{-}A_{7,2}
\frac{8(D-2)(D^4-28D^3+220D^2-696D+784)}{(5D-18)(D-1)(5D-16)} 
\end{eqnarray*}
}

{\allowdisplaybreaks
\begin{eqnarray*}
\lefteqn{X^q_{C_FC_A N_F}=}\nonumber \\
&&{+}B_{4,1} 
\left(+\frac{24D^2}{5}-\frac{396D}{25}-\frac{250}{3(3D-10)}+\frac{330910}{2673(D-1)}-\frac{170}{3(D-2)}\right.\nonumber \\
&&\left.\qquad+\frac{34408}{243(D-4)}+\frac{92064}{1375(5D-16)}+\frac{32}{3(D-2)^2}+\frac{14624}{81(D-4)^2}+\frac{3904}{27(D-4)^3}\right.\nonumber \\
&&\left.\qquad+\frac{256}{9(D-4)^4}+\frac{4364}{125}\right)\nonumber \\
&&{+}A_{5,1} 
\left(+\frac{208D^2}{75}-\frac{3656D}{125}+\frac{25352}{441(3D-10)}-\frac{4990592}{63063(D-1)}+\frac{184}{3(D-2)}\right.\nonumber \\
&&\left.\qquad-\frac{992}{9(D-4)}+\frac{148512}{6875(5D-16)}+\frac{1849344}{8125(5D-18)}-\frac{6976}{189(D-1)^2}-\frac{16}{(D-2)^2}\right.\nonumber \\
&&\left.\qquad+\frac{256}{27(D-4)^2}-\frac{512}{9(D-4)^3}+\frac{205952}{5625}\right)\nonumber \\
&&{+}A_{5,2} 
\left(+\frac{84D^2}{25}-\frac{6984D}{125}+\frac{460}{63(3D-10)}-\frac{20721640}{243243(D-1)}+\frac{20}{(D-2)}\right.\nonumber \\
&&\left.\qquad-\frac{24568}{243(D-4)}-\frac{24312}{6875(5D-16)}+\frac{792576}{8125(5D-18)}-\frac{16}{3(D-2)^2}-\frac{5024}{81(D-4)^2}\right.\nonumber \\
&&\left.\qquad-\frac{1024}{27(D-4)^3}+\frac{700768}{5625}\right)\nonumber \\
&&{+}A_{6,1}
\left(+\frac{3D^2}{2}-\frac{47D}{4}-\frac{3073}{72(D-1)}+\frac{86}{3(D-2)}+\frac{11}{8(D-3)}\right.\nonumber \\
&&\left.\qquad+\frac{8}{9(D-4)}-\frac{109}{4(D-1)^2}-\frac{8}{(D-2)^2}+\frac{16}{3(D-4)^2}+\frac{133}{4}\right)\nonumber \\
&&{+}A_{6,2}
\frac{90D^7-1803D^6+15301D^5-70848D^4+191676D^3-299024D^2+242976D-74880}{9(5D-16)(D-3)(D-2)^2(D-1)(D-4)}\nonumber \\
&&{-}A_{6,3}
\frac{42D^7-656D^6+3854D^5-11430D^4+24896D^3-65144D^2+134560D-113856}{3(D-1)(D-2)^2(5D-18)(5D-16)}\nonumber \\
&&{+}A_{7,2}
\frac{4(D-2)(D^4-28D^3+220D^2-696D+784)}{(5D-18)(D-1)(5D-16)} 
\end{eqnarray*}
}

{\allowdisplaybreaks
\begin{eqnarray*}
\lefteqn{X^q_{C_FN_F^2}=}\nonumber \\
&&{+}A_{6,1}
\frac{(6D^3-65D^2+238D-288)(D-2)^2}{6(D-4)(D-3)(D-1)^2} 
\end{eqnarray*}
}

{\allowdisplaybreaks
\begin{eqnarray*}
\lefteqn{X^q_{C_FN_{F,V}}=}\nonumber \\
&&{+}B_{4,1} 
\left(+\frac{119132D}{625}+\frac{380}{9(2D-7)}-\frac{50245888}{975(2D-9)}+\frac{280}{3(3D-10)}+\frac{9088443}{242(3D-14)}\right.\nonumber \\
&&\left.\qquad+\frac{9269061200}{722007(D-1)}-\frac{4169543}{675(D-2)}+\frac{182}{3(D-3)}+\frac{1194157}{54(D-4)}-\frac{57818921783}{265443750(5D-16)}\right.\nonumber \\
&&\left.\qquad+\frac{808885693}{243750(5D-18)}+\frac{221659776}{53125(5D-22)}+\frac{15748}{9(D-2)^2}+\frac{35}{3(D-3)^2}+\frac{750554}{45(D-4)^2}\right.\nonumber \\
&&\left.\qquad-\frac{608}{3(D-2)^3}+\frac{25148}{3(D-4)^3}+\frac{2048}{(D-4)^4}+\frac{12730856}{9375}\right)\nonumber \\
&&{+}A_{5,1} 
\left(+\frac{12243D}{625}+\frac{665}{9(2D-7)}-\frac{448624}{325(2D-9)}-\frac{56}{3(3D-10)}+\frac{92219200}{65637(D-1)}\right.\nonumber \\
&&\left.\qquad-\frac{83479}{225(D-2)}+\frac{10}{3(D-3)}+\frac{148888}{135(D-4)}-\frac{14475728}{446875(5D-16)}-\frac{29074423}{40625(5D-18)}\right.\nonumber \\
&&\left.\qquad-\frac{292095986}{53125(5D-22)}-\frac{1496}{3(D-2)^2}-\frac{14608}{45(D-4)^2}+\frac{256}{(D-2)^3}-\frac{2048}{15(D-4)^3}\right.\nonumber \\
&&\left.\qquad-\frac{2436926}{9375}\right)\nonumber \\
&&{-}A_{5,2} 
\left(+\frac{14644D}{1875}-\frac{76}{9(2D-7)}+\frac{40784}{325(2D-9)}+\frac{56}{9(3D-10)}+\frac{215775}{121(3D-14)}\right.\nonumber \\
&&\left.\qquad+\frac{4075274240}{2166021(D-1)}-\frac{1082473}{675(D-2)}+\frac{34666}{405(D-4)}-\frac{176454986}{132721875(5D-16)}-\frac{831538}{40625(5D-18)}\right.\nonumber \\
&&\left.\qquad-\frac{99059576}{53125(5D-22)}+\frac{8024}{9(D-2)^2}+\frac{17912}{135(D-4)^2}-\frac{640}{3(D-2)^3}+\frac{2176}{45(D-4)^3}\right.\nonumber \\
&&\left.\qquad+\frac{2370412}{28125}\right)\nonumber \\
&&{+}B_{5,1}
\left(+\frac{3474D}{625}-\frac{196273}{325(2D-9)}-\frac{25123840}{65637(D-1)}+\frac{144542}{225(D-2)}-\frac{2}{(D-3)}\right.\nonumber \\
&&\left.\qquad+\frac{93338}{135(D-4)}+\frac{1437614}{4021875(5D-16)}-\frac{9370564}{40625(5D-18)}-\frac{39796848}{53125(5D-22)}-\frac{1984}{3(D-2)^2}\right.\nonumber \\
&&\left.\qquad+\frac{2900}{9(D-4)^2}+\frac{224}{(D-2)^3}+\frac{1664}{15(D-4)^3}+\frac{190269}{3125}\right)\nonumber \\
&&{-}B_{5,2} 
\left(+\frac{37136D}{1875}-\frac{285488}{325(2D-9)}+\frac{80}{9(3D-10)}+\frac{28564480}{65637(D-1)}+\frac{163559}{450(D-2)}\right.\nonumber \\
&&\left.\qquad+\frac{150403}{135(D-4)}-\frac{25678283}{4021875(5D-16)}-\frac{13477879}{81250(5D-18)}+\frac{10007466}{53125(5D-22)}-\frac{2972}{3(D-2)^2}\right.\nonumber \\
&&\left.\qquad+\frac{32612}{45(D-4)^2}+\frac{480}{(D-2)^3}+\frac{3328}{15(D-4)^3}+\frac{4577488}{28125}\right)\nonumber \\
&&{-}A_{6,1}
\left(+\frac{49D}{625}-\frac{14385}{242(3D-14)}-\frac{58221088}{722007(D-1)}+\frac{931}{18(D-2)}+\frac{1}{2(D-3)}\right.\nonumber \\
&&\left.\qquad+\frac{418}{135(D-4)}-\frac{3759722}{3403125(5D-16)}-\frac{432134}{40625(5D-18)}+\frac{3537842}{53125(5D-22)}-\frac{16}{(D-2)^2}\right.\nonumber \\
&&\left.\qquad-\frac{64}{45(D-4)^2}+\frac{100399}{18750}\right)\nonumber \\
&&{-}A_{6,2}
\left(+\frac{17533D}{3000}+\frac{19}{24(2D-7)}+\frac{10196}{975(2D-9)}+\frac{1382570}{1683(D-1)}-\frac{411274}{675(D-2)}\right.\nonumber \\
&&\left.\qquad-\frac{1623}{56(D-3)}+\frac{1088}{45(D-4)}+\frac{18324568}{2413125(5D-16)}+\frac{1024224}{74375(5D-22)}+\frac{3680}{9(D-2)^2}\right.\nonumber \\
&&\left.\qquad-\frac{11}{6(D-3)^2}-\frac{256}{3(D-2)^3}+\frac{42883}{1875}\right)\nonumber \\
&&{+}A_{6,3}
\left(+\frac{26253D}{2500}-\frac{95}{36(2D-7)}+\frac{41793584}{21879(D-1)}-\frac{13967}{9(D-2)}+\frac{3}{(D-3)}\right.\nonumber \\
&&\left.\qquad+\frac{3548}{45(D-4)}-\frac{359662}{103125(5D-16)}+\frac{883413}{40625(5D-18)}-\frac{12129744}{53125(5D-22)}+\frac{2048}{3(D-2)^2}\right.\nonumber \\
&&\left.\qquad+\frac{352}{15(D-4)^2}-\frac{64}{(D-2)^3}-\frac{243387}{12500}\right)\nonumber \\
&&{+}A_{7,1}
\left(+\frac{9D}{100}+\frac{37}{18(2D-7)}+\frac{2549}{1560(2D-9)}+\frac{280}{143(D-1)}+\frac{37}{90(D-2)}\right.\nonumber \\
&&\left.\qquad-\frac{144076}{160875(5D-16)}+\frac{6313}{9750(5D-18)}+\frac{4}{(D-2)^2}+\frac{3927}{1000}\right)\nonumber \\
&&{-}A_{7,2}
\frac{5D^6-93D^5+892D^4-4656D^3+12528D^2-15472D+6080}{(D-1)(D-2)^2(5D-18)(5D-16)}\nonumber \\
&&{-}A_{7,3}
\left(+\frac{211D}{625}+\frac{37}{18(2D-7)}+\frac{2549}{5850(2D-9)}-\frac{5424}{187(D-1)}+\frac{2566}{75(D-2)}\right.\nonumber \\
&&\left.\qquad+\frac{2}{3(D-3)}+\frac{26995456}{4021875(5D-16)}-\frac{79688}{28125(5D-18)}-\frac{56582}{53125(5D-22)}-\frac{328}{3(D-2)^2}\right.\nonumber \\
&&\left.\qquad-\frac{2}{(D-3)^2}+\frac{128}{(D-2)^3}+\frac{4576}{3125}\right)\nonumber \\
&&{+}A_{7,4}
\left(+\frac{162D}{625}-\frac{37}{72(2D-7)}+\frac{2549}{1170(2D-9)}-\frac{22736}{2431(D-1)}+\frac{754}{45(D-2)}\right.\nonumber \\
&&\left.\qquad-\frac{323}{84(D-3)}-\frac{2696828}{4021875(5D-16)}+\frac{97442}{365625(5D-18)}-\frac{1877744}{1115625(5D-22)}-\frac{8}{(D-2)^2}\right.\nonumber \\
&&\left.\qquad-\frac{1}{12(D-3)^2}+\frac{273763}{75000}\right)\nonumber \\
&&{+}A_{7,5}
\left(+\frac{162D}{625}-\frac{37}{72(2D-7)}+\frac{2549}{1170(2D-9)}-\frac{22736}{2431(D-1)}+\frac{754}{45(D-2)}\right.\nonumber \\
&&\left.\qquad-\frac{323}{84(D-3)}-\frac{2696828}{4021875(5D-16)}+\frac{97442}{365625(5D-18)}-\frac{1877744}{1115625(5D-22)}-\frac{8}{(D-2)^2}\right.\nonumber \\
&&\left.\qquad-\frac{1}{12(D-3)^2}+\frac{273763}{75000}\right)\nonumber \\
&&{+}A_{8,1}
\left(+\frac{1107D}{80000}+\frac{1643}{768(2D-7)}-\frac{8085}{884(D-1)}+\frac{349}{45(D-2)}+\frac{1305}{448(D-3)}\right.\nonumber \\
&&\left.\qquad+\frac{22363}{28125(5D-16)}-\frac{77484}{40625(5D-18)}+\frac{85352}{1115625(5D-22)}+\frac{133}{384(2D-7)^2}-\frac{8}{(D-2)^2}\right.\nonumber \\
&&\left.\qquad-\frac{1511991}{800000}\right)\nonumber \\
&&{-}A_{9,1}
\left(+\frac{243D}{5000}+\frac{24255}{884(D-1)}-\frac{24}{(D-2)}-\frac{75}{56(D-3)}-\frac{10076}{9375(5D-16)}\right.\nonumber \\
&&\left.\qquad-\frac{83136}{40625(5D-18)}+\frac{252703}{1115625(5D-22)}+\frac{16}{(D-2)^2}+\frac{5103}{25000}\right)\nonumber \\
&&{+}A_{9,2}
\frac{3(3D-14)(D-4)(D^5+25D^4-290D^3+1036D^2-1560D+928)}{20(D-1)(D-2)(D-3)(5D-22)(5D-18)(5D-16)} 
\end{eqnarray*}
}

{\allowdisplaybreaks
\begin{eqnarray*}
\lefteqn{X^g_{C_A^3}=}\nonumber \\
&&+B_{4,1}
\left(+\frac{141\,D^2}{2}-\frac{2592693\,D}{500}-\frac{1120}{D}-\frac{223915}{126\,(2D-7)}+\frac{11621367296}{16575\,(2D-9)}\right.\nonumber \\
&&\left.\qquad+\frac{26512}{21\,(3D-10)}-\frac{17781309}{110\,(3D-14)}+\frac{32}{5\,(3D-8)}-\frac{9391405657}{1216215\,(D-1)}+\frac{2004296}{225\,(D-2)}\right.\nonumber \\
&&\left.\qquad+\frac{280}{3\,(D-3)}-\frac{158773009}{1215\,(D-4)}+\frac{360}{(D-6)}-\frac{7497336208}{30121875\,(5D-14)}+\frac{1340410849}{4021875\,(5D-16)}\right.\nonumber \\
&&\left.\qquad-\frac{3469841417}{243750\,(5D-18)}-\frac{2529983456}{3125\,(5D-22)}-\frac{55039}{432\,(D-1)^2}-\frac{25736}{5\,(D-2)^2}-\frac{1787}{16\,(D-3)^2}\right.\nonumber \\
&&\left.\qquad-\frac{32740532}{405\,(D-4)^2}+\frac{832}{(D-2)^3}-\frac{5426656}{135\,(D-4)^3}-\frac{89824}{9\,(D-4)^4}+\frac{512}{3\,(D-4)^5}\right.\nonumber \\
&&\left.\qquad+\frac{300276143}{18750}\right)\nonumber \\
&&-A_{5,1}
\left(+\frac{528664\,D}{375}+\frac{152}{5\,(2D-5)}+\frac{3380}{9\,(2D-7)}-\frac{103762208}{5525\,(2D-9)}-\frac{64048}{2205\,(3D-10)}\right.\nonumber \\
&&\left.\qquad-\frac{1621951952}{567567\,(D-1)}+\frac{2189086}{225\,(D-2)}+\frac{8}{(D-3)}+\frac{1487996}{81\,(D-4)}+\frac{1440}{7\,(D-6)}\right.\nonumber \\
&&\left.\qquad+\frac{65552704}{4303125\,(5D-14)}+\frac{1985701772}{28153125\,(5D-16)}+\frac{194176682}{40625\,(5D-18)}-\frac{140342384}{3125\,(5D-22)}-\frac{5504}{189\,(D-1)^2}\right.\nonumber \\
&&\left.\qquad-\frac{23952}{5\,(D-2)^2}+\frac{40016}{3\,(D-4)^2}+\frac{768}{(D-2)^3}+\frac{315872}{45\,(D-4)^3}+\frac{5888}{3\,(D-4)^4}\right.\nonumber \\
&&\left.\qquad-\frac{248426212}{28125}\right)\nonumber \\
&&-A_{5,2}
\left(+\frac{797251\,D}{2250}-\frac{144}{D}-\frac{9424}{1485\,(2D-5)}+\frac{2704}{63\,(2D-7)}-\frac{9432928}{5525\,(2D-9)}\right.\nonumber \\
&&\left.\qquad-\frac{34088}{945\,(3D-10)}-\frac{253295}{33\,(3D-14)}+\frac{112}{15\,(3D-8)}-\frac{734044771}{810810\,(D-1)}+\frac{1047656}{225\,(D-2)}\right.\nonumber \\
&&\left.\qquad+\frac{2009948}{1215\,(D-4)}+\frac{576}{7\,(D-6)}+\frac{857312704}{90365625\,(5D-14)}+\frac{1909981852}{28153125\,(5D-16)}+\frac{402543587}{446875\,(5D-18)}\right.\nonumber \\
&&\left.\qquad+\frac{35345856}{3125\,(5D-22)}-\frac{497}{54\,(D-1)^2}-\frac{3184}{(D-2)^2}+\frac{1267888}{405\,(D-4)^2}+\frac{768}{(D-2)^3}\right.\nonumber \\
&&\left.\qquad+\frac{333248}{135\,(D-4)^3}+\frac{7040}{9\,(D-4)^4}-\frac{225667646}{84375}\right)\nonumber \\
&&-B_{5,1}
\left(+ {51\,D^2} -\frac{495373\,D}{1000}+\frac{5681}{288\,(2D-5)}+\frac{13125}{32\,(2D-7)}-\frac{45395966}{5525\,(2D-9)}\right.\nonumber \\
&&\left.\qquad-\frac{317395}{162\,(D-1)}+\frac{235954}{75\,(D-2)}+\frac{5}{2\,(D-3)}+\frac{859702}{135\,(D-4)}+\frac{51714784}{4303125\,(5D-14)}\right.\nonumber \\
&&\left.\qquad-\frac{103936}{40625\,(5D-16)}-\frac{3138226}{3125\,(5D-18)}+\frac{101873408}{9375\,(5D-22)}-\frac{9056}{5\,(D-2)^2}+\frac{47116}{5\,(D-4)^2}\right.\nonumber \\
&&\left.\qquad+\frac{320}{(D-2)^3}+\frac{20464}{3\,(D-4)^3}+\frac{1920}{(D-4)^4}+\frac{14613393}{12500}\right)\nonumber \\
&&+B_{5,2}
\left(+\frac{218919\,D}{250}+\frac{769576}{10395\,(2D-5)}-\frac{66030496}{5525\,(2D-9)}+\frac{4784}{105\,(3D-10)}+\frac{32}{5\,(3D-8)}\right.\nonumber \\
&&\left.\qquad-\frac{1921435783}{810810\,(D-1)}+\frac{582929}{75\,(D-2)}+\frac{565904}{135\,(D-4)}+\frac{72}{7\,(D-6)}-\frac{24448544}{4303125\,(5D-14)}\right.\nonumber \\
&&\left.\qquad+\frac{25102328}{1340625\,(5D-16)}-\frac{415352077}{446875\,(5D-18)}+\frac{81364304}{3125\,(5D-22)}+\frac{4267}{54\,(D-1)^2}-\frac{23272}{5\,(D-2)^2}\right.\nonumber \\
&&\left.\qquad+\frac{791264}{135\,(D-4)^2}+\frac{960}{(D-2)^3}+\frac{33184}{9\,(D-4)^3}+\frac{2432}{3\,(D-4)^4}-\frac{45656086}{9375}\right)\nonumber \\
&&-A_{6,1}
\left(+\frac{D^2}{4}+\frac{1581\,D}{125}+\frac{50659}{198\,(3D-14)}+\frac{593}{162\,(D-1)}+\frac{179}{3\,(D-2)}\right.\nonumber \\
&&\left.\qquad+\frac{199}{3\,(D-4)}+\frac{1676696}{253125\,(5D-14)}-\frac{155971}{34375\,(5D-16)}-\frac{19461}{6250\,(5D-18)}-\frac{1262352}{3125\,(5D-22)}\right.\nonumber \\
&&\left.\qquad-\frac{787}{36\,(D-1)^2}-\frac{96}{5\,(D-2)^2}+\frac{132}{5\,(D-4)^2}-\frac{5861159}{112500}\right)\nonumber \\
&&+A_{6,2}
\left(+\frac{28093\,D}{225}-\frac{160}{D}-\frac{551}{42\,(2D-5)}+\frac{169}{42\,(2D-7)}+\frac{2358232}{16575\,(2D-9)}\right.\nonumber \\
&&\left.\qquad-\frac{25856}{297\,(D-1)}+\frac{91264}{75\,(D-2)}+\frac{4136}{105\,(D-3)}-\frac{5894}{135\,(D-4)}-\frac{16}{7\,(D-6)}\right.\nonumber \\
&&\left.\qquad-\frac{265141864}{2008125\,(5D-14)}-\frac{1316274}{625625\,(5D-16)}+\frac{960816}{4375\,(5D-22)}-\frac{832}{(D-2)^2}-\frac{10}{(D-3)^2}\right.\nonumber \\
&&\left.\qquad-\frac{320}{9\,(D-4)^2}-\frac{3682658}{5625}\right)\nonumber \\
&&-A_{6,3}
\left(+ {D^2} +\frac{33372\,D}{125}-\frac{845}{63\,(2D-7)}+\frac{304}{15\,(3D-8)}+\frac{4688143}{38610\,(D-1)}\right.\nonumber \\
&&\left.\qquad+\frac{131356}{45\,(D-2)}+\frac{25}{2\,(D-3)}+\frac{24319}{45\,(D-4)}-\frac{29688032}{590625\,(5D-14)}+\frac{2863027}{309375\,(5D-16)}\right.\nonumber \\
&&\left.\qquad+\frac{19412}{40625\,(5D-18)}-\frac{4328064}{3125\,(5D-22)}-\frac{10688}{5\,(D-2)^2}+\frac{3124}{15\,(D-4)^2}+\frac{384}{(D-2)^3}\right.\nonumber \\
&&\left.\qquad+\frac{32}{(D-4)^3}-\frac{18671174}{9375}\right)\nonumber \\
&&-B_{6,1}
\left(+ {D^3} - {30\,D^2} + {300\,D} +\frac{528}{(D-2)}-\frac{192}{(D-4)}\right.\nonumber \\
&&\left.\qquad-\frac{288}{(D-2)^2}-\frac{192}{(D-4)^2}+\frac{64}{(D-2)^3}-\frac{64}{(D-4)^3}- {1024} \right)\nonumber \\
&&-B_{6,2}
\left(+\frac{27\,D^2}{2}-\frac{1689\,D}{8}+\frac{2261}{288\,(2D-5)}+\frac{625}{32\,(2D-7)}-\frac{350}{9\,(D-1)}\right.\nonumber \\
&&\left.\qquad-\frac{788}{(D-2)}-\frac{610}{3\,(D-4)}+\frac{512}{(D-2)^2}-\frac{700}{3\,(D-4)^2}-\frac{128}{(D-2)^3}\right.\nonumber \\
&&\left.\qquad-\frac{96}{(D-4)^3}+\frac{1765}{2}\right)\nonumber \\
&&-C_{6,1}
\left(+\frac{27\,D^2}{2}-\frac{1689\,D}{8}+\frac{2261}{288\,(2D-5)}+\frac{625}{32\,(2D-7)}-\frac{350}{9\,(D-1)}\right.\nonumber \\
&&\left.\qquad-\frac{788}{(D-2)}-\frac{610}{3\,(D-4)}+\frac{512}{(D-2)^2}-\frac{700}{3\,(D-4)^2}-\frac{128}{(D-2)^3}\right.\nonumber \\
&&\left.\qquad-\frac{96}{(D-4)^3}+\frac{1765}{2}\right)\nonumber \\
&&-A_{7,1}
\left(+\frac{27\,D}{10}+\frac{76}{77\,(2D-5)}-\frac{1198}{63\,(2D-7)}+\frac{294779}{13260\,(2D-9)}-\frac{2268}{715\,(D-1)}\right.\nonumber \\
&&\left.\qquad-\frac{349}{45\,(D-2)}+\frac{20}{(D-4)}+\frac{12}{35\,(D-6)}+\frac{72}{119\,(5D-14)}-\frac{56684}{32175\,(5D-16)}\right.\nonumber \\
&&\left.\qquad+\frac{139433}{10725\,(5D-18)}+\frac{8}{(D-2)^2}+\frac{873}{100}\right)\nonumber \\
&&+A_{7,2}
\left(+\frac{196\,D}{25}-\frac{3098}{715\,(D-1)}+\frac{26}{(D-2)}-\frac{40}{(D-4)}-\frac{48}{35\,(D-6)}\right.\nonumber \\
&&\left.\qquad-\frac{176}{125\,(5D-14)}+\frac{5304}{1925\,(5D-16)}+\frac{122054}{1625\,(5D-18)}-\frac{16}{(D-2)^2}-\frac{4618}{125}\right)\nonumber \\
&&+A_{7,3}
\left(+\frac{6516\,D}{125}+\frac{57}{(2D-5)}-\frac{1198}{63\,(2D-7)}+\frac{294779}{49725\,(2D-9)}-\frac{502}{(D-1)}\right.\nonumber \\
&&\left.\qquad+\frac{361018}{225\,(D-2)}+\frac{4258}{35\,(D-3)}-\frac{14232368}{74375\,(5D-14)}-\frac{5003392}{73125\,(5D-16)}+\frac{176798}{5625\,(5D-18)}\right.\nonumber \\
&&\left.\qquad-\frac{164176}{13125\,(5D-22)}-\frac{1840}{(D-2)^2}-\frac{10}{(D-3)^2}+\frac{512}{(D-2)^3}-\frac{328828}{625}\right)\nonumber \\
&&-A_{7,4}
\left(+\frac{8401\,D}{1000}-\frac{19}{16\,(2D-5)}+\frac{599}{126\,(2D-7)}+\frac{294779}{9945\,(2D-9)}+\frac{43}{60\,(D-1)}\right.\nonumber \\
&&\left.\qquad+\frac{1234}{45\,(D-2)}-\frac{717}{140\,(D-3)}-\frac{4}{5\,(D-6)}+\frac{11198396}{1115625\,(5D-14)}-\frac{1234996}{365625\,(5D-16)}\right.\nonumber \\
&&\left.\qquad-\frac{119686}{28125\,(5D-18)}-\frac{1761496}{65625\,(5D-22)}-\frac{16}{(D-2)^2}+\frac{1}{4\,(D-3)^2}-\frac{1219407}{50000}\right)\nonumber \\
&&-A_{7,5}
\left(+\frac{8401\,D}{1000}-\frac{19}{16\,(2D-5)}+\frac{599}{126\,(2D-7)}+\frac{294779}{9945\,(2D-9)}+\frac{43}{60\,(D-1)}\right.\nonumber \\
&&\left.\qquad+\frac{1234}{45\,(D-2)}-\frac{717}{140\,(D-3)}-\frac{4}{5\,(D-6)}+\frac{11198396}{1115625\,(5D-14)}-\frac{1234996}{365625\,(5D-16)}\right.\nonumber \\
&&\left.\qquad-\frac{119686}{28125\,(5D-18)}-\frac{1761496}{65625\,(5D-22)}-\frac{16}{(D-2)^2}+\frac{1}{4\,(D-3)^2}-\frac{1219407}{50000}\right)\nonumber \\
&&-A_{8,1}
\left(+\frac{837\,D}{1000}-\frac{463}{84\,(2D-7)}+\frac{844}{45\,(D-2)}+\frac{15}{14\,(D-3)}-\frac{4}{(D-4)}\right.\nonumber \\
&&\left.\qquad-\frac{79508}{21875\,(5D-14)}+\frac{112651}{28125\,(5D-16)}+\frac{95172}{3125\,(5D-18)}+\frac{80068}{65625\,(5D-22)}+\frac{169}{96\,(2D-7)^2}\right.\nonumber \\
&&\left.\qquad-\frac{16}{(D-2)^2}-\frac{807543}{100000}\right)\nonumber \\
&&-B_{8,1}
\frac{3\,(D-3)\,(3D-8)\,(D^3-16\,D^2+68\,D-88)}{4\,(2D-5)\,(D-2)\,(2D-7)\,(D-4)}\nonumber \\
&&-C_{8,1}
\frac{3\,(D-3)\,(3D-8)\,(D^3-16\,D^2+68\,D-88)}{4\,(2D-5)\,(D-2)\,(2D-7)\,(D-4)}\nonumber \\
&&+A_{9,1}
\left(+\frac{729\,D}{500}+\frac{132}{5\,(D-2)}+\frac{15}{28\,(D-3)}-\frac{2}{(D-4)}-\frac{16464}{3125\,(5D-14)}\right.\nonumber \\
&&\left.\qquad+\frac{13794}{3125\,(5D-16)}+\frac{1872}{3125\,(5D-18)}+\frac{30056}{21875\,(5D-22)}-\frac{96}{5\,(D-2)^2}-\frac{156321}{12500}\right)\nonumber \\
&&-A_{9,2}
\frac{3\,(3D-14)\,(75\,D^6-1048\,D^5+5956\,D^4-17776\,D^3+30208\,D^2-29440\,D+13952)}{10\,(D-2)\,(D-3)\,(5D-22)\,(5D-14)\,(5D-16)\,(5D-18)}\nonumber \\
\end{eqnarray*}
}

{\allowdisplaybreaks
\begin{eqnarray*}
\lefteqn{X^g_{C_A^2N_F}=}\nonumber \\
&&-B_{4,1}
\left(+\frac{37168\,D^2}{125}-\frac{5809119\,D}{1250}+\frac{2240}{3\,D}+\frac{34925}{126\,(2D-7)}+\frac{104123558}{3315\,(2D-9)}\right.\nonumber \\
&&\left.\qquad+\frac{11162}{21\,(3D-10)}-\frac{2318004}{385\,(3D-14)}+\frac{64}{45\,(3D-8)}+\frac{77206856351}{4864860\,(D-1)}-\frac{657533}{90\,(D-2)}\right.\nonumber \\
&&\left.\qquad-\frac{41}{12\,(D-3)}-\frac{8729134}{1215\,(D-4)}+\frac{180}{(D-6)}-\frac{563997052}{6024375\,(5D-14)}+\frac{538277353}{4021875\,(5D-16)}\right.\nonumber \\
&&\left.\qquad+\frac{16666937}{81250\,(5D-18)}-\frac{1207808264}{28125\,(5D-22)}+\frac{7637}{36\,(D-1)^2}+\frac{6740}{3\,(D-2)^2}-\frac{151}{4\,(D-3)^2}\right.\nonumber \\
&&\left.\qquad+\frac{783268}{405\,(D-4)^2}-\frac{896}{3\,(D-2)^3}+\frac{313448}{135\,(D-4)^3}+\frac{29312}{45\,(D-4)^4}+\frac{27982342}{5625}\right)\nonumber \\
&&-A_{5,1}
\left(+\frac{6881\,D^2}{125}-\frac{7596253\,D}{11250}+\frac{608}{5\,(2D-5)}+\frac{1075}{18\,(2D-7)}+\frac{27315013}{22100\,(2D-9)}\right.\nonumber \\
&&\left.\qquad-\frac{212536}{6615\,(3D-10)}-\frac{3290734688}{567567\,(D-1)}+\frac{239912}{75\,(D-2)}+\frac{8}{(D-3)}-\frac{3016}{3\,(D-4)}\right.\nonumber \\
&&\left.\qquad-\frac{720}{7\,(D-6)}-\frac{19049176}{4303125\,(5D-14)}-\frac{160235632}{5630625\,(5D-16)}+\frac{67577636}{40625\,(5D-18)}+\frac{10234588}{9375\,(5D-22)}\right.\nonumber \\
&&\left.\qquad-\frac{11008}{189\,(D-1)^2}-\frac{4928}{3\,(D-2)^2}-\frac{62704}{135\,(D-4)^2}+\frac{256}{(D-2)^3}-\frac{19264}{45\,(D-4)^3}\right.\nonumber \\
&&\left.\qquad+\frac{183292489}{67500}\right)\nonumber \\
&&+A_{5,2}
\left(+\frac{7688\,D^2}{375}-\frac{822412\,D}{1875}+\frac{96}{D}+\frac{37696}{1485\,(2D-5)}-\frac{430}{63\,(2D-7)}\right.\nonumber \\
&&\left.\qquad-\frac{2489693}{5525\,(2D-9)}-\frac{20648}{945\,(3D-10)}-\frac{66040}{231\,(3D-14)}+\frac{224}{135\,(3D-8)}+\frac{247020002}{135135\,(D-1)}\right.\nonumber \\
&&\left.\qquad-\frac{111043}{75\,(D-2)}+\frac{377408}{1215\,(D-4)}+\frac{288}{7\,(D-6)}+\frac{67345168}{3614625\,(5D-14)}+\frac{73779424}{9384375\,(5D-16)}\right.\nonumber \\
&&\left.\qquad-\frac{137313893}{446875\,(5D-18)}+\frac{19692512}{28125\,(5D-22)}+\frac{994}{27\,(D-1)^2}+\frac{504}{(D-2)^2}+\frac{125248}{405\,(D-4)^2}\right.\nonumber \\
&&\left.\qquad-\frac{256}{3\,(D-2)^3}+\frac{4480}{27\,(D-4)^3}+\frac{3858677}{3375}\right)\nonumber \\
&&+B_{5,1}
\left(+\frac{1397\,D^2}{1000}+\frac{646933\,D}{10000}-\frac{5681}{72\,(2D-5)}-\frac{6125}{96\,(2D-7)}-\frac{2169937}{5525\,(2D-9)}\right.\nonumber \\
&&\left.\qquad+\frac{317395}{81\,(D-1)}-\frac{541766}{225\,(D-2)}-\frac{1}{2\,(D-3)}-\frac{1862}{45\,(D-4)}+\frac{49642912}{4303125\,(5D-14)}\right.\nonumber \\
&&\left.\qquad-\frac{488704}{365625\,(5D-16)}+\frac{1246}{125\,(5D-18)}+\frac{14295008}{9375\,(5D-22)}+\frac{3136}{3\,(D-2)^2}+\frac{3136}{45\,(D-4)^2}\right.\nonumber \\
&&\left.\qquad-\frac{128}{(D-2)^3}+\frac{1312}{15\,(D-4)^3}-\frac{11790061}{20000}\right)\nonumber \\
&&+B_{5,2}
\left(+\frac{871\,D^2}{75}-\frac{1716962\,D}{5625}+\frac{3078304}{10395\,(2D-5)}+\frac{3156272}{5525\,(2D-9)}-\frac{256}{945\,(3D-10)}\right.\nonumber \\
&&\left.\qquad-\frac{64}{45\,(3D-8)}-\frac{1836706258}{405405\,(D-1)}+\frac{1384157}{450\,(D-2)}-\frac{15388}{405\,(D-4)}-\frac{36}{7\,(D-6)}\right.\nonumber \\
&&\left.\qquad+\frac{1683416}{53125\,(5D-14)}+\frac{137885032}{28153125\,(5D-16)}+\frac{88774237}{893750\,(5D-18)}-\frac{6784148}{3125\,(5D-22)}+\frac{502}{3\,(D-1)^2}\right.\nonumber \\
&&\left.\qquad-\frac{5348}{3\,(D-2)^2}-\frac{21104}{135\,(D-4)^2}+\frac{224}{(D-2)^3}-\frac{4672}{45\,(D-4)^3}+\frac{18770368}{16875}\right)\nonumber \\
&&-A_{6,1}
\left(+\frac{197\,D^2}{50}-\frac{2993\,D}{50}-\frac{6604}{693\,(3D-14)}-\frac{4333}{81\,(D-1)}-\frac{159}{5\,(D-2)}\right.\nonumber \\
&&\left.\qquad+\frac{44}{45\,(D-4)}-\frac{3850072}{354375\,(5D-14)}+\frac{4564}{4125\,(5D-16)}-\frac{11563}{625\,(5D-18)}+\frac{17696}{1875\,(5D-22)}\right.\nonumber \\
&&\left.\qquad-\frac{443}{9\,(D-1)^2}-\frac{8}{(D-4)^2}+\frac{1203449}{5625}\right)\nonumber \\
&&+A_{6,2}
\left(+\frac{119\,D^2}{50}-\frac{84531\,D}{1000}-\frac{320}{3\,D}-\frac{1102}{21\,(2D-5)}+\frac{215}{336\,(2D-7)}\right.\nonumber \\
&&\left.\qquad-\frac{93023}{7800\,(2D-9)}-\frac{51712}{297\,(D-1)}-\frac{19264}{75\,(D-2)}-\frac{4183}{210\,(D-3)}-\frac{1084}{135\,(D-4)}\right.\nonumber \\
&&\left.\qquad+\frac{8}{7\,(D-6)}+\frac{38037128}{118125\,(5D-14)}-\frac{161431412}{5630625\,(5D-16)}-\frac{331172}{39375\,(5D-22)}+\frac{272}{3\,(D-2)^2}\right.\nonumber \\
&&\left.\qquad+\frac{4}{(D-3)^2}-\frac{64}{3\,(D-2)^3}+\frac{40336667}{90000}\right)\nonumber \\
&&-A_{6,3}
\left(+\frac{921\,D^2}{100}-\frac{1571239\,D}{9000}-\frac{1075}{504\,(2D-7)}-\frac{4757193}{88400\,(2D-9)}-\frac{608}{135\,(3D-8)}\right.\nonumber \\
&&\left.\qquad+\frac{4688143}{19305\,(D-1)}-\frac{116434}{225\,(D-2)}-\frac{3}{(D-3)}+\frac{112}{45\,(D-4)}+\frac{139041824}{2008125\,(5D-14)}\right.\nonumber \\
&&\left.\qquad-\frac{1488052}{804375\,(5D-16)}-\frac{2138}{1625\,(5D-18)}+\frac{20224}{625\,(5D-22)}+\frac{32}{(D-2)^2}-\frac{24}{5\,(D-4)^2}\right.\nonumber \\
&&\left.\qquad+\frac{128}{(D-2)^3}+\frac{182032831}{270000}\right)\nonumber \\
&&-B_{6,2}
\left(+\frac{7\,D^2}{8}-\frac{185\,D}{16}+\frac{2261}{72\,(2D-5)}+\frac{875}{288\,(2D-7)}-\frac{700}{9\,(D-1)}\right.\nonumber \\
&&\left.\qquad+\frac{592}{9\,(D-2)}+\frac{62}{9\,(D-4)}-\frac{160}{3\,(D-2)^2}+\frac{8}{(D-4)^2}+\frac{793}{32}\right)\nonumber \\
&&-C_{6,1}
\left(+\frac{7\,D^2}{8}-\frac{185\,D}{16}+\frac{2261}{72\,(2D-5)}+\frac{875}{288\,(2D-7)}-\frac{700}{9\,(D-1)}\right.\nonumber \\
&&\left.\qquad+\frac{592}{9\,(D-2)}+\frac{62}{9\,(D-4)}-\frac{160}{3\,(D-2)^2}+\frac{8}{(D-4)^2}+\frac{793}{32}\right)\nonumber \\
&&-A_{7,1}
\left(+\frac{4\,D^2}{25}-\frac{1437\,D}{500}+\frac{304}{77\,(2D-5)}-\frac{451}{126\,(2D-7)}-\frac{28181}{26520\,(2D-9)}\right.\nonumber \\
&&\left.\qquad-\frac{9248}{2145\,(D-1)}-\frac{263}{30\,(D-2)}-\frac{6}{35\,(D-6)}+\frac{2253164}{223125\,(5D-14)}+\frac{38046376}{5630625\,(5D-16)}\right.\nonumber \\
&&\left.\qquad-\frac{7073303}{536250\,(5D-18)}-\frac{20}{3\,(D-2)^2}+\frac{10761}{1000}\right)\nonumber \\
&&+A_{7,2}
\left(+\frac{17\,D^2}{80}-\frac{2237\,D}{400}-\frac{71585}{8064\,(2D-7)}-\frac{226533}{141440\,(2D-9)}-\frac{6196}{715\,(D-1)}\right.\nonumber \\
&&\left.\qquad+\frac{142}{45\,(D-2)}+\frac{24}{35\,(D-6)}+\frac{63088}{14875\,(5D-14)}+\frac{105104}{375375\,(5D-16)}+\frac{55686}{1625\,(5D-18)}\right.\nonumber \\
&&\left.\qquad-\frac{32}{3\,(D-2)^2}+\frac{35181}{1600}\right)\nonumber \\
&&+A_{7,3}
\left(+\frac{11\,D^2}{125}-\frac{4966\,D}{625}+\frac{228}{(2D-5)}-\frac{451}{126\,(2D-7)}-\frac{28181}{99450\,(2D-9)}\right.\nonumber \\
&&\left.\qquad-\frac{1004}{(D-1)}+\frac{141554}{225\,(D-2)}-\frac{202}{5\,(D-3)}+\frac{63075448}{371875\,(5D-14)}+\frac{1714432}{73125\,(5D-16)}\right.\nonumber \\
&&\left.\qquad+\frac{129298}{28125\,(5D-18)}+\frac{2948}{9375\,(5D-22)}-\frac{480}{(D-2)^2}+\frac{2}{(D-3)^2}+\frac{128}{(D-2)^3}\right.\nonumber \\
&&\left.\qquad+\frac{52438}{625}\right)\nonumber \\
&&-A_{7,4}
\left(+\frac{21\,D^2}{1000}-\frac{15043\,D}{5000}-\frac{19}{4\,(2D-5)}+\frac{6805}{4032\,(2D-7)}-\frac{93023}{37440\,(2D-9)}\right.\nonumber \\
&&\left.\qquad+\frac{43}{30\,(D-1)}-\frac{779}{45\,(D-2)}+\frac{467}{210\,(D-3)}+\frac{2}{5\,(D-6)}+\frac{1100828}{65625\,(5D-14)}\right.\nonumber \\
&&\left.\qquad-\frac{2621912}{365625\,(5D-16)}-\frac{4403}{5625\,(5D-18)}+\frac{393514}{196875\,(5D-22)}+\frac{1}{4\,(D-3)^2}+\frac{916069}{60000}\right)\nonumber \\
&&-A_{7,5}
\left(+\frac{21\,D^2}{1000}-\frac{15043\,D}{5000}-\frac{19}{4\,(2D-5)}+\frac{6805}{4032\,(2D-7)}-\frac{93023}{37440\,(2D-9)}\right.\nonumber \\
&&\left.\qquad+\frac{43}{30\,(D-1)}-\frac{779}{45\,(D-2)}+\frac{467}{210\,(D-3)}+\frac{2}{5\,(D-6)}+\frac{1100828}{65625\,(5D-14)}\right.\nonumber \\
&&\left.\qquad-\frac{2621912}{365625\,(5D-16)}-\frac{4403}{5625\,(5D-18)}+\frac{393514}{196875\,(5D-22)}+\frac{1}{4\,(D-3)^2}+\frac{916069}{60000}\right)\nonumber \\
&&-A_{8,1}
\left(+\frac{113787\,D}{80000}-\frac{2377}{1792\,(2D-7)}+\frac{712}{45\,(D-2)}-\frac{45}{28\,(D-3)}-\frac{169858}{21875\,(5D-14)}\right.\nonumber \\
&&\left.\qquad+\frac{105754}{28125\,(5D-16)}+\frac{12804}{3125\,(5D-18)}-\frac{3356}{65625\,(5D-22)}+\frac{215}{768\,(2D-7)^2}-\frac{920019}{80000}\right)\nonumber \\
&&-B_{8,1}
\frac{(2\,D^3-25\,D^2+94\,D-112)\,(D^3-16\,D^2+68\,D-88)}{8\,(2D-7)\,(D-2)^2\,(2D-5)}\nonumber \\
&&-C_{8,1}
\frac{(2\,D^3-25\,D^2+94\,D-112)\,(D^3-16\,D^2+68\,D-88)}{8\,(2D-7)\,(D-2)^2\,(2D-5)}\nonumber \\
&&+A_{9,1}
\left(+\frac{81\,D^2}{1000}-\frac{9369\,D}{5000}-\frac{44}{3\,(D-2)}+\frac{45}{56\,(D-3)}+\frac{8232}{625\,(5D-14)}\right.\nonumber \\
&&\left.\qquad-\frac{40172}{9375\,(5D-16)}+\frac{2928}{3125\,(5D-18)}+\frac{3106}{65625\,(5D-22)}+\frac{54801}{5000}\right)\nonumber \\
&&+A_{9,2}
\frac{(3D-14)\,(D-4)\,(135\,D^5-2200\,D^4+15156\,D^3-54336\,D^2+99776\,D-74112)}{20\,(D-3)\,(5D-22)\,(5D-14)\,(5D-16)\,(5D-18)\,(D-2)}\nonumber \\
&&-A_{9,4}
\left(+\frac{81\,D^2}{16000}-\frac{513\,D}{10000}-\frac{973}{9216\,(2D-7)}-\frac{75511}{5657600\,(2D-9)}-\frac{4}{75\,(D-2)}\right.\nonumber \\
&&\left.\qquad+\frac{756}{10625\,(5D-14)}-\frac{14036}{121875\,(5D-16)}+\frac{1168}{3125\,(5D-18)}+\frac{2234}{28125\,(5D-22)}+\frac{36117}{320000}\right)\nonumber \\
\end{eqnarray*}
}

{\allowdisplaybreaks
\begin{eqnarray*}
\lefteqn{X^g_{C_AC_FN_F}=}\nonumber \\
&&+B_{4,1}
\left(+\frac{169208\,D^2}{125}-\frac{28533686\,D}{1875}-\frac{1120}{3\,D}+\frac{4465}{7\,(2D-7)}+\frac{1070886586}{16575\,(2D-9)}\right.\nonumber \\
&&\left.\qquad-\frac{56648}{63\,(3D-10)}+\frac{18168111}{770\,(3D-14)}-\frac{1664}{45\,(3D-8)}+\frac{19856072}{405405\,(D-1)}-\frac{3342712}{675\,(D-2)}\right.\nonumber \\
&&\left.\qquad-\frac{1138}{3\,(D-3)}+\frac{480653}{135\,(D-4)}+\frac{2161801556}{30121875\,(5D-14)}+\frac{868101691}{2413125\,(5D-16)}+\frac{669068729}{48750\,(5D-18)}\right.\nonumber \\
&&\left.\qquad-\frac{1607924552}{9375\,(5D-22)}+\frac{13376}{9\,(D-2)^2}-\frac{19}{3\,(D-3)^2}+\frac{554888}{15\,(D-4)^2}+\frac{471544}{15\,(D-4)^3}\right.\nonumber \\
&&\left.\qquad+\frac{46784}{5\,(D-4)^4}+\frac{876803866}{28125}\right)\nonumber \\
&&+A_{5,1}
\left(+\frac{25803\,D^2}{125}-\frac{21754033\,D}{11250}+\frac{1216}{5\,(2D-5)}+\frac{2390}{9\,(2D-7)}+\frac{64410511}{22100\,(2D-9)}\right.\nonumber \\
&&\left.\qquad-\frac{63976}{945\,(3D-10)}-\frac{13760}{567\,(D-1)}+\frac{295132}{225\,(D-2)}+\frac{40}{3\,(D-3)}-\frac{711848}{405\,(D-4)}\right.\nonumber \\
&&\left.\qquad-\frac{11375744}{478125\,(5D-14)}+\frac{6633272}{73125\,(5D-16)}+\frac{9607724}{3125\,(5D-18)}-\frac{7186816}{3125\,(5D-22)}-\frac{3712}{3\,(D-2)^2}\right.\nonumber \\
&&\left.\qquad-\frac{5776}{135\,(D-4)^2}-\frac{19328}{45\,(D-4)^3}+\frac{1255139429}{337500}\right)\nonumber \\
&&-A_{5,2}
\left(+\frac{27472\,D^2}{375}-\frac{6814294\,D}{5625}-\frac{48}{D}+\frac{75392}{1485\,(2D-5)}-\frac{1912}{63\,(2D-7)}\right.\nonumber \\
&&\left.\qquad-\frac{7070567}{5525\,(2D-9)}-\frac{27436}{945\,(3D-10)}+\frac{258805}{231\,(3D-14)}-\frac{992}{27\,(3D-8)}-\frac{2189468}{81081\,(D-1)}\right.\nonumber \\
&&\left.\qquad-\frac{1407278}{675\,(D-2)}+\frac{1712}{(D-4)}+\frac{1146502544}{30121875\,(5D-14)}-\frac{482461664}{12065625\,(5D-16)}-\frac{464255741}{446875\,(5D-18)}\right.\nonumber \\
&&\left.\qquad-\frac{10192}{375\,(5D-22)}+\frac{6544}{9\,(D-2)^2}+\frac{71776}{45\,(D-4)^2}+\frac{10496}{15\,(D-4)^3}+\frac{124446769}{28125}\right)\nonumber \\
&&+B_{5,1}
\left(+\frac{7331\,D^2}{500}+\frac{205803\,D}{5000}+\frac{5681}{36\,(2D-5)}+\frac{6125}{48\,(2D-7)}+\frac{270116}{325\,(2D-9)}\right.\nonumber \\
&&\left.\qquad+\frac{40704}{25\,(D-2)}+\frac{19}{(D-3)}+\frac{10708}{15\,(D-4)}-\frac{31328}{5625\,(5D-14)}-\frac{10336}{40625\,(5D-16)}\right.\nonumber \\
&&\left.\qquad-\frac{3005968}{3125\,(5D-18)}-\frac{11927552}{1875\,(5D-22)}-\frac{928}{(D-2)^2}-\frac{224}{3\,(D-4)^2}-\frac{192}{5\,(D-4)^3}\right.\nonumber \\
&&\left.\qquad-\frac{52885663}{50000}\right)\nonumber \\
&&-B_{5,2}
\left(+\frac{7409\,D^2}{75}-\frac{5308534\,D}{5625}+\frac{6156608}{10395\,(2D-5)}+\frac{392896}{325\,(2D-9)}+\frac{28528}{945\,(3D-10)}\right.\nonumber \\
&&\left.\qquad+\frac{16}{9\,(3D-8)}+\frac{1921904}{27027\,(D-1)}+\frac{423956}{225\,(D-2)}+\frac{129704}{81\,(D-4)}+\frac{13379128}{253125\,(5D-14)}\right.\nonumber \\
&&\left.\qquad+\frac{1042664096}{28153125\,(5D-16)}-\frac{184123036}{446875\,(5D-18)}-\frac{4877264}{625\,(5D-22)}-\frac{5344}{3\,(D-2)^2}+\frac{118976}{135\,(D-4)^2}\right.\nonumber \\
&&\left.\qquad+\frac{10624}{45\,(D-4)^3}+\frac{112377904}{84375}\right)\nonumber \\
&&+A_{6,1}
\left(+\frac{5321\,D^2}{375}-\frac{340129\,D}{1875}+\frac{51761}{1386\,(3D-14)}-\frac{430}{27\,(D-1)}-\frac{2273}{15\,(D-2)}\right.\nonumber \\
&&\left.\qquad+\frac{178}{3\,(D-4)}-\frac{1645052}{590625\,(5D-14)}-\frac{823378}{103125\,(5D-16)}-\frac{121841}{6250\,(5D-18)}-\frac{148988}{3125\,(5D-22)}\right.\nonumber \\
&&\left.\qquad+\frac{272}{15\,(D-4)^2}+\frac{17076368}{28125}\right)\nonumber \\
&&-A_{6,2}
\left(+\frac{1489\,D^2}{100}-\frac{334171\,D}{1500}+\frac{160}{3\,D}-\frac{2204}{21\,(2D-5)}+\frac{239}{84\,(2D-7)}\right.\nonumber \\
&&\left.\qquad-\frac{3947149}{132600\,(2D-9)}-\frac{461624}{675\,(D-2)}+\frac{8131}{210\,(D-3)}+\frac{304}{9\,(D-4)}+\frac{169118132}{669375\,(5D-14)}\right.\nonumber \\
&&\left.\qquad-\frac{123983528}{1535625\,(5D-16)}-\frac{232964}{13125\,(5D-22)}+\frac{544}{9\,(D-2)^2}+\frac{5}{12\,(D-3)^2}+\frac{14824063}{15000}\right)\nonumber \\
&&+A_{6,3}
\left(+\frac{24791\,D^2}{500}-\frac{26566007\,D}{45000}-\frac{1195}{126\,(2D-7)}-\frac{14271579}{88400\,(2D-9)}+\frac{1520}{27\,(3D-8)}\right.\nonumber \\
&&\left.\qquad-\frac{91192}{225\,(D-2)}-\frac{2}{(D-3)}+\frac{13466}{45\,(D-4)}-\frac{146291696}{3346875\,(5D-14)}+\frac{4052666}{365625\,(5D-16)}\right.\nonumber \\
&&\left.\qquad+\frac{200184}{3125\,(5D-18)}-\frac{510816}{3125\,(5D-22)}+\frac{1552}{15\,(D-4)^2}+\frac{2491573703}{1350000}\right)\nonumber \\
&&+B_{6,2}
\left(+\frac{23\,D^2}{4}-\frac{729\,D}{8}+\frac{2261}{36\,(2D-5)}+\frac{875}{144\,(2D-7)}-\frac{1264}{9\,(D-2)}\right.\nonumber \\
&&\left.\qquad+\frac{380}{9\,(D-4)}-\frac{32}{3\,(D-2)^2}+\frac{80}{3\,(D-4)^2}+\frac{5401}{16}\right)\nonumber \\
&&+C_{6,1}
\left(+\frac{23\,D^2}{4}-\frac{729\,D}{8}+\frac{2261}{36\,(2D-5)}+\frac{875}{144\,(2D-7)}-\frac{1264}{9\,(D-2)}\right.\nonumber \\
&&\left.\qquad+\frac{380}{9\,(D-4)}-\frac{32}{3\,(D-2)^2}+\frac{80}{3\,(D-4)^2}+\frac{5401}{16}\right)\nonumber \\
&&+A_{7,1}
\left(+\frac{16\,D^2}{25}-\frac{771\,D}{125}+\frac{608}{77\,(2D-5)}-\frac{139}{21\,(2D-7)}-\frac{877}{390\,(2D-9)}\right.\nonumber \\
&&\left.\qquad+\frac{1744}{429\,(D-1)}-\frac{1192}{45\,(D-2)}+\frac{10112}{525\,(5D-14)}+\frac{72675488}{5630625\,(5D-16)}-\frac{9210328}{268125\,(5D-18)}\right.\nonumber \\
&&\left.\qquad-\frac{32}{3\,(D-2)^2}+\frac{21831}{1250}\right)\nonumber \\
&&-A_{7,2}
\left(+\frac{83\,D^2}{80}-\frac{1307\,D}{80}-\frac{71585}{2688\,(2D-7)}-\frac{679599}{141440\,(2D-9)}-\frac{246}{5\,(D-2)}\right.\nonumber \\
&&\left.\qquad+\frac{226392}{14875\,(5D-14)}+\frac{7712}{4875\,(5D-16)}+\frac{12886}{125\,(5D-18)}+\frac{645603}{8000}\right)\nonumber \\
&&-A_{7,3}
\left(+\frac{169\,D^2}{125}-\frac{24\,D}{5}+\frac{456}{(2D-5)}-\frac{139}{21\,(2D-7)}-\frac{1754}{2925\,(2D-9)}\right.\nonumber \\
&&\left.\qquad+\frac{15704}{225\,(D-2)}+\frac{472}{15\,(D-3)}+\frac{1432656}{21875\,(5D-14)}-\frac{7709696}{365625\,(5D-16)}+\frac{47912}{5625\,(5D-18)}\right.\nonumber \\
&&\left.\qquad+\frac{176}{1875\,(5D-22)}-\frac{1792}{3\,(D-2)^2}+\frac{1}{(D-3)^2}-\frac{238436}{3125}\right)\nonumber \\
&&+A_{7,4}
\left(+\frac{313\,D^2}{1000}-\frac{10751\,D}{3000}-\frac{19}{2\,(2D-5)}+\frac{1807}{448\,(2D-7)}-\frac{3947149}{636480\,(2D-9)}\right.\nonumber \\
&&\left.\qquad-\frac{1681}{45\,(D-2)}+\frac{1817}{420\,(D-3)}+\frac{12096842}{371875\,(5D-14)}-\frac{6605536}{365625\,(5D-16)}-\frac{57247}{28125\,(5D-18)}\right.\nonumber \\
&&\left.\qquad+\frac{111166}{21875\,(5D-22)}+\frac{32}{3\,(D-2)^2}+\frac{5}{12\,(D-3)^2}+\frac{5654087}{300000}\right)\nonumber \\
&&+A_{7,5}
\left(+\frac{313\,D^2}{1000}-\frac{10751\,D}{3000}-\frac{19}{2\,(2D-5)}+\frac{1807}{448\,(2D-7)}-\frac{3947149}{636480\,(2D-9)}\right.\nonumber \\
&&\left.\qquad-\frac{1681}{45\,(D-2)}+\frac{1817}{420\,(D-3)}+\frac{12096842}{371875\,(5D-14)}-\frac{6605536}{365625\,(5D-16)}-\frac{57247}{28125\,(5D-18)}\right.\nonumber \\
&&\left.\qquad+\frac{111166}{21875\,(5D-22)}+\frac{32}{3\,(D-2)^2}+\frac{5}{12\,(D-3)^2}+\frac{5654087}{300000}\right)\nonumber \\
&&+A_{8,1}
\left(+\frac{37071\,D}{20000}-\frac{1853}{448\,(2D-7)}+\frac{1132}{45\,(D-2)}-\frac{15}{7\,(D-3)}-\frac{368628}{21875\,(5D-14)}\right.\nonumber \\
&&\left.\qquad+\frac{202312}{28125\,(5D-16)}+\frac{31416}{3125\,(5D-18)}-\frac{1468}{13125\,(5D-22)}+\frac{239}{192\,(2D-7)^2}-\frac{1622997}{100000}\right)\nonumber \\
&&+B_{8,1}
\frac{(2\,D^3-25\,D^2+94\,D-112)\,(D^3-16\,D^2+68\,D-88)}{4\,(2D-7)\,(D-2)^2\,(2D-5)}\nonumber \\
&&+C_{8,1}
\frac{(2\,D^3-25\,D^2+94\,D-112)\,(D^3-16\,D^2+68\,D-88)}{4\,(2D-7)\,(D-2)^2\,(2D-5)}\nonumber \\
&&-A_{9,1}
\left(+\frac{81\,D^2}{250}-\frac{6939\,D}{1250}-\frac{144}{5\,(D-2)}+\frac{15}{14\,(D-3)}+\frac{49392}{3125\,(5D-14)}\right.\nonumber \\
&&\left.\qquad-\frac{5808}{3125\,(5D-16)}-\frac{12864}{3125\,(5D-18)}+\frac{1752}{4375\,(5D-22)}+\frac{175239}{6250}\right)\nonumber \\
&&+A_{9,2}
\frac{3\,(D-4)\,(3D-14)\,(15\,D^5-266\,D^4+1708\,D^3-5016\,D^2+6624\,D-2944)}{5\,(D-3)\,(5D-22)\,(5D-14)\,(5D-16)\,(5D-18)\,(D-2)}\nonumber \\
&&+A_{9,4}
\left(+\frac{243\,D^2}{16000}-\frac{1539\,D}{10000}-\frac{973}{3072\,(2D-7)}-\frac{226533}{5657600\,(2D-9)}-\frac{4}{25\,(D-2)}\right.\nonumber \\
&&\left.\qquad+\frac{2268}{10625\,(5D-14)}-\frac{14036}{40625\,(5D-16)}+\frac{3504}{3125\,(5D-18)}+\frac{2234}{9375\,(5D-22)}+\frac{108351}{320000}\right)\nonumber \\
\end{eqnarray*}
}

{\allowdisplaybreaks
\begin{eqnarray*}
\lefteqn{X^g_{C_F^2 N_F}=}\nonumber \\
&&-B_{4,1}
\left(+\frac{188119\,D^2}{125}-\frac{61815901\,D}{3750}-\frac{23257388}{5525\,(2D-9)}-\frac{280}{9\,(3D-10)}+\frac{13405743}{385\,(3D-14)}\right.\nonumber \\
&&\left.\qquad-\frac{160}{9\,(3D-8)}-\frac{23276}{25\,(D-2)}-\frac{499}{6\,(D-3)}+\frac{773374}{45\,(D-4)}-\frac{1533843736}{3346875\,(5D-14)}\right.\nonumber \\
&&\left.\qquad+\frac{4855107866}{4021875\,(5D-16)}+\frac{79707397}{9375\,(5D-18)}-\frac{2286575824}{28125\,(5D-22)}+\frac{64}{(D-2)^2}-\frac{53}{6\,(D-3)^2}\right.\nonumber \\
&&\left.\qquad+\frac{573968}{15\,(D-4)^2}+\frac{135408}{5\,(D-4)^3}+\frac{36096}{5\,(D-4)^4}+\frac{296368802}{9375}\right)\nonumber \\
&&-A_{5,1}
\left(+\frac{27082\,D^2}{125}-\frac{1697483\,D}{625}+\frac{7475589}{11050\,(2D-9)}+\frac{56}{9\,(3D-10)}-\frac{89768}{75\,(D-2)}\right.\nonumber \\
&&\left.\qquad-\frac{32}{3\,(D-3)}+\frac{282832}{135\,(D-4)}-\frac{16480952}{286875\,(5D-14)}+\frac{1500528}{40625\,(5D-16)}+\frac{2180808}{3125\,(5D-18)}\right.\nonumber \\
&&\left.\qquad-\frac{48975472}{9375\,(5D-22)}+\frac{94688}{45\,(D-4)^2}+\frac{8704}{15\,(D-4)^3}+\frac{87298177}{11250}\right)\nonumber \\
&&+A_{5,2}
\left(+\frac{21317\,D^2}{375}-\frac{1283138\,D}{1875}-\frac{4077594}{5525\,(2D-9)}-\frac{344}{27\,(3D-10)}+\frac{127310}{77\,(3D-14)}\right.\nonumber \\
&&\left.\qquad-\frac{400}{27\,(3D-8)}-\frac{46972}{225\,(D-2)}+\frac{29872}{45\,(D-4)}+\frac{141597536}{3346875\,(5D-14)}+\frac{73031344}{1340625\,(5D-16)}\right.\nonumber \\
&&\left.\qquad-\frac{1607982}{3125\,(5D-18)}-\frac{6282976}{5625\,(5D-22)}+\frac{32}{(D-2)^2}+\frac{8576}{15\,(D-4)^2}+\frac{256}{(D-4)^3}\right.\nonumber \\
&&\left.\qquad+\frac{152810662}{84375}\right)\nonumber \\
&&-B_{5,1}
\left(+\frac{7239\,D^2}{125}-\frac{446671\,D}{625}-\frac{1784}{5\,(D-2)}-\frac{11}{(D-3)}+\frac{184976}{135\,(D-4)}\right.\nonumber \\
&&\left.\qquad-\frac{2031568}{84375\,(5D-14)}+\frac{2576}{3125\,(5D-16)}-\frac{1879752}{3125\,(5D-18)}-\frac{2550912}{625\,(5D-22)}+\frac{37696}{45\,(D-4)^2}\right.\nonumber \\
&&\left.\qquad+\frac{4736}{15\,(D-4)^3}+\frac{6406811}{3125}\right)\nonumber \\
&&+B_{5,2}
\left(+\frac{11609\,D^2}{75}-\frac{726482\,D}{375}-\frac{728}{27\,(3D-10)}-\frac{400}{27\,(3D-8)}-\frac{4948}{5\,(D-2)}\right.\nonumber \\
&&\left.\qquad+\frac{55288}{27\,(D-4)}+\frac{117104}{3375\,(5D-14)}+\frac{12584}{625\,(5D-16)}-\frac{133092}{625\,(5D-18)}-\frac{515088}{125\,(5D-22)}\right.\nonumber \\
&&\left.\qquad+\frac{13280}{9\,(D-4)^2}+\frac{6272}{15\,(D-4)^3}+\frac{3837572}{675}\right)\nonumber \\
&&-A_{6,1}
\left(+\frac{4732\,D^2}{375}-\frac{300428\,D}{1875}+\frac{12731}{231\,(3D-14)}-\frac{5458}{45\,(D-2)}+\frac{2524}{45\,(D-4)}\right.\nonumber \\
&&\left.\qquad-\frac{1669664}{196875\,(5D-14)}-\frac{289948}{61875\,(5D-16)}-\frac{733}{125\,(5D-18)}-\frac{665896}{9375\,(5D-22)}+\frac{64}{3\,(D-4)^2}\right.\nonumber \\
&&\left.\qquad+\frac{4892596}{9375}\right)\nonumber \\
&&+A_{6,2}
\left(+\frac{1013\,D^2}{50}-\frac{215473\,D}{750}-\frac{226533}{22100\,(2D-9)}-\frac{154016}{225\,(D-2)}+\frac{98}{3\,(D-3)}\right.\nonumber \\
&&\left.\qquad+\frac{208}{15\,(D-4)}+\frac{1662152}{6375\,(5D-14)}-\frac{5141248}{73125\,(5D-16)}+\frac{8936}{5625\,(5D-22)}-\frac{64}{(D-2)^2}\right.\nonumber \\
&&\left.\qquad-\frac{5}{6\,(D-3)^2}+\frac{8774231}{7500}\right)\nonumber \\
&&-A_{6,3}
\left(+\frac{14331\,D^2}{250}-\frac{1856633\,D}{2500}-\frac{4757193}{44200\,(2D-9)}-\frac{60536}{75\,(D-2)}+\frac{9}{(D-3)}\right.\nonumber \\
&&\left.\qquad+\frac{4396}{15\,(D-4)}+\frac{2026304}{159375\,(5D-14)}+\frac{56236}{24375\,(5D-16)}+\frac{33752}{625\,(5D-18)}-\frac{761024}{3125\,(5D-22)}\right.\nonumber \\
&&\left.\qquad+\frac{512}{5\,(D-4)^2}+\frac{64132247}{25000}\right)\nonumber \\
&&-A_{7,1}
\frac{16\,(D-3)\,(D^2-7\,D+16)\,(5\,D^3-62\,D^2+236\,D-288)}{(D-2)\,(5D-14)\,(5D-16)\,(5D-18)}\nonumber \\
&&+A_{7,2}
\left(+\frac{49\,D^2}{40}-\frac{3181\,D}{200}-\frac{71585}{4032\,(2D-7)}-\frac{226533}{70720\,(2D-9)}-\frac{1556}{45\,(D-2)}\right.\nonumber \\
&&\left.\qquad+\frac{215664}{14875\,(5D-14)}+\frac{2752}{2925\,(5D-16)}+\frac{6548}{125\,(5D-18)}+\frac{50413}{800}\right)\nonumber \\
&&+A_{7,3}
\left(+\frac{419\,D^2}{125}-\frac{33136\,D}{625}-\frac{328}{(D-2)}+\frac{1126}{105\,(D-3)}+\frac{1169616}{3125\,(5D-14)}\right.\nonumber \\
&&\left.\qquad+\frac{60416}{3125\,(5D-16)}-\frac{16616}{9375\,(5D-18)}-\frac{3504}{4375\,(5D-22)}-\frac{7}{(D-3)^2}+\frac{787461}{3125}\right)\nonumber \\
&&-A_{7,4}
\left(+\frac{271\,D^2}{500}-\frac{48887\,D}{7500}+\frac{3197}{2016\,(2D-7)}-\frac{75511}{35360\,(2D-9)}-\frac{622}{45\,(D-2)}\right.\nonumber \\
&&\left.\qquad+\frac{3}{(D-3)}-\frac{694636}{74375\,(5D-14)}-\frac{2925056}{365625\,(5D-16)}-\frac{1314}{3125\,(5D-18)}+\frac{49148}{28125\,(5D-22)}\right.\nonumber \\
&&\left.\qquad-\frac{1}{3\,(D-3)^2}+\frac{47263}{2000}\right)\nonumber \\
&&-A_{7,5}
\left(+\frac{271\,D^2}{500}-\frac{48887\,D}{7500}+\frac{3197}{2016\,(2D-7)}-\frac{75511}{35360\,(2D-9)}-\frac{622}{45\,(D-2)}\right.\nonumber \\
&&\left.\qquad+\frac{3}{(D-3)}-\frac{694636}{74375\,(5D-14)}-\frac{2925056}{365625\,(5D-16)}-\frac{1314}{3125\,(5D-18)}+\frac{49148}{28125\,(5D-22)}\right.\nonumber \\
&&\left.\qquad-\frac{1}{3\,(D-3)^2}+\frac{47263}{2000}\right)\nonumber \\
&&+A_{9,1}
\left(+\frac{81\,D^2}{250}-\frac{6939\,D}{1250}-\frac{144}{5\,(D-2)}+\frac{15}{14\,(D-3)}+\frac{49392}{3125\,(5D-14)}\right.\nonumber \\
&&\left.\qquad-\frac{5808}{3125\,(5D-16)}-\frac{12864}{3125\,(5D-18)}+\frac{1752}{4375\,(5D-22)}+\frac{175239}{6250}\right)\nonumber \\
&&-A_{9,4}
\left(+\frac{81\,D^2}{8000}-\frac{513\,D}{5000}-\frac{973}{4608\,(2D-7)}-\frac{75511}{2828800\,(2D-9)}-\frac{8}{75\,(D-2)}\right.\nonumber \\
&&\left.\qquad+\frac{1512}{10625\,(5D-14)}-\frac{28072}{121875\,(5D-16)}+\frac{2336}{3125\,(5D-18)}+\frac{4468}{28125\,(5D-22)}+\frac{36117}{160000}\right)\nonumber \\
\end{eqnarray*}
}

{\allowdisplaybreaks
\begin{eqnarray*}
\lefteqn{X^g_{C_A N_F^2}=}\nonumber \\
&&+B_{4,1}
\left(+\frac{968\,D}{25}+\frac{715}{21\,(3D-10)}-\frac{1693529}{162162\,(D-1)}-\frac{55}{3\,(D-2)}-\frac{37}{6\,(D-3)}\right.\nonumber \\
&&\left.\qquad-\frac{12496}{81\,(D-4)}+\frac{12056}{3375\,(5D-14)}+\frac{20832}{1375\,(5D-16)}+\frac{299376}{1625\,(5D-18)}+\frac{9217}{108\,(D-1)^2}\right.\nonumber \\
&&\left.\qquad+\frac{4}{(D-2)^2}+\frac{1}{4\,(D-3)^2}-\frac{1504}{9\,(D-4)^2}-\frac{896}{9\,(D-4)^3}-\frac{46756}{375}\right)\nonumber \\
&&-A_{5,2}
\left(+\frac{238\,D}{75}-\frac{68}{63\,(3D-10)}-\frac{2806094}{81081\,(D-1)}+\frac{80}{3\,(D-2)}-\frac{256}{81\,(D-4)}\right.\nonumber \\
&&\left.\qquad+\frac{21296}{10125\,(5D-14)}-\frac{2496}{1375\,(5D-16)}-\frac{23328}{1625\,(5D-18)}-\frac{994}{27\,(D-1)^2}-\frac{8}{(D-2)^2}\right.\nonumber \\
&&\left.\qquad-\frac{256}{27\,(D-4)^2}-\frac{4096}{1125}\right)\nonumber \\
&&+B_{5,2}
\left(+\frac{154\,D}{75}-\frac{32}{63\,(3D-10)}-\frac{278326}{81081\,(D-1)}-\frac{128}{81\,(D-4)}-\frac{17072}{10125\,(5D-14)}\right.\nonumber \\
&&\left.\qquad-\frac{2208}{1375\,(5D-16)}-\frac{11664}{1625\,(5D-18)}+\frac{502}{27\,(D-1)^2}-\frac{128}{27\,(D-4)^2}-\frac{10408}{1125}\right)\nonumber \\
&&+A_{6,1}
\frac{(5D-16)\,(D-4)}{3\,(D-1)^2\,(D-2)^2}\nonumber \\
&&-A_{7,1}
\frac{8\,(D-3)\,(D-4)\,(5\,D^3-62\,D^2+236\,D-288)}{(D-1)\,(5D-14)\,(5D-16)\,(5D-18)}\nonumber \\
\end{eqnarray*}
}

{\allowdisplaybreaks
\begin{eqnarray*}
\lefteqn{X^g_{C_F N_F^2}=}\nonumber \\
&&-B_{4,1}
\left(+\frac{2336\,D}{25}-\frac{1360}{21\,(3D-10)}-\frac{64}{15\,(3D-8)}+\frac{20262752}{135135\,(D-1)}-\frac{4256}{9\,(D-4)}\right.\nonumber \\
&&\left.\qquad+\frac{24112}{3375\,(5D-14)}+\frac{41664}{1375\,(5D-16)}+\frac{598752}{1625\,(5D-18)}-\frac{576}{(D-4)^2}-\frac{256}{(D-4)^3}\right.\nonumber \\
&&\left.\qquad-\frac{158512}{375}\right)\nonumber \\
&&+A_{5,2}
\left(+\frac{1976\,D}{75}+\frac{160}{63\,(3D-10)}-\frac{32}{9\,(3D-8)}+\frac{4378936}{81081\,(D-1)}-\frac{1024}{27\,(D-4)}\right.\nonumber \\
&&\left.\qquad+\frac{42592}{10125\,(5D-14)}-\frac{4992}{1375\,(5D-16)}-\frac{46656}{1625\,(5D-18)}-\frac{256}{9\,(D-4)^2}-\frac{49064}{375}\right)\nonumber \\
&&-B_{5,2}
\left(+\frac{136\,D}{25}+\frac{160}{63\,(3D-10)}+\frac{32}{9\,(3D-8)}+\frac{1481384}{81081\,(D-1)}-\frac{512}{27\,(D-4)}\right.\nonumber \\
&&\left.\qquad-\frac{34144}{10125\,(5D-14)}-\frac{4416}{1375\,(5D-16)}-\frac{23328}{1625\,(5D-18)}-\frac{128}{9\,(D-4)^2}-\frac{35816}{1125}\right)\nonumber \\
&&+A_{7,1}
\frac{16\,(D-3)\,(D-4)\,(5\,D^3-62\,D^2+236\,D-288)}{(D-1)\,(5D-14)\,(5D-16)\,(5D-18)}\nonumber \\
\end{eqnarray*}
}

\end{appendix}

\end{document}